\documentclass[12pt,preprint]{aastex}
\usepackage{natbib}

\begin{document}
\title{The Stellar Population of h and $\chi$ Persei: \\Cluster Properties, Membership, and the
Intrinsic Colors and Temperatures of Stars}
\author{Thayne Currie\altaffilmark{1,7}, Jesus Hernandez\altaffilmark{2}, Jonathan Irwin\altaffilmark{1,3}
Scott J. Kenyon\altaffilmark{1}, Susan Tokarz\altaffilmark{1}, Zoltan Balog\altaffilmark{4},   
Ann Bragg\altaffilmark{5},Perry Berlind\altaffilmark{6}, and Mike Calkins\altaffilmark{6}}
\altaffiltext{1}{Harvard-Smithsonian Center for Astrophysics, 60 Garden St. Cambridge, MA 02140}
\altaffiltext{2}{Centro de Investigaciones de Astronomica (CIDA), Apdo. Postal 264, Merida 5101-A, Venezuela}
\altaffiltext{3}{Institute of Astronomy, University of Cambridge}
\altaffiltext{4}{MPIA-Heidelberg}
\altaffiltext{5}{Department of Physics, Marietta College}
\altaffiltext{6}{Fred Lawrence Whipple Observatory}
\altaffiltext{7}{NASA-Goddard Space Flight Center}
\email{tcurrie@cfa.harvard.edu, jhernan@umich.edu, jirwin@cfa.harvard.edu}
\begin{abstract}
From photometric observations of $\sim$ 47,000 stars and spectroscopy of $\sim$ 11,000 stars, 
we describe the first extensive study of the stellar population of the famous Double Cluster, 
h and $\chi$ Persei, down to subsolar masses.  By analyzing optical spectra and optical/infrared photometry, we constrain 
the distance moduli (dM), reddening (E(B-V)), and ages for h Persei, $\chi$ Persei and the low-density halo population 
surrounding both cluster cores.  With the exception of mass and spatial distribution, the clusters are nearly identical in 
every measurable way.   Both clusters have E(B-V) $\sim$ 0.52--0.55 and dM = 11.8--11.85; 
the halo population, while more poorly constrained, likely has identical properties.  
As determined from the main sequence turnoff, the luminosity of M supergiants, and pre-main sequence 
isochrones, ages for h Persei, $\chi$ Persei and the halo population 
 all converge on $\approx$ 14 Myr, thus showing a stunning agreement between estimates based on entirely 
different physics.   From these data, we establish the first spectroscopic and photometric membership lists 
of cluster stars down to early/mid M dwarfs.  At minimum, there are
$\sim$ 5,000 members within 10' of the cluster centers, while the entire h and $\chi$ Persei region 
has at least $\sim$ 13,000 and as many as 20,000 members.  The Double Cluster contains 
$\approx$ 8,400 M$_{\odot}$ of stars within 10' of the cluster centers.  
We estimate a total mass of at least 20,000 M$_{\odot}$.  We conclude our study by outlining 
outstanding questions regarding the past and present properties of h and $\chi$ Persei.  
From comparing recent work, we compile a list of intrinsic colors and derive a new
effective temperature scale for O--M dwarfs, giants, and supergiants.
\end{abstract}
\section{Introduction}
Known and studied for the past two centuries and perhaps since ancient times, the 
Double Cluster -- h and $\chi$ Persei -- presents a rare opportunity to 
precisely study multiple stages of stellar evolution.  The Double Cluster is located in the Perseus spiral arm and 
contains an exceptionally high density of evolved stars -- red supergiants and B giants/supergiants -- 
and early B-type dwarfs, indicating that the cluster is young, $\lesssim$ 30 Myr old \citep[cf.][]{Humphreys1978}.  This huge 
sample of high-mass stars probes post-main sequence stellar evolution, specifically the main sequence turnoff and 
the evolution of M supergiants.  For any reasonable initial mass function (e.g. Miller-Scalo), 
h and $\chi$ Persei must also have thousands of solar/subsolar-mass members, which allow probes 
of pre-main sequence evolution.

Derived post main sequence and pre-main sequence \textit{ages} for stars in h and $\chi$ Persei
are an acid test for stellar evolution models and critically 
affect conclusions about planet formation.  Because all stars in Gyr-old clusters have 
reached the main sequence, their ages are derived from comparing the main sequence turnoff 
and the luminosities of giants/supergiants with predictions from post-main sequence isochrones.
In young clusters, high-mass stars are typically too few in number to derive post-main sequence
 ages.  Thus, ages for young, $\lesssim$ 50 Myr-old clusters are almost always derived 
from fitting pre-main sequence isochrones to color-magnitude diagrams \citep[e.g.][]{Mayne07}.
  However, if post and pre-main sequence age estimates for young, populous 
clusters significantly disagree, it is not clear that either estimate for any cluster should be considered 
accurate.  Furthermore, the lack of a reliable absolute age calibration impedes attempts to 
use circumstellar disk population statistics for different clusters to constrain planet formation empirically
and prevents strong comparisons with solar system chronology and models 
of planet formation \citep[e.g.][]{He07, Currie2010, jcast07, KenyonBromley2009}.

An extensive study of h and $\chi$ Persei from early type, high-mass stars to late type, low-mass stars 
also addresses issues specific to the Double Cluster.  Chief among these issues is membership: 
 formal lists are currently limited to B type dwarfs, giants, and supergiants; 
A--K supergiants; some stars showing evidence for circumstellar gas accretion or luminous debris disk emission; 
and x-ray luminous stars \citep[][]{Uribe2002, Sl02, Bk05, Cu07b, Cu07c, Cu08a, CurrieEvans2009}.  A more uniform 
membership list would yield better estimates for the cluster mass function, constrain the spatial distribution of 
stars, and provide a reliable census from which to constrain models of star and planet formation.

In this paper, we describe the first exhaustive photometric and spectroscopic survey of h and $\chi$ Persei down to 
subsolar masses.  Comprised of optical photometry for $\sim$ 47,000 stars and spectra for $\sim$ 11,000, our survey 
strongly constrains cluster properties -- reddening, distance, age, and structure -- and produces the first membership list 
for the Double Cluster that includes main sequence and pre-main sequence stars.  \S 2 describes our data acquisition, 
image processing, and basic photometry/spectroscopic analysis.  \S 3 yields distance and pre/post-main sequence 
age estimates for each component of the Double Cluster: the h Persei core, the $\chi$ Persei core, 
and the low-density halo population surrounding both cores.  In \S 4, we use the results of previous sections to
identify h and $\chi$ Persei members based on spectroscopy and identify probable members based on photometry.  
With a membership catalog, we investigate other cluster properties in \S 5 such as the spatial distribution of 
members, the cluster mass function, and mass segregation.  In \S 6, we summarize our results,
 discuss outstanding questions regarding h and $\chi$ Per, and suggest fruitful future research 
programs.  The Appendix lists a new effective temperature scale and intrinsic colors for O--M dwarfs, 
giants, and supergiants.
\section{Data}
\subsection{Optical Photometry}
\subsubsection{Observations, Image Processing, and Photometry}
 Optical VI photometry of h and $\chi$ Persei were taken with the Mosaic Imager at the
4-meter Mayall telescope at the Kitt Peak National Observatory on October 13-16 and
27-30, 2006, as a part of the MONITOR project \citep{Ai07}.
Observing conditions were photometric with $\sim$ 0.9"--1.0" natural seeing in V and I.  
Exposures in Harris V band and Sloan i' band 
were taken using 2, 75, and 300 (2 and 75) second
integrations with a 36'x36' field of view centered on 
2$^{h}$18$^{m}$ 55.9$^{s}$, 57$^{o}$8'25" for h Persei and 
2$^{h}$22$^{m}$5.8$^{s}$, 57$^{o}$ 8'43" for $\chi$ Persei.  
These positions are comparable to the cluster centers determined by \citet{Bk05}, which 
we will adopt in this paper: h Persei$_{center}$ = 2$^{h}$18$^{m}$ 56.4$^{s}$, 57$^{o}$8'25", $\chi$ Persei$_{center}$ = 
2$^{h}$22$^{m}$4.3$^{s}$, 57$^{o}$ 8'35".  Because the clusters are separated by $\approx$ 27 arc-minutes, 
the total effective area of coverage is slightly smaller, $\approx$ 0.6 square degrees.  
The 75-second i' band exposures were repeated 24 times, yielding 
a total integration time of 1800s, and combined together into an averaged frame 
with outlier rejection to match the i' band sensitivity to V band.  
The data were reduced using the pipeline for the INT wide-field survey \citep{Ir01, Ir07}, corrected for the effects of
cross-talk, bias subtracted, and flatfielded  using median-combined twilight flats.

For source detection, we used the crowded field algorithm of \citet{Irwin1985}, requiring 
a minimum of 4 connected pixels lying each 1.5 $\sigma$ above the sky.
  Photometry was performed following the method of \citet{Ir07}.  
Aperture photometry is used for unblended sources;  a series of 
aperture radii ranging from $\sim$ 1" to 2" are used to compute the source flux.
Aperture corrections were derived from a curve-of-growth analysis as described 
in \citet{Ir01}.  Photometry for blended sources (sources with overlapping isophotes) 
is done by simultaneously fitting circular top-hat functions to the overlapping sources. 
  Background estimation is made by computing the 
background level binned in a 64x64 pixel grid using a robust background estimator in each bin, which is then 
filtered using 2D bilinear and median filters as described in \citet{Irwin1985} and
\citet{Ir07} and references therein.

Initial photometric calibration was made by observing Landolt standard 
stars before and after h and $\chi$ Per observations.  
Our method for converting from instrumental to 
standard magnitudes in the Johnson-Cousins system follows that of \citet{Mayne07}.
As an initial guess for color transformations,
we chose the V and $\textit{i}$ color transformations for the 
Mosaice II imager, which has an identical Sloan \textit{i'} filter and 
similar V-band filter.  To fine-tune the color transformations, we compared photometry 
taken with M34, observed immediately after h and $\chi$ Per, 
with published VI$_{c}$ M34 photometry taken at the Isaac Newton Telescope 
\citep{Irwin06} and VI photometry of h and $\chi$ Persei from \citet{Mayne07}.  
The transformations from \textit{$i_{instr.}$} to I were identical to 
those for the Mosaic II imager; those for \textit{$V_{instr.}$} to V differed 
slightly, producing a change in V-I of $\approx$ 0.5--5\% for stars with 
V-I $\sim$ -1--3.  Comparisons between our data and both the \citet{Mayne07} 
h and $\chi$ Per data and the \citet{Irwin06} M34 data fail to identify any 
residual color-dependent offsets indicative of errors in the Harris/Sloan to 
Johnson-Cousins color transformations for either set.

To identify and correct $\sim$ 1\% level chip-to-chip offsets,
two of us (J.I. and T.C.) cross-correlated photometry for a subsample of 11,000
stars 
 as a function of chip position and exposure time with deep VI photometry 
of h and $\chi$ Persei stars from \citet{Mayne07} with 5--10$\sigma$ detections.  
While their photometry is slightly shallower than ours (V(5$\sigma$) $\sim$ 22), 
their photometry draws from CCD observations with multiple exposure times using
a larger chip area, which allows us to identify small zero-point uncertainties 
in each exposure time and any systematic offsets due to chip-to-chip 
variations.  The chip-to-chip level offsets were identified and corrected for 
using a reduced $\chi$$^{2}$ fit.

\subsubsection{Saturation and Completeness}
The initial photometric catalog has roughly
 52,000 candidate sources detected in at least 
one filter for at least one exposure time.  
To remove sources with bad photometry, we check each star 
for blending, check stars for contamination from image artifacts (e.g. diffraction spikes 
from nearby bright stars), and identify stars with point-spread functions 
indicative of saturation.  By comparing 
the photometry from different exposure times, we empirically 
identify a saturation limit of V $\sim$ 15 and 16.5 for t=75s and 300s exposures 
and I $\sim$ 15.5 for t=75s.  The saturation limit for the shortest V band exposures (t = 2s)
is V $\sim$ 11 (see \S 2.1.3).  Photometry for saturated 
stars were then flagged and removed from the catalog, while blends and contaminated 
stars were identified.  
Priority was given to stars lacking contamination or blending flags regardless 
of rms errors.  Otherwise, the photometry for the final catalog was 
selected from the exposure that yielded the smallest rms errors.  

The final catalog lists 47,060 stars with detections in both filters.  Of these, we use 
photometry derived from the longest exposures (300s for V band, 75s for I band) for $\sim$ 90\% of the 
detections in either filter.  Other sources are either bright stars saturated in the 
long exposures or are near bright stars and masked from view in the long exposures.  
Photometry is chosen from shorter exposures for these sources.  Table \ref{opticalcatalog} lists 
all stars detected at both V and I band.

\begin{figure}
\plottwo{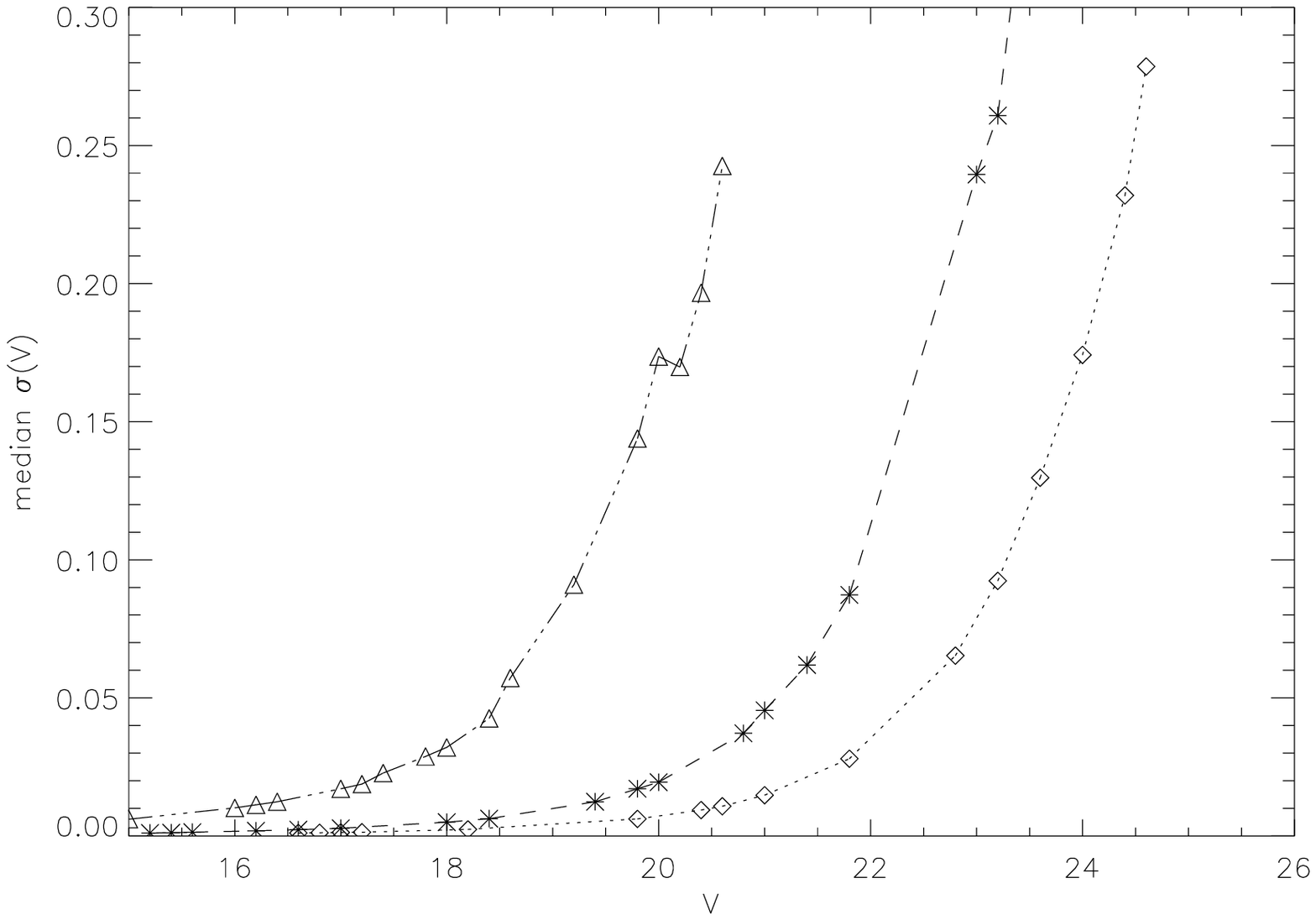}{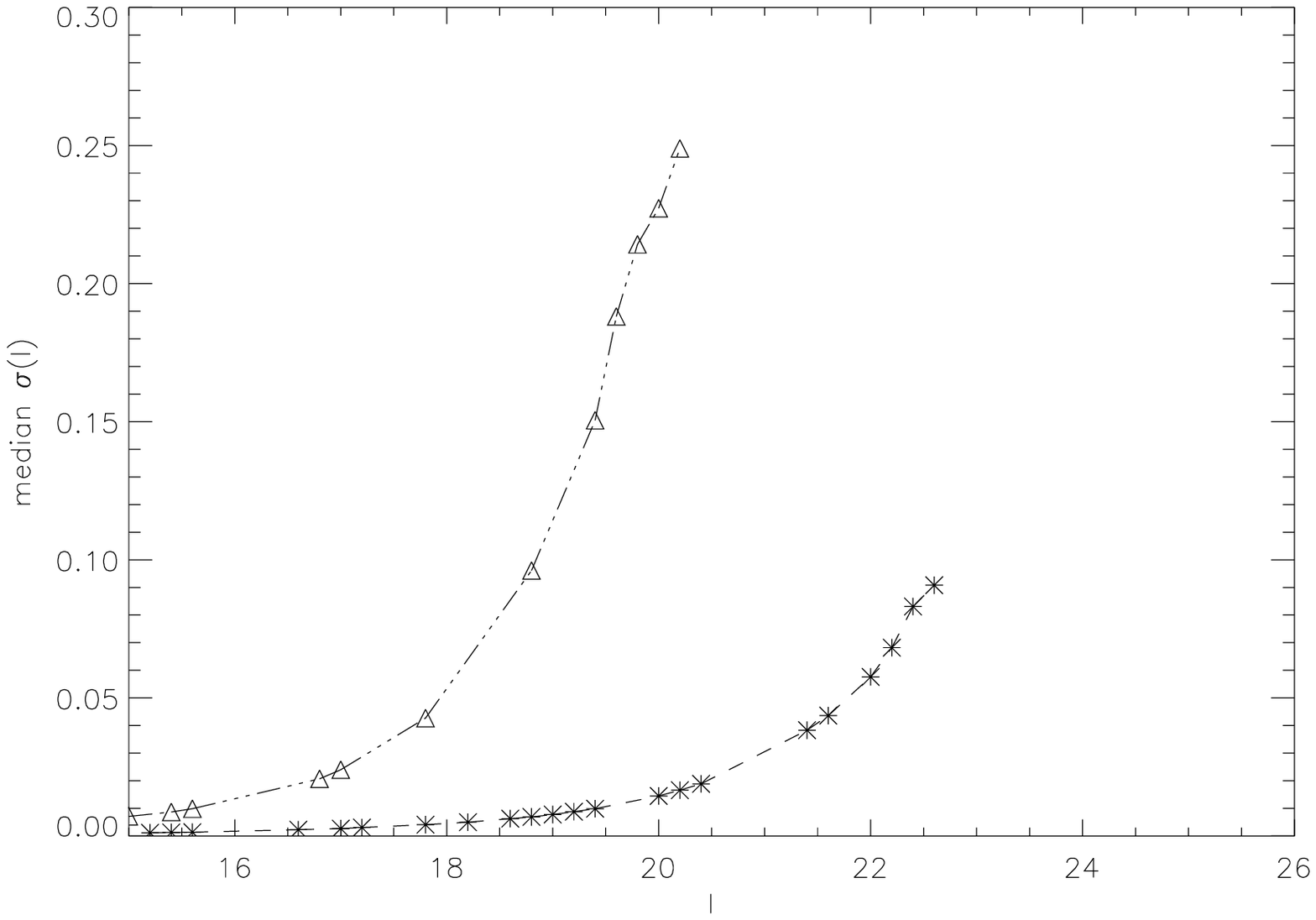}
\caption{Median V and I magnitude uncertainties as a function of V and I.  The median 
uncertainties were computed in bins of 0.2 magnitudes.}
\label{vandidist}
\end{figure}

Figure \ref{vandidist} shows the distribution of V and I magnitudes versus their 
photometric uncertainties.  There are clearly three (two) distributions of 
 V vs. $\sigma$(V) (I vs. $\sigma$(I)), which result from the three (two) 
separate exposure times used.  The 5$\sigma$ (10$\sigma$) limits in 
V band are V=21, 23, and 24 (19.5, 22, 23.25) for 2, 75, and 300s exposure times.
The corresponding limits in I band are I=19.5, 23.5 (19, 23) for 2s and 75s exposures.
The number counts in V band plateau at V $\sim$ 22--24 and fall rapidly for fainter 
magnitudes.  The number counts in I band peak at I $\sim$ 21. 

\subsubsection{Comparisons with Previous Photometric Surveys}
To provide a context for our photometry,
we cross correlate our source list with catalogs from three major 
photometric surveys: \citet{Mayne07}, \citet{Sl02}, and \citet{Ke01}.
\citet{Mayne07} is by far the deepest of the three but has a slightly smaller 
coverage area than our photometry ($\sim$ 0.45 deg$^{2}$). \citet{Sl02} is the shallowest but has 
a larger coverage ($\sim$ 1 deg$^{2}$), and \citet{Ke01} is intermediate in depth and has 
the smallest coverage ($\sim$ 0.37 deg$^{2}$).  
The \citet{Mayne07} and \citet{Ke01} catalogs include I-band photometry;
\citet{Sl02} lacks I-band photometry.

\begin{figure}
\centering
\plottwo{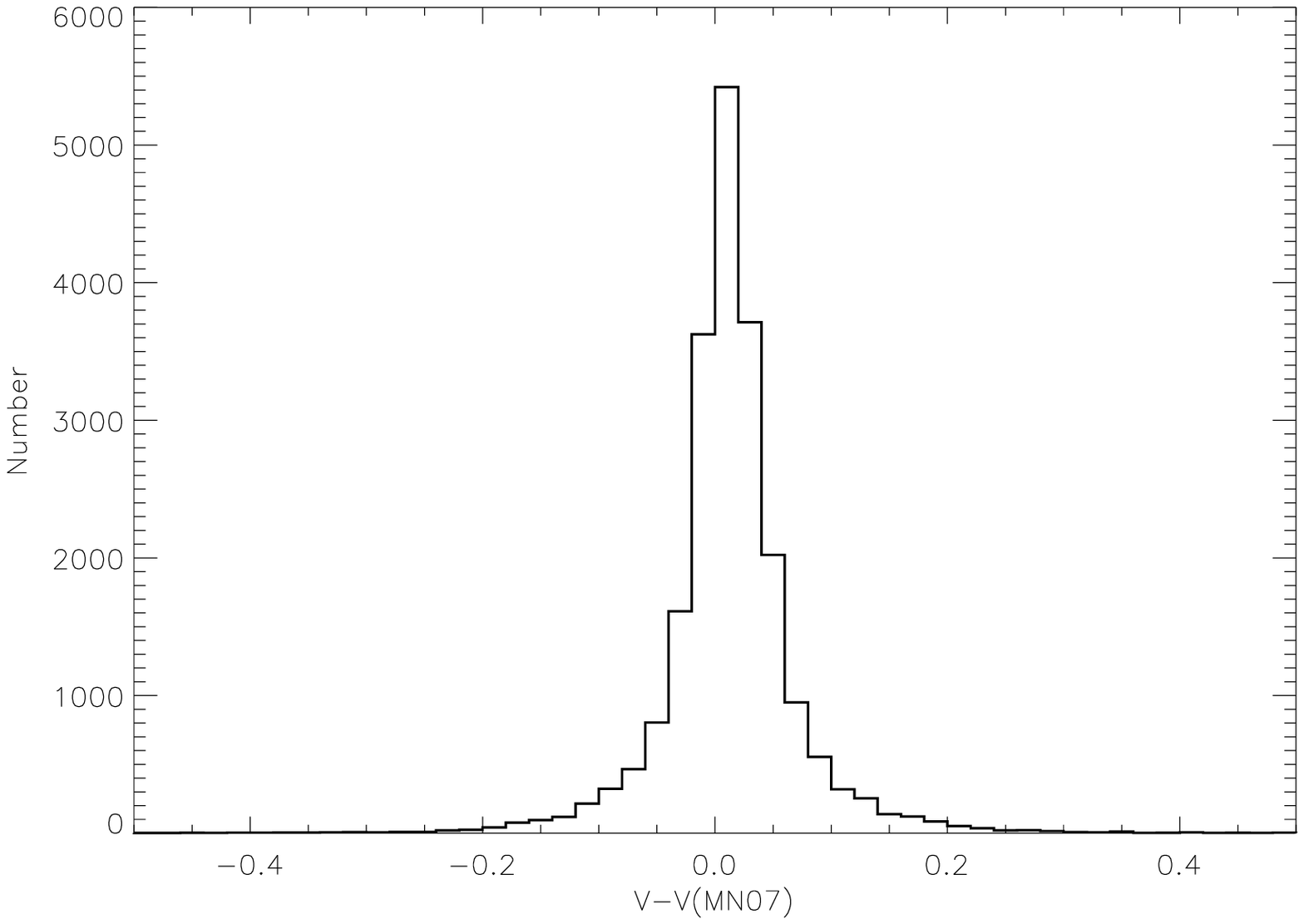}{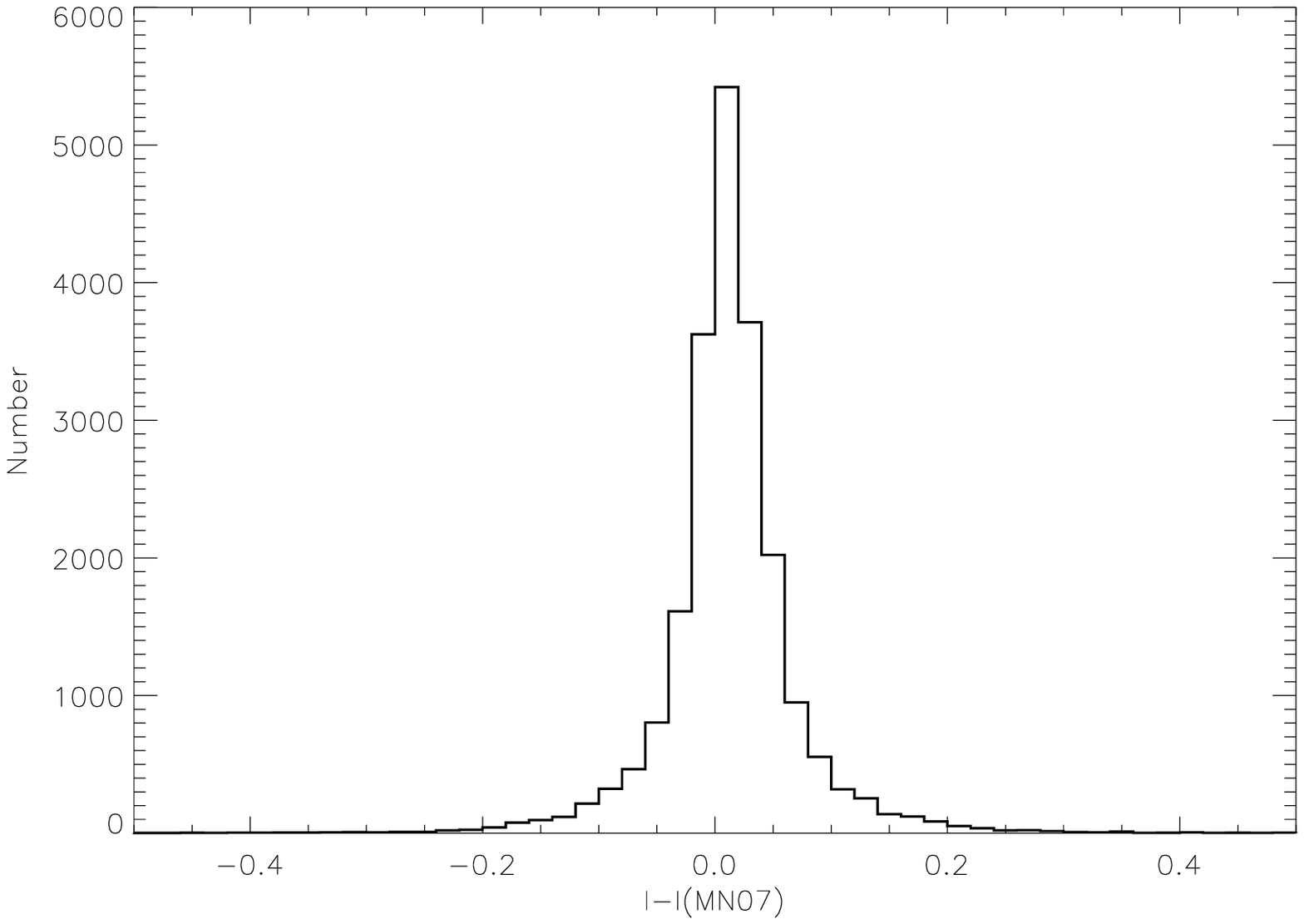}
\plottwo{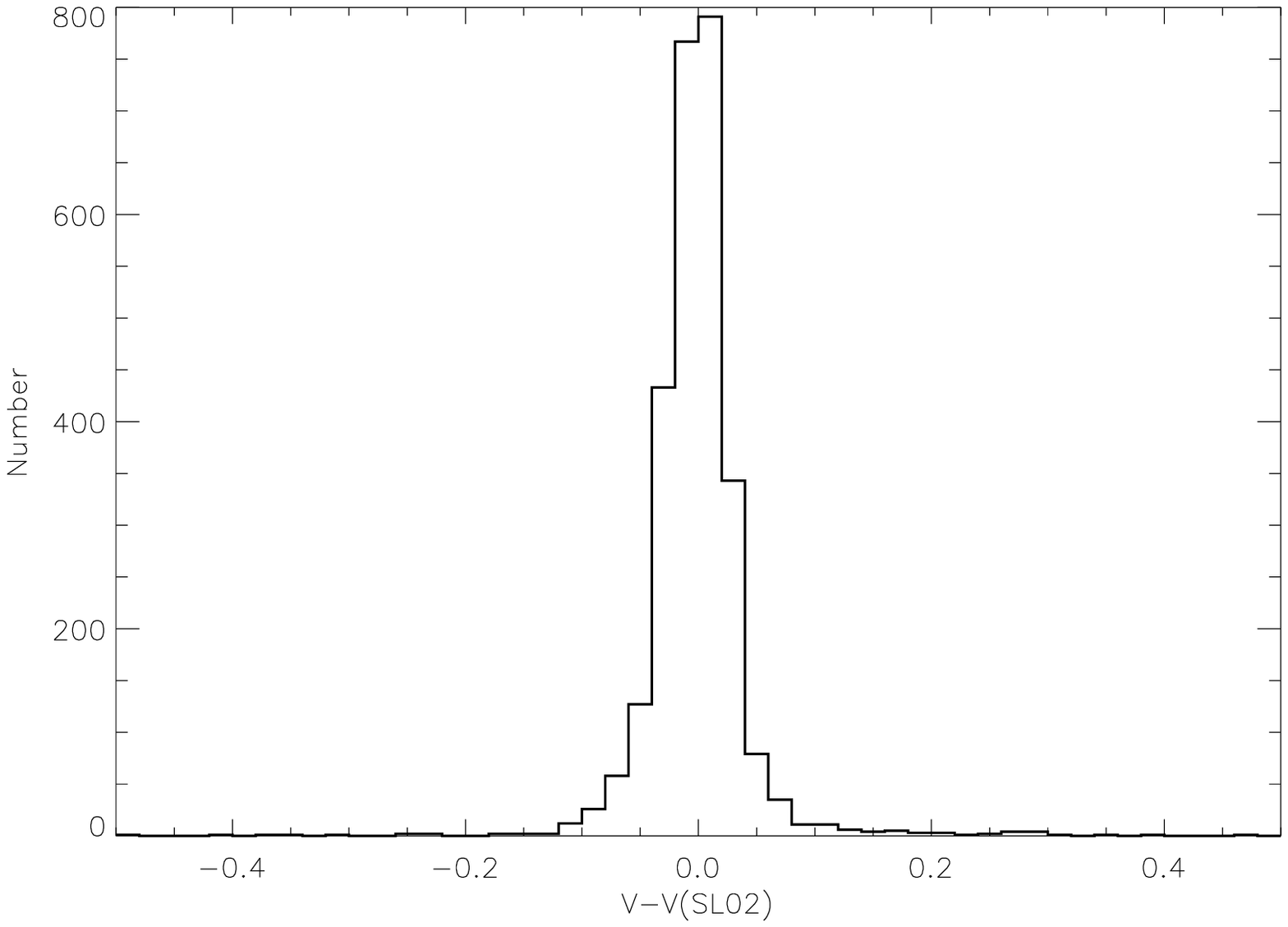}{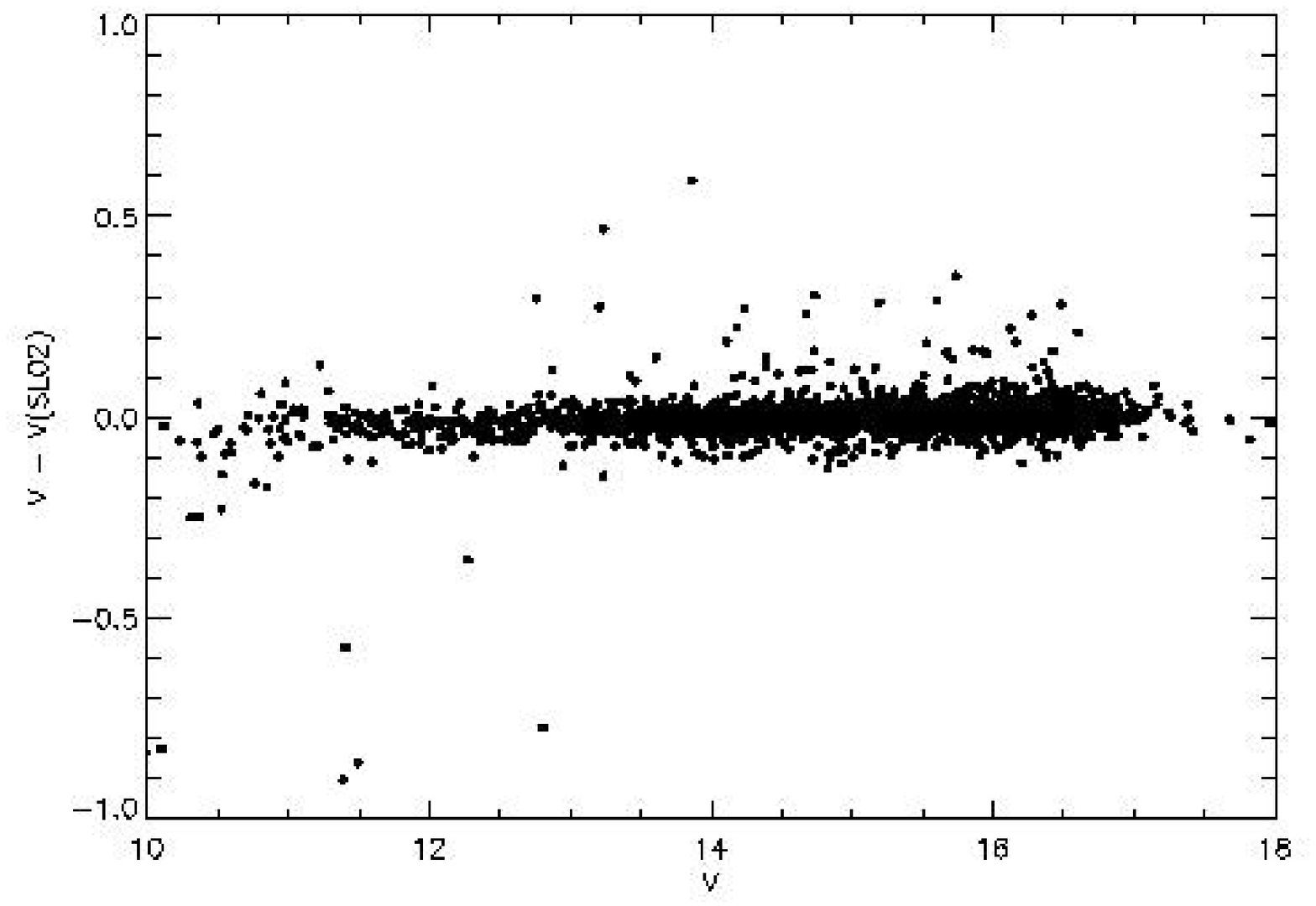}
\plottwo{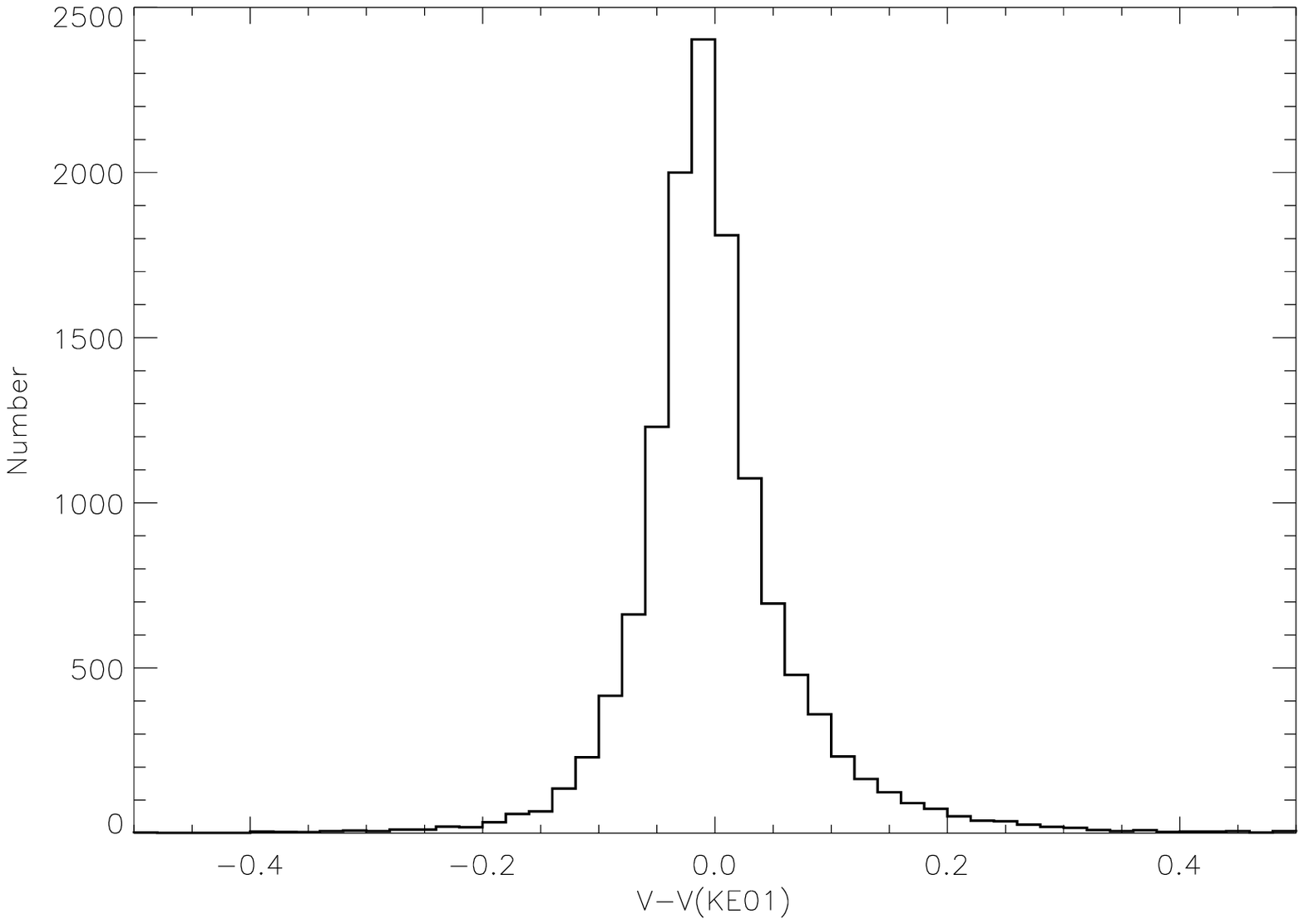}{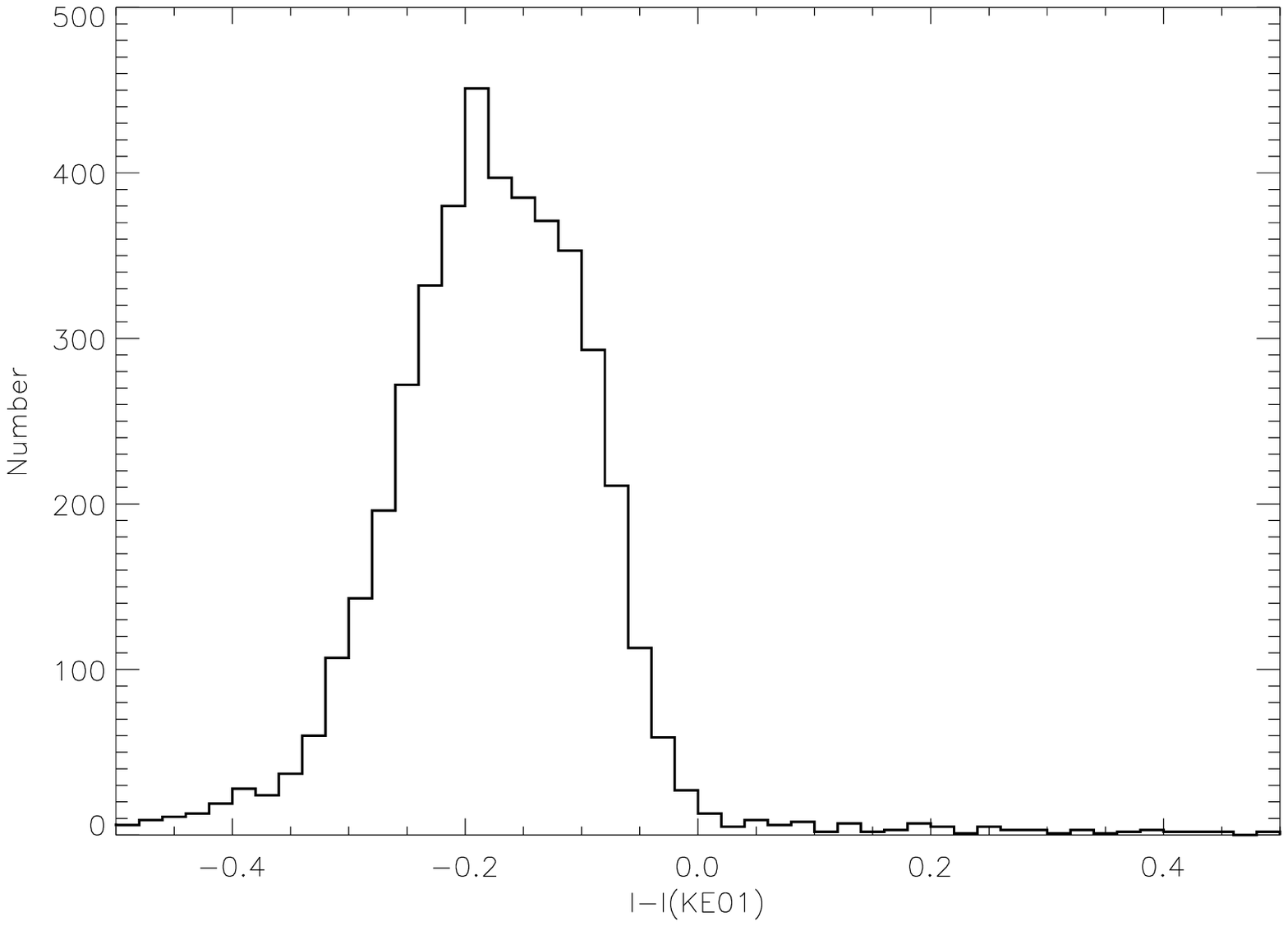}
\caption{Differences between our optical photometry and previous work by 
\citet{Mayne07}, \citet{Sl02}, and \citet{Ke01}.  (Top) Histogram plots of the 
magnitude differences in V and I band between our photometry and that from 
\citet{Mayne07}.  (Middle Row) Histogram plot (left) comparing our V band photometry 
with \citet{Sl02} and the photometric difference plotted against our V magnitudes (right).  
(Bottom) Histogram plots comparing our V and I photometry with that from \citet{Ke01}.}
\label{optphotcomp}
\end{figure}

To limiting magnitudes of V=22 and I = 19.5, our photometry 
 agrees very well with data from \citet{Mayne07}.
To make this comparison, we consider a broad range of V=11--22 where 
our photometry is unsaturated and where the \citet{Mayne07} data have
errors $\lesssim$ 0.1 mag.  For this range of V, 
the V-I colors for most stars range from $\sim$ 0 to 2.5.
Using a 1.0" matching radius, we find 13,806 common sources brighter than 
V=22.  Figure \ref{optphotcomp} (top panels) shows histogram plots of the differences in V and I magnitudes binned 
in units of 0.02 magnitudes.  The differences 
in magnitudes are consistent with gaussian distributions that are
 strongly peaked about zero with full-width half-maxima of $\approx$ 0.04 and 0.06
magnitudes for V and I.

The agreement between our V-band photometry and that from \citet{Sl02} is also excellent.  
The distribution of V magnitude differences among 1307 common sources 
has a full-width half-maximum of $\sim$ 0.04 magnitudes (Figure \ref{optphotcomp}, middle panels).
For nearly all sources with photometric differences greater than $\sim$ 0.1 magnitudes, 
we find fainter magnitudes from our photometry.  \citet{Sl02} measure stars' brightnesses using 
aperture photometry whereas as our method deblends overlapping sources.  Therefore, 
the few cases with slightly discrepant photometry were plausibly blended sources in \citet{Sl02} that were 
deblended with our photometry.

Our V-band photometry shows good agreement with that from \citet[][Figure \ref{optphotcomp}, bottom-left panel]{Ke01}.  
The distribution of magnitude differences from 5355 common sources (V $\lesssim$ 21) has a FWHM of $\sim$ 0.06 magnitudes.
  However, we find substantial disagreement 
between the \citet{Ke01} photometry and our I band photometry (Figure \ref{optphotcomp}, bottom-right panel): 
there is a $\sim$ 0.2 magnitude zero-point offset and an additional, possibly color-dependent 
term which results in a wide dispersion (FWHM $\sim$ 0.2 magnitudes).  
We speculate that this difference may arise because \citet{Ke01} used a nonstandard I-band filter or had 
larger uncertainties in their color transformations.

\subsection{Optical Spectroscopy}
To supplement the optical photometry, we acquired low-resolution 
optical spectroscopy of 2MASS-detected stars within one square-degree 
of the cluster centers.  For faint (J $\ge$ 14.25; V $\ge$ 16) stars, we used the 
 multiobject, fiber-fed spectrograph Hectospec \citep{Fa05} on the 6.5m MMT.  Brighter 
stars were observed with the fiber-fed spectrograph Hydra \citep{Ba93} on the 
3.5m WIYN telescope at Kitt Peak National Observatory and single-slit 
FAST spectrograph \citep{Fa98} on the 1.5m Tillinghast telescope at the Fred Lawrence Whipple Observatory.

\subsubsection{Spectroscopy Reduction, Survey Coverage, and Completeness}
We obtained Hectospec spectra of 9,373 stars with V $\sim$ 16-19, J$\sim$ 14.25-16.25, 
and J-H $\sim$ 0-1.5 during the Fall 2006 and Fall 2007 observing trimesters 
in queue mode.  Each source was observed in three 10-minute exposures
using the 270 mm$^{-1}$ grating.  This configuration yields spectra at
4000-9000 \AA\ with $\sim$ 6 \AA\ resolution.   The data were
processed using the standard Hectospec reduction pipeline 
\citep{Fa05} and typically have S/N $\gtrsim$ 30-50 at 5000 \AA.  

We acquired additional spectra of 610 sources with
the Hydra multifiber spectrograph on the WIYN 3.5 m telescope at the Kitt Peak National
Observatory.  Hydra spectra were obtained by A. Bragg and S. 
Kenyon during two observing runs in November 2000 and October 2001
and include stars with V $\sim$ 14-17 and J $\sim$ 12-14.5.  Exposure times ranged from 
30 minutes to 90 minutes, depending on the source brightness and seeing conditions, and 
yielded spectra with S/N $\approx$ 10-30.  
We used the 400 g mm$^{-1}$ setting blazed at 42$^{o}$, with a resolution of 7 \AA\
 and a coverage of 3600-6700 \AA.  The standard IRAF task \textit{dohydra} was
used to reduce the spectra.  We also obtained spectra of 257 bright 
(J $\le$ 14.25) stars surrounding h and $\chi$ Per with the 
FAST spectrograph on the 1.5m Tillinghast telescope at Whipple Observatory.
Finally, we added archived data of 1,025 h and $\chi$ Persei sources from \citet{Bk05} and \citet{Bragg2004}.

The combined Hectospec, FAST, and Hydra observations yield 
 11,265 spectra of stars on the h and $\chi$ Per field.  
The spatial coverage of the Hectospec, FAST, and Hydra 
observations are shown in Figure \ref{coverage}.  Hectospec and FAST 
observations cover about one square degree; Hydra observations and 
archival data focus on regions within $\approx$ 15' of the cluster centers 
(black dots).  

Our survey comprises a significant fraction of all 2MASS-detected 
sources on the field, especially those within $\sim$ 10--15' of the 
cluster centers.  Within the $\approx$ 1 square-degree area surrounding 
h and $\chi$ Persei, we obtained spectra for 75\% of all the 2MASS 
detections (11,265/14,951).  Because many bright stars with colors 
suggestive of foreground M stars were not selected for observations 
with Hydra and FAST, completeness is better at faint 
survey limits (J $\ge$ 15; 95\%) than for brighter stars (J $<$ 15; 63\%)

Completeness in the cluster-dominated regions (r $\le$ 5--15') is good 
for stars of all magnitudes.  Within 5' of both h Persei and $\chi$ Persei, we 
have spectra for nearly all stars ($\gtrsim$ 98\%).  Within 10', 
we have spectra for $>$ 80\% of bright stars (J $<$ 14), $>$ 95\% of faint (J $\ge$ 15) stars, 
and $>$ 90\% of all stars.  The majority (60\%--70\%) of bright stars at 10--15' away from the cluster 
centers are detected.  Thus, our spectroscopic survey comprises the vast majority of stars 
in the cluster-dominated regions, is unbiased within 10' of the cluster centers, and is 
not biased for faint stars regardless of radial distance from the cluster centers.

\subsubsection{Spectral Classification}
The large number of spectra observed and the large number of stars 
earlier than K/M - which can be difficult to spectral type - makes manual, 
 spectral classification inefficient. To spectral type stars,  we employ the semi-automatic 
quantitative spectral-typing code \textbf{SPTCLASS}\footnote{See 
www.astro.lsa.umich.edu/~hernandj/SPTclass/sptclass.html for more information}, 
an IRAF/IDL code based on the method described in \citet{He04}.  

\textbf{SPTCLASS} calculates the spectral types of stars using spectral indices, comparing 
  the line flux of spectral features which are 
sensitive to stellar effective temperature \citep[][and later references]{Payne1924, Payne1925}.  
Three independent spectral typing modules are included in SPTCLASS: 
indices  that  characterize early (OBA, 44 indices), intermediate 
(FG, 11 indices),  and late spectral type (KM, 16 indices) stars. 
Each index is based  on the equivalent width for each spectral feature 
which is obtained by measuring the decrease in flux due to line absorption 
from the continuum that is expected when interpolating between two adjacent 
bands. Indices measured by this procedure are largely insensitive to 
reddening as long as the wavelength coverage of each band is relatively 
small. Three spectral type estimates are then calculated by averaging 
the indices characterizing early, intermediate, and late-type stars. 
In general, visual inspection of the spectrum and the dispersion of 
individual results indicate which of these three results is the correct 
one for the star.  The errors in these estimates are computed 
from the dispersion in spectral types for individual indices weighted 
by their correlation functions.

Typically, the most important lines were Ca II (3933\AA), the G
band (4305 \AA), and Na I (5890 \AA) for mid A to mid G stars and 
the TiO bands for later stars \citep{Payne1929,Gray2009}.
The spectral types as determined from individual indices for each of 
the three modules usually showed strong agreement. In cases where 
scatter was large (e.g., greater than 2-3 subclasses), we manually 
smoothed the spectra with splot and measured the indices for highly 
correlated lines listed above as a check on the computed spectral type.

For O, B, and early A stars, the strongest lines are the Balmer lines and the He I lines \citep[e.g.][and later references]{Payne1929}.  
Because accreting pre-main sequence stars and Be stars can have Balmer line emission,  
spectral types based on the Balmer lines are produced by SPTCLASS but are not explicitly 
included in the semiautomated estimates. However, no B to mid A stars show accretion signatures; 
 the Be star population is well known \citep[e.g.,][]{Sl02,Br02,Cu08a}.  The absence of Balmer lines in the final spectral type 
determination may then introduce errors, especially for stars earlier than B5, where strengths of the 
important Mn I and Fe I lines become uncorrelated with spectral type.  Therefore, we slightly modified our approach 
for early-type stars (B1--A2, as identified by SPTCLASS), basing the spectral types solely on the Balmer indices and 
He I indices.  We selected lines that yielded a minimal scatter in the empirical HR diagram (V vs. spectral type), 
specifically H$_{\beta}$, H$_{\gamma}$, H$_{\delta}$, He I 4026 \AA, and He I 6678 \AA.  The median average of the 
spectral types determined from each of these indices was chosen as the star's spectral type; the standard deviation 
in these spectral types is identified as the uncertainty in spectral type.  While this method produced essentially 
identical results for B5--A2 stars compared to the nominal SPTCLASS calculation,  it significantly tightened 
the HR diagram locus for earlier stars, thus improving precision. 
Of the 11,265 stars with spectra, 10,934 were assigned spectral types.  Stars without 
spectral types either had extremely low signal-to-noise or were cases where the Hectospec or Hydra fibers 
failed to center on the stars, possibly because of small astrometric errors.  

We applied a final correction for the spectral types of the earliest stars by cross-correlating our 
source list with the spectroscopic survey of \citet{Sl02}.  There are 108 stars in common.  Agreement in spectral types 
determined from \citet{Sl02} and those determined from SPTCLASS is typically good, within 1 subclass. 
  However, as noted by many authors \citep{Sl02, Bk05, Schild1965, Schild1967}, h and $\chi$ 
Persei contains a large population of B supergiants, giants, and subgiants.  Differences in luminosity classes for the earliest 
stars imply differences in surface gravities.  Surface gravities affect line strengths and thus induce scatter in determining 
spectral types from these line strengths if the index-spectral type relationship is derived primarily from main sequence/pre-main sequence 
dwarfs.  Therefore, we identify all stars listed by \citet{Sl02} as having luminosity class I--IV and replaced our spectral types
 with those determined by \citet{Sl02}.  
Aside from shifting several stars initially classified as O9.5--O9.9 to slightly 
later types (B0--B1.5) \citep[see also][]{Bragg2004}, this correction yields inconsequential changes in the spectral type distribution 
and thus has a negligible impact on our later analysis.  Figure \ref{stypesequence} displays a 
spectral type sequence from our sample.

\begin{figure}
\centering
\plottwo{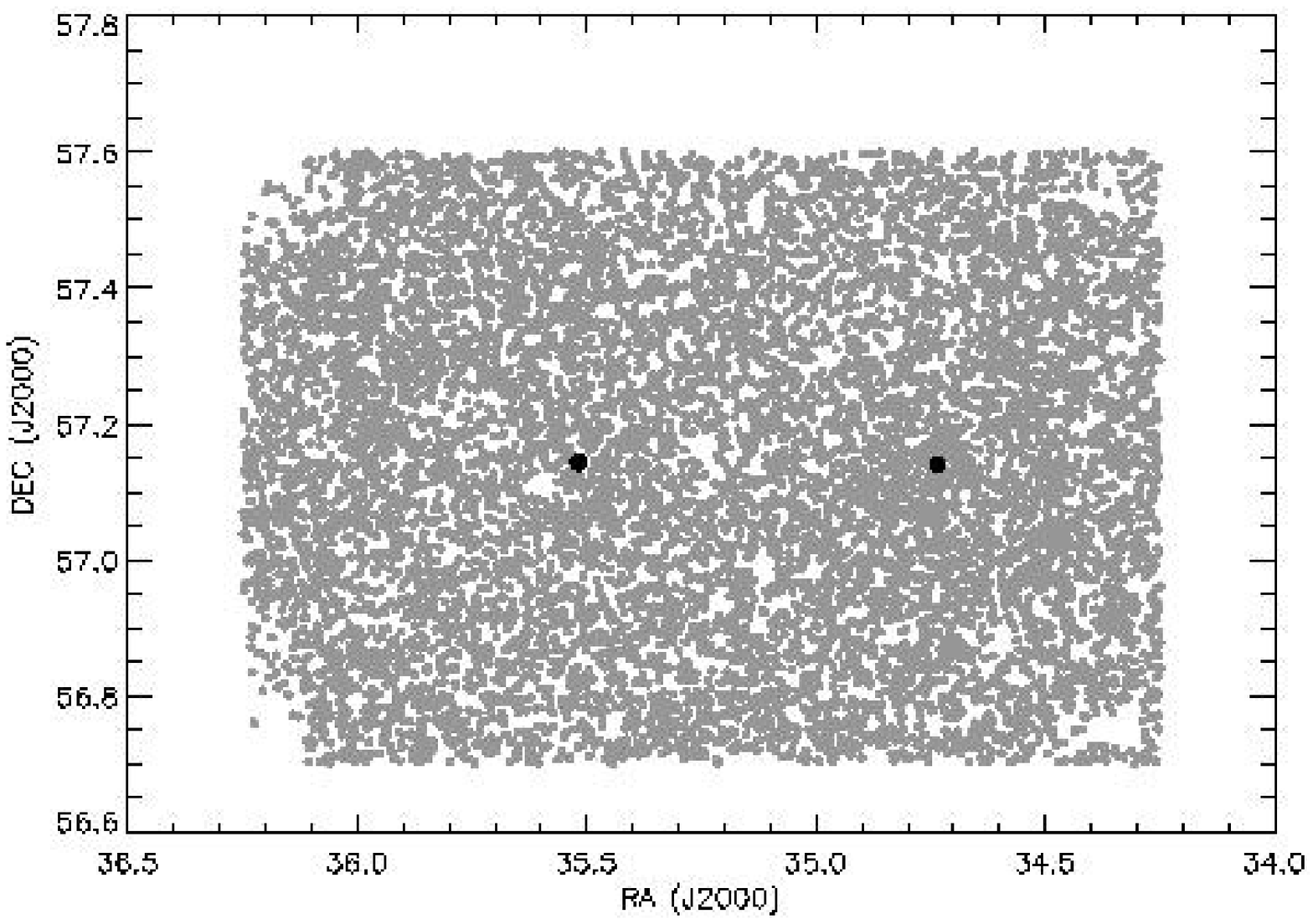}{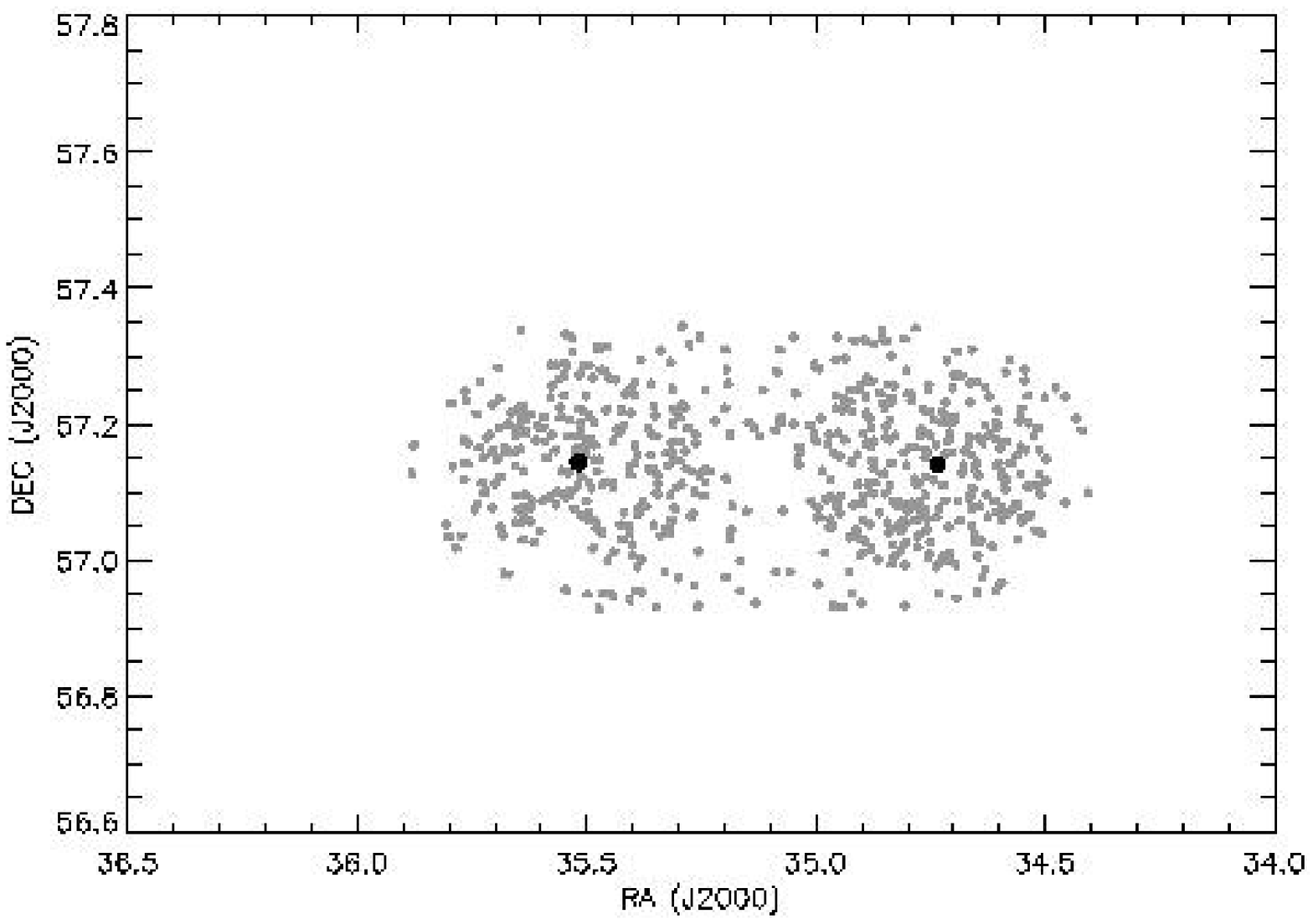}
\plottwo{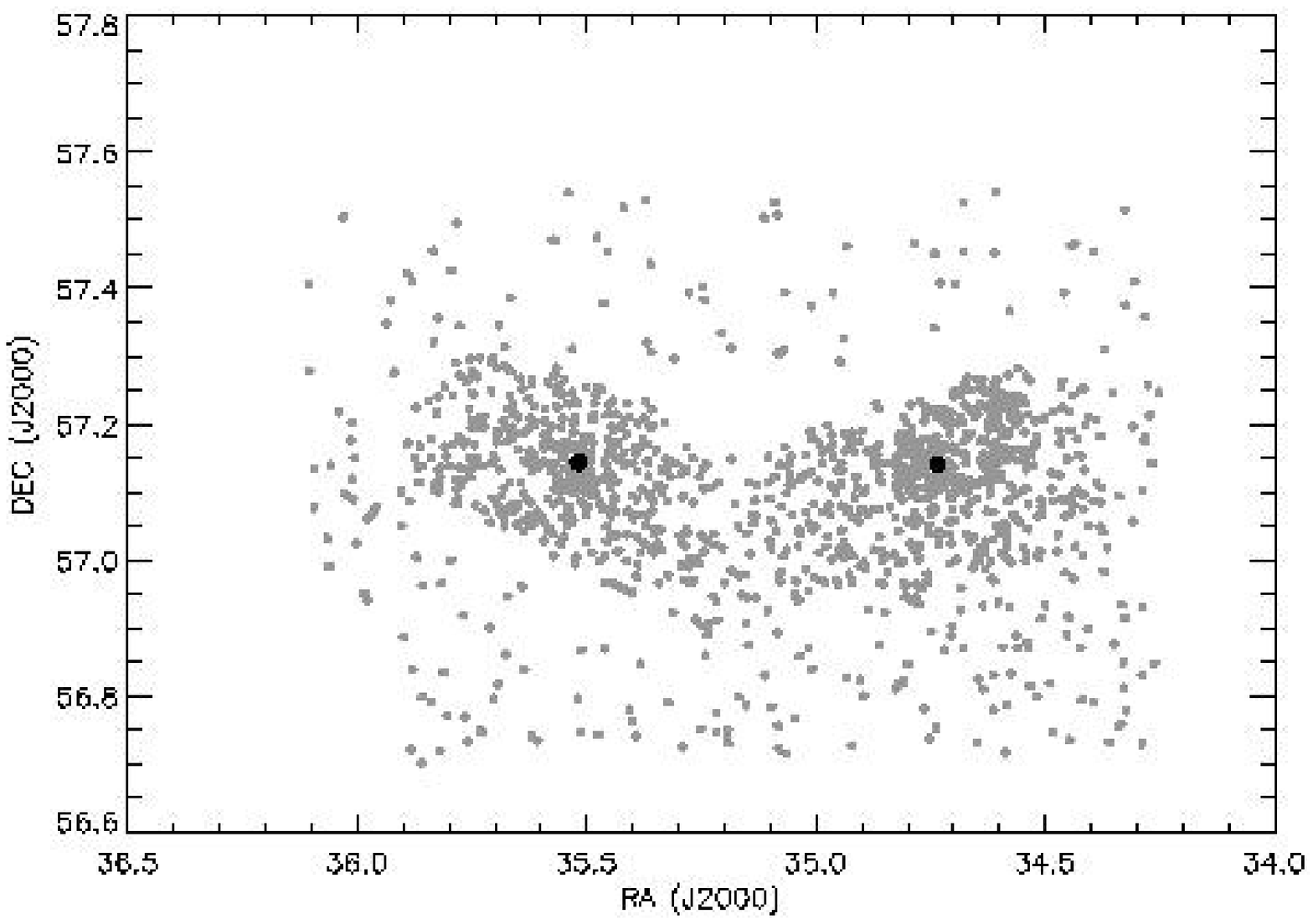}{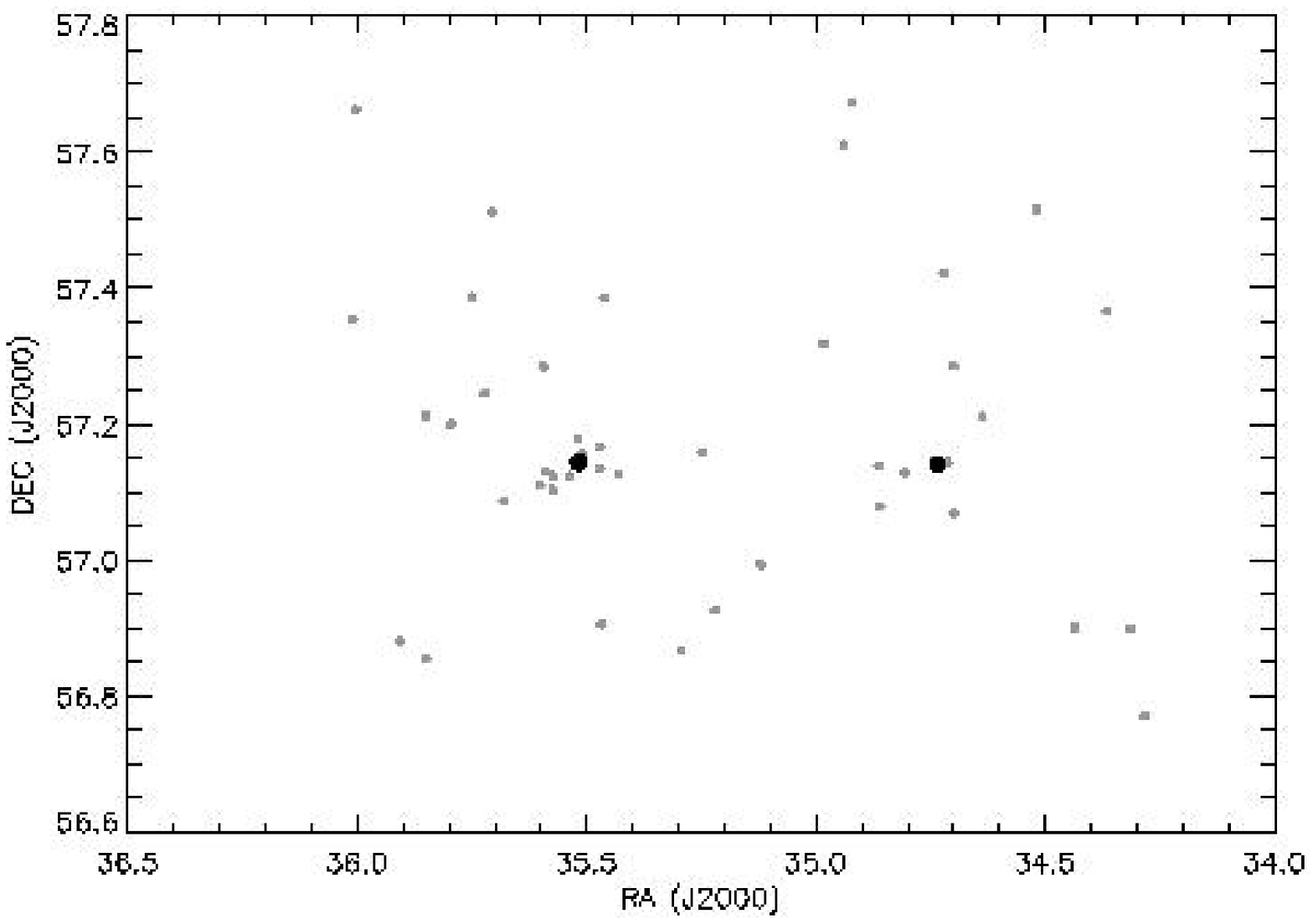}
\caption{Coverage map of spectroscopy surveys from Hectospec (top-left), FAST (bottom-left),
Hydra (top-right), and \citet{Sl02} (bottom-right).  The centers of
h Persei and $\chi$ Persei are shown as the right and left dots, respectively}
\label{coverage}
\end{figure}

\begin{figure}
\epsscale{0.75}
\plotone{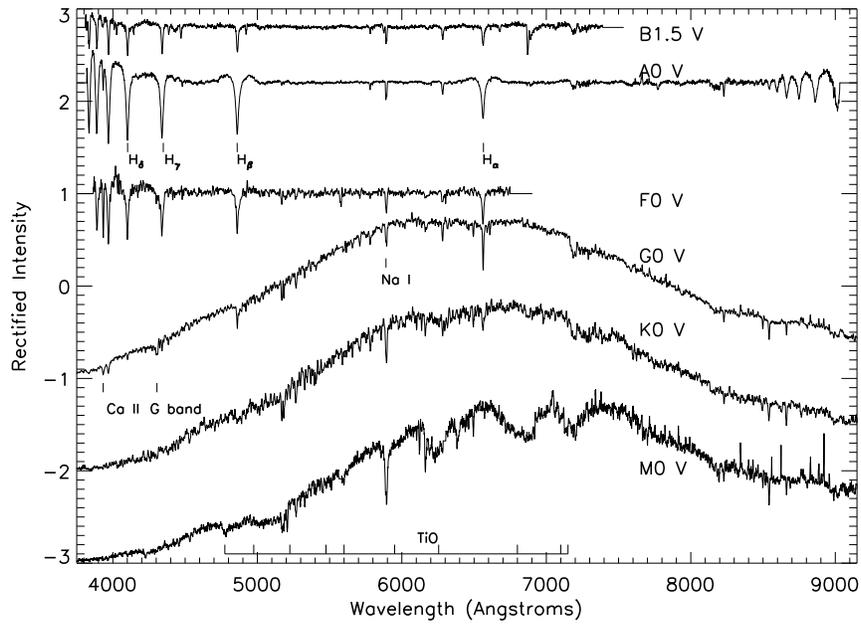}
\caption{Spectral type sequence from B1.5 V (top) to M0 (bottom) showing 
the change in line strengths with spectral type/effective temperature.  
The IDs of these stars in Table 3 are (from top to bottom) ID-82, 
ID-912, ID-2056, ID-4905, ID-6457, and ID-7180.  We follow the convention 
of \citet{Gray2009} by displaying the continuum flux of BAF stars 
as a constant, while displaying the normalized spectra of GKM stars.
}
\label{stypesequence}
\end{figure}

Finally, we added 44 stars in the \citet{Sl02} spectroscopic catalog that were not targeted with our survey (see Figure \ref{coverage} 
for spatial distribution).  These stars were predominantly optically and infrared bright B giants and supergiants, infrared bright red supergiants, 
and likely foreground dwarfs.   Spectral type uncertainties for these 44 stars were set to two subclasses, consistent with the 
maximum dispersion in spectral types for early-type stars implied by comparing the \citet{Sl02} results to SPTCLASS and previous literature measurements.

\begin{figure}
\plotone{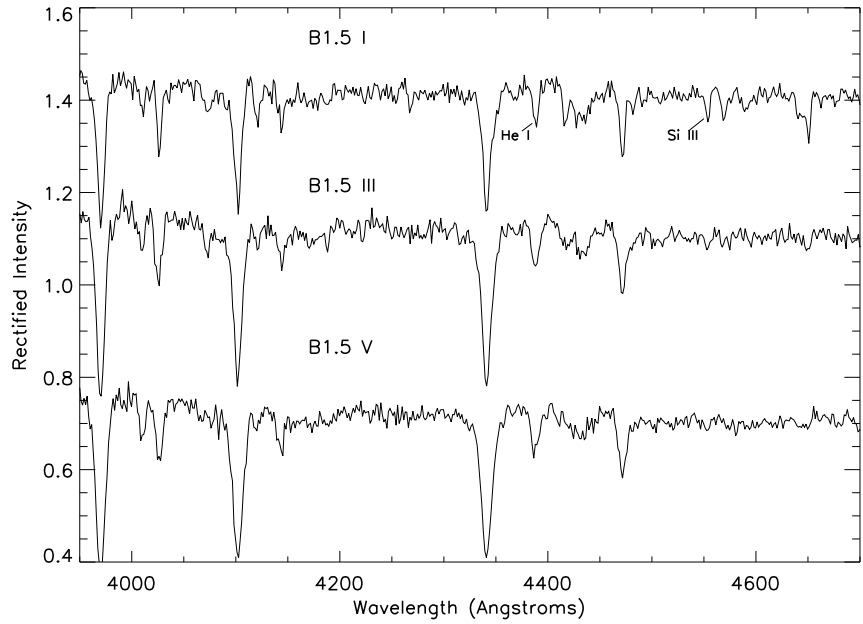}
\caption{Luminiosity sequence of B1.5 stars -- class I supergiant (top), class III giant (middle), and class V dwarf (bottom) -- showing 
the normalized continuum intensity vs. wavelength.  The B1.5 I star is ID-61 in Table 3, the B1.5 III star is ID-67 in Table 3, and the 
B1.5 V star is ID-82 in Table 3.  The spectra exhibit a clear evolution in their Si4552 to He4387 line ratios: Si4552 is strongest for 
the supergiant, weak but detectable for the giant, and almost nonexistent for the dwarf.  
}
\label{lumsequence}
\end{figure}

Because SPTCLASS does not determine luminosity classes, we manually analyzed the spectra of stars to 
identify giants and supergiants.  Our identification criteria follow \citet{Gray2009} and 
\citet{Walborn1971}.  For early-type stars, we measured
 the Si III 4552/He 4387 I line ratio as a primary indicator with the Si 4116 IV/He 4121 I line ratio and the 
O II 4415--4417 line strength as secondary indicators, using spectra from \citet{Gray2009} 
as standards.  For M stars, we use the strength of the Ca I 4226 line and the shape of the TiO lines
 to identify giants and supergiants.  Our analysis confirms the identification of giants and supergiants 
from \citet{Sl02} and adds 30 more for a total of 103 giants and supergiants.  Our primary spectra selection criteria 
(J $>$ 14) removes many foreground FGK stars (likely giants/supergiants) from our target list.  Moreover, 
our spectroscopic survey is most complete for regions close to the cluster centers, where we expect the 
ratio of cluster stars to background/foreground stars to be the highest.  Thus, the overwhelmingly large number 
of dwarfs assuredly overestimates the true ratio of dwarfs to giant stars on the field.  Figure 
\ref{lumsequence} displays a luminosity class sequence for early B stars.

\begin{figure}
\centering
\epsscale{0.9}
\plotone{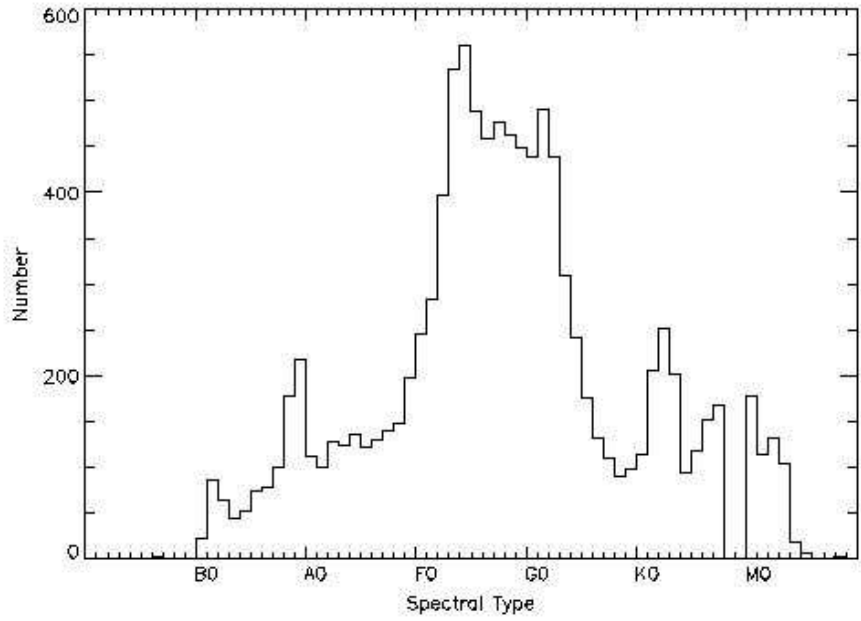}
\caption{Spectral type distribution of sources.  The rise in number from B stars to F stars is 
likely a consequence of the cluster mass function.  The drop in the number of G5--M0 stars 
is due to survey incompleteness.} 
\label{totalvspec}
\end{figure}

Our final catalog contains 11,309 stars of which 10,983 have spectral types.
Figure \ref{totalvspec} shows the distribution of spectral types.  The sample contains 1 O6.5 dwarf star,
914 B stars, 1,330 A stars, 4,362 F stars, 2,525 G stars, 1,300 K stars, and 551 M stars.
Thus, a large fraction of our sample are either F stars or G stars.  Because later-type 
stars at a given age correspond to lower-mass stars, the large number of FG stars compared to 
OBA stars is likely a consequence of the initial mass function.  Only 2MASS sources were 
selected for our Hectospec observations.  Cluster sources with JHK$_{s}$ magnitudes near the 
2MASS sensitivity limit (J $\sim$ 15.7) likely have spectral types between G0 and K0 (\citealt{Cu07a}, 
see also \citealt{Ba98}).  Therefore, the 2MASS sensitivity limits are 
likely responsible for the low number of detected mid G, K, and M stars.

\begin{figure}
\centering
\plottwo{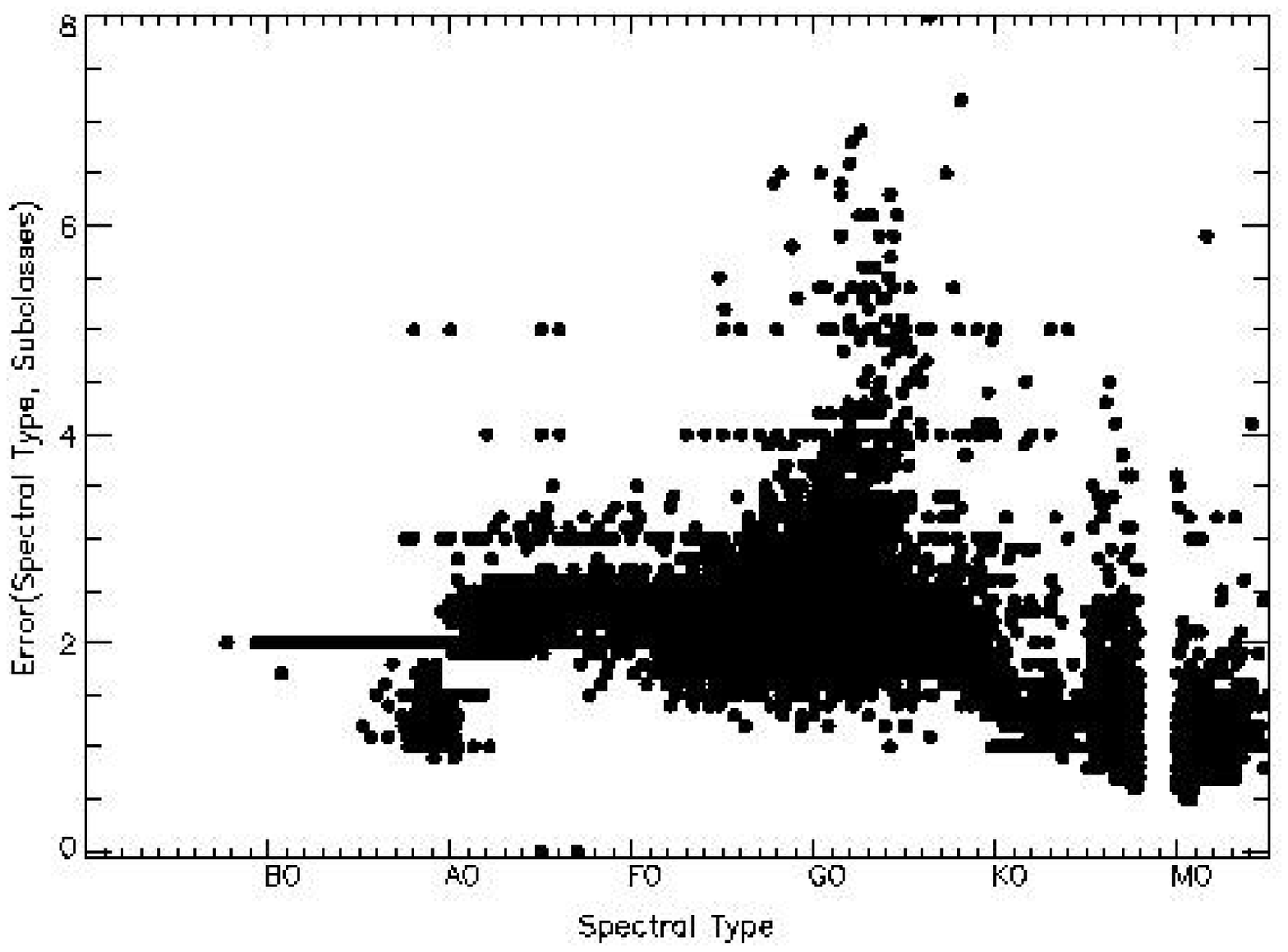}{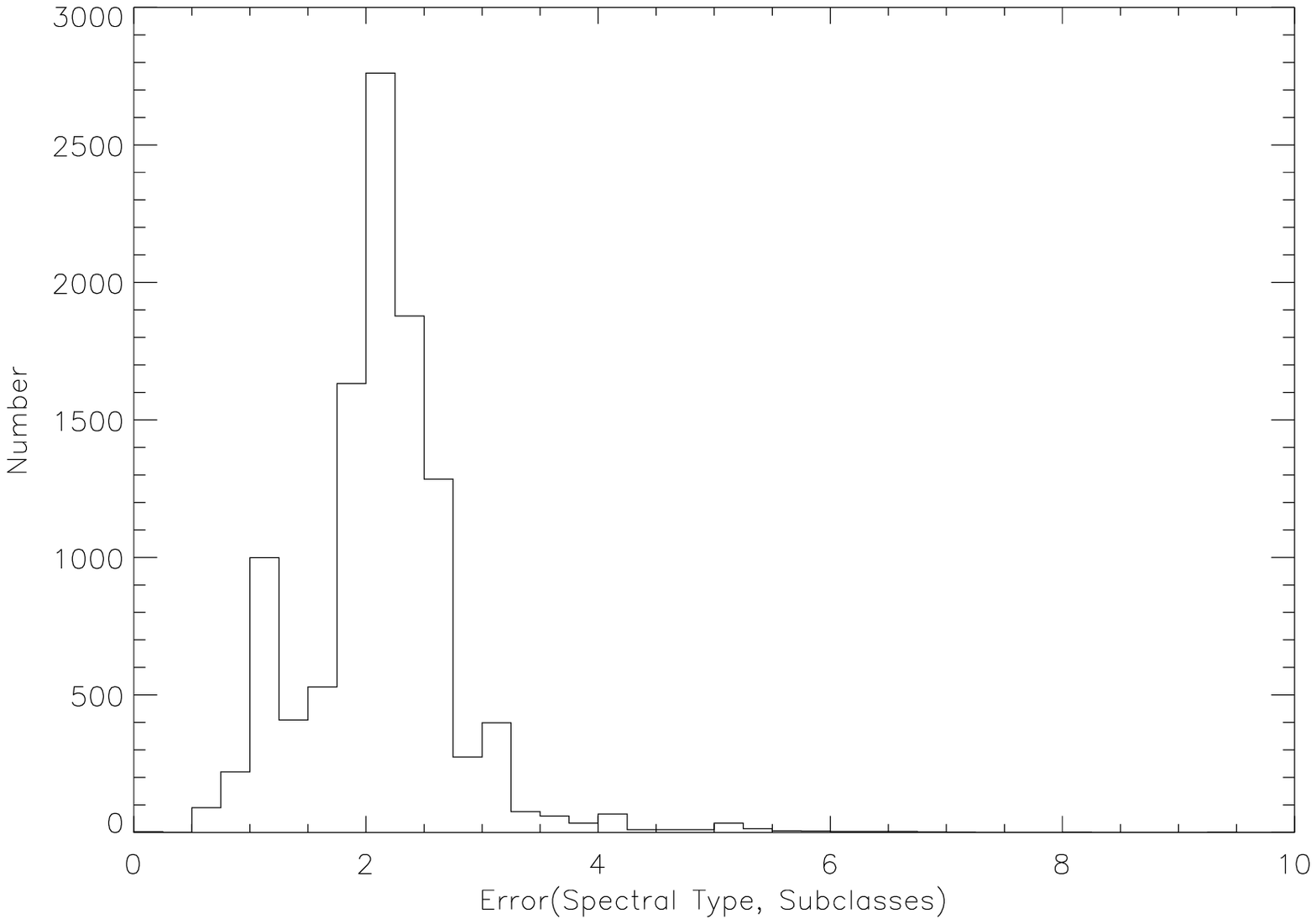}
\caption{Uncertainty in spectral type vs. spectral type (left) and histogram of uncertainties 
in spectral types.  The uncertainties in spectral types are given in subclasses.  The gap between 
K7 and M0 occurs because there are no K8 or K9 stars in the standard MK spectral classification system.}
\label{specerr}
\end{figure}

The distribution of uncertainties in spectral types is shown in Figure \ref{specerr}.  
Most stars have $\sigma$(ST) $\sim$ 2--2.5 subclasses, which indicates that the dispersion in 
spectral types determined from individual indices is small for the vast majority of 
our sample.  Stars with spectral types between O5 and F0 and K0--M5 have the 
smallest uncertainties ($\sim$ 1 subclass); stars with 
spectral types between F5 and G5 have the largest uncertainties ($\sim$ 2.5--3 subclasses).  
Intermediate spectral type stars have higher uncertainties because they have fewer lines 
whose indices strongly correlate with spectral type.
The deep Balmer lines and He I lines make spectral typing early-type 
stars robust; spectral typing later-type stars is aided by the TiO lines, whose 
strengths correlate extremely well with spectral type for stars later than K0.  
Table \ref{speccatalog} lists the properties of stars with spectra.  Each 
spectra taken with Hectospec, Hydra, and FAST is downloadable from the Telescope Data Center 
website: http://tdc-www.cfa.harvard.edu/instruments/hectospec/progs/HXP/ for Hectospec, 
http://tdc-www.cfa.harvard.edu/instruments/hectospec/progs/Hydra/ for Hydra,
and http://tdc-www.cfa.harvard.edu/instruments/fast/progs/ (Programs 83 and 170).

\subsubsection{Optical Color-Magnitude Diagrams, HR Diagrams,  and the Combined Photometric 
and Spectroscopic Catalog}
Figure \ref{Hess} shows the V/V-I color-magnitude diagram for all stars in our optical 
catalog.  To better identify trends in the V/V-I distribution, we use a Hess diagram, which 
plots the density of stars in color-magnitude diagrams, using bin sizes of 0.03 magnitudes 
in V-I and 0.06 magnitudes in V.  The densest (darkest) region 
is located within the range of colors and magnitudes consistent with main sequence 
field stars with a wide range of distances, spectral types, and reddenings \citep[e.g.][]{Mayne07}.

Exceptional among nearly all young open clusters, the locus of h and $\chi$ Persei 
stars is obviously distinguishable from the background star population by eye.  
An extremely narrow ($\delta$(V-I) $\sim$ 0.25) distribution of 
stars with bluer colors than the background field star population  
defines the Hess diagram for V $<$ 17.  Equally striking is the distribution of stars 
extending from V $\sim$ 19, V-I $\sim$ 1.75 to V $\sim$ 24, V-I $\sim$ 3.5,  
which has a far more narrow dispersion in color for a given magnitude 
than the field star population: $\delta$(V-I) $\sim$ 0.5 at V = 19 to 
$\delta$(V-I) $\sim$ 0.75 at V = 24.  These distributions are consistent with 
a single narrow locus of young cluster stars.

Using a 1" matching radius, there are 7,465 sources with spectral types, near-IR photometry, 
and optical photometry from either our VI survey or from \citet{Sl02}.
Thus, a substantial percentage ($\sim$ 68\%) of stars with spectra have optical data.  
For our combined photometric and spectroscopic catalog, we adopt our optical photometry in a first iteration.  Our V-band photometry show 
excellent agreement with that of \citet{Sl02} over a wide magnitude range; our V-band data saturates 
at V $<$ 11-12.  Therefore, we adopt the \citet{Sl02} V-band photometry for stars brighter than 
V=12.  

Sources with both optical photometry and spectroscopy show a clear distribution 
from early to late-type stars (Figure \ref{HSdiagram}).  Nearly all stars 
brighter than 16th magnitude and blueward of V-I = 1 have spectral types 
between B0 and A5.  Similarly, stars from V = 16 to 19 become progressively later 
 (A5 to K0).  Comparing both panels of Figure \ref{HSdiagram} clearly reveals 
foreground and interloping stars, whose spectral types are discrepant compared to 
other stars located in the same regions in V/V-I space.  Table \ref{specphotcatalog} 
lists the 7,465 stars with optical photometry and spectroscopy.  Our study 
focuses on these stars and on the $\sim$ 47,000 with optical photometry.

\subsubsection{Determining Effective Temperature Scale and Intrinsic Colors of Dwarfs, Giants, and Supergiants}
Deriving the physical properties of cluster stars from spectroscopy requires 
adopting a spectral type-effective temperature (T$_{e}$) scale and identifying the intrinsic colors of 
stars as a function of T$_{e}$.  Our adopted effective temperature scales and intrinsic colors are described 
and justified in Appendix A.  Briefly, the effective temperature scale for 
dwarfs is drawn from analyses of O5--B0 stars \citep{Massey2005}; B0.2--A0 stars \citep{Humphreys1984, deJager1987, Bessell1998}; 
and cooler main sequence stars \citep{Kh95,Gray2009}.  For O giants 
and supergiants we adopt the relations from \citet{Massey2005}.  
For B--M giants we adopt the scale from \citet{Gray2009}; we use a combination of \citet{Humphreys1984}, \citet{deJager1987} and \citet{Gray2009} 
for B supergiants, \citet{Gray2009} alone for A--K supergiants, and \citet{Gray2009} and \citet{Levesque2005} for M supergiants.

To derive intrinisic colors of stars as a function of effective temperature, we follow 
the methods of the Padova stellar evolution group \citep[e.g.][]{Girardi2002, Marigo2008}.
We use colors derived from the corrected ATLAS9 spectra \citep{CastelliKurucz2003} for O--K dwarfs 
and supergiants/giants (see also \citealt{Kurucz1992, CastelliKurucz1994}).  
For cooler dwarfs, we use colors from the PHOENIX "BDDUSTY99" stellar atmosphere models \citep{Allard2000, 
Allard2000b}.  We supplement the ATLAS9/PHOENIX grid for the coolest dwarfs (T$_{e}$ $<$ 3400 K) with empirically calibrated 
dwarf colors from the BaSeL stellar library \citep{Lejeune1998}.
For M supergiants, we adopt the corrected colors from the empirical M giant spectra of \citet{Fluks1994}.
Table \ref{dwarfcolors}, \ref{giantcolors}, and \ref{sgiantcolors} list the intrinsic UBVI$_{c}$JHK 
Johnson-Cousins-Glass colors of stars as a function of T$_{e}$, spectral type, and luminosity class.

\subsubsection{Measuring Reddening and Extinction}
To derive the reddening (E(B-V)) for each source, we compare the 
observed optical/near-IR colors with intrinsic colors for the 
star's spectral type.  We first convert our 2MASS photometry into the Johnson-Cousins-Glass system 
as formulated by \citet{Bessell1990} using an updated version\footnote{http://www.astro.caltech.edu/~jmc/2mass/v3/transformations/ .  
These values are similar to but supercede those published in \citet{Carpenter2001}.  Their republication 
in \citet{Bessell2005} contains a typographical error for the J--K color transformation.}
 of the color transformations from \citet{Carpenter2001}.  From our effective temperature scale, we identify T$_{e}$ 
for each star based on its spectral type.  We calculate the star's intrinsic colors by interpolating 
between values on the ATLAS9/BDDUST99/BaSeL/Fluks grid for a given T$_{e}$.

To constrain reddening for dwarfs and hot giants/supergiants, we use the long-baseline V-J, V-H, and V-K colors.
From the equations of \citet{Cardelli1989}, the wavelength zeropoints from \citet{Bessell1988}
 and \citet{Bessell1990}, and for R$_{V}$ = A$_{V}$/E(B-V) = 3.12 we use:
\begin{eqnarray}
E(B-V)_{J} = E(V-J)/2.20\\
E(B-V)_{H} = E(V-H)/2.55\\
E(B-V)_{K} = E(V-K)/2.76.
\end{eqnarray}
Because our I band data saturate at much fainter magnitudes than either the \citet{Sl02} V band photometry or 2MASS,
we do not derive E(B-V) from I band photometry. 
The difference between the reddening derived from individual colors and the mean reddening has a 
gaussian distribution with a full-width half maximum $\lesssim$ 0.03 mags centered on 0--0.01.
To minimize photometric errors whose propogation leads to errors in reddening, 
we median combine the three values together to derive a final E(B-V) for 
each star\footnote{We choose the median reddening instead of the mean because it is less sensitive to 
photometric errors in one filter.}

For M supergiants, we determine reddening solely from the J-K colors.  Most M supergiants in h and $\chi$ Persei 
undergo significant radial pulsations \citep[e.g.][]{Levesque2005} which change their effective temperatures by tens to 100s of K.  
Because the Johnson BVI bands fall along the Wien tail of M stars' spectral energy distributions, these 
temperature changes produce large changes in the stars' V-J, H, and K colors.  Furthermore, the long-baseline 
colors (in particular, V-J) are very sensitive to surface gravity.  The J-K colors
are only weakly affected by pulsation and photometric errors: 
\begin{equation}
E(B-V) = E(J-K)/0.56. 
\end{equation}

Thus, measuring E(B-V) for all stars on the field, we deredden their photometry in each passband 
using extinction relations from \citet{Cardelli1989}:
\begin{eqnarray}
V_{o} = V_{obs} - 3.12\times E(B-V)\\
I_{o} = I_{obs} - 1.87\times E(B-V)\\
J_{o} = J_{obs} - 0.92\times E(B-V)\\
H_{o} = H_{obs} - 0.57\times E(B-V)\\
K_{o} = K_{obs} - 0.36\times E(B-V)
\end{eqnarray}
These relations agree with those from \citet{Bessell1988} to within $\sim$ 3\%.
To verify that these equations deredden the data well, we compare the dereddened and 
intrinsic colors for our B--M dwarfs.  The systematic offsets in V-I, V-J, V-H, and V-K vs. 
spectral type and T$_{e}$ are 0.005--0.02 magnitudes in all cases, less than the uncertainties in 
intrinsic colors and reddening.

Figure \ref{ebmvdist} shows the distributions of reddening vs. spectral type for 
stars in the core-dominated regions (r$_{core}$ $<$ 10') and the low-density halo population.
 Most stars on the field have E(B-V) $\approx$ 0.45--0.65 (top-left panel), but closer analyses 
reveal systematic differences in the reddening distribution as a function of location (top-right panel).  Specifically, 
the h Persei core has a slightly higher median reddening among B3--F9 stars\footnote{We determined the median 
cluster reddening from spectral type range because it puts us safely out of the range of Be stars and nearly all supergiants on the bright, early end 
and nearly all of the background field stars on the late, faint end.} than the $\chi$ Persei core, though its dispersion in 
reddening is essentially identical:
\begin{eqnarray}
 E(B-V)_{h Per} \sim 0.55 \pm 0.10\\ 
E(B-V)_{\chi Per} \sim 0.52 \pm 0.10.  
\end{eqnarray}
The low-density halo has approximately the same median reddening for B3--A9 stars, E(B-V) $\sim$ 0.52 $\pm$ 0.1, though the 
far heavier field star contamination prevents robust conclusions about the median reddening for stars earlier than G0 
as with the cluster-dominated regions.  These results show exceptional agreement with those from \citet{Sl02} who 
determined E(B-V) for stars earlier than B3--B5. 

In contrast to results from \citet{Bk05}, there is no evidence for 
a spectral type-dependent reddening in either core region through G0 (bottom panels).  The population of G0--K0 stars with 
high reddening is almost completely drawn from stars whose positions on both observed and dereddened 
color-magnitude diagrams place them foreground to the clusters. The core region stellar population later than G0
is diluted by a combination of survey incompleteness and mass segregation.  A substantially non-solar metallicity 
would be the most likely cause of any spectral-type dependent reddening, because the colors for a given T$_{e}$ would be 
different, especially for stars later than A spectral type.  Thus, our data show no clear evidence that h and $\chi$ Persei are significantly 
metal poor or metal rich.

As shown by Figure \ref{dredall}, dereddening stars makes an already narrow and populous upper main sequence 
even better defined.  Both the dereddened V/V-J color-magnitude diagram and the V vs. spectral type 
Hertzsprung-Russell diagram show a thin, dense locus of probable cluster stars from V$_{o}$ = 10 to 
V$_{o}$ = 14.75 that is clearly separable from a second, low-density distribution that presumably 
contains mostly field stars.  Comparing Figure \ref{dredall} with Figure \ref{Hess} indicates that the 
distribution of spectroscopically examined stars widens at V$_{o}$ = 15.5--16.75, F5--G5 as the 
cluster locus runs above the background field star distribution.  The distribution thins out 
for stars later than G0--G5 because many late-type stars were too faint to be selected for spectroscopic 
observations.

\section{Analysis: The Distance Moduli and Ages of h and $\chi$ Persei}
We now combine the optical photometric and spectroscopic data of all stars on the field 
 to constrain the mean distance moduli and ages of h and $\chi$ Persei. 
First, we analyze the dereddened V-band magnitudes of main sequence stars in cluster-dominated regions 
as a function of spectral type and V-J, V-H, and V-K colors to infer distance moduli for both clusters.  
We then constrain their post-main sequences ages by identifying 
the main sequence turnoffs and analyzing the luminosities of the clusters' M supergiants.
Finally, we analyze the deep optical data alone to measure ages derived from fitting isochrones to
pre-main sequence stars and compare these estimates with post-main sequence ages to 
pinpoint the true mean age and dispersion in age for h and $\chi$ Persei.

For both clusters, we successfully derive robust and self-consistent values for 
cluster properties.  Even though $\chi$ Persei is less populous \citep[see Section 5 and][]{Sl02, Bk05,Currie2008}, 
its parameters have the smallest uncertainties \citep[see also][]{Mayne07}.  Stars within the h Persei core ($<$ 10' of the 
center) exhibit a wider dispersion in properties.  Analysis of stars in the low-density halo regions surrounding both cores is 
plagued by sample incompleteness and slightly higher levels of field star contamination.  Thus, in each
 section we discuss properties of $\chi$ Persei first, h Persei second, and the low-density halo region third to  
illustrate how our analysis constrains cluster properties.

\subsection{The Distance to h and $\chi$ Persei}
We now measure the distance moduli to h and $\chi$ Persei by "main sequence fitting". 
This approach slightly differs from that by other authors \citep[e.g.][]{Sl02} who argue that 
main sequence fitting must be done with post-main sequence and pre-main sequence isochrones.  
Although certainly true with previous, shallower h and $\chi$ Per spectroscopic surveys,
 the colors and magnitudes of dereddened stars from our larger dataset define a 
tight locus for much later spectral types, where stars are on the main 
sequence for a wide range of ages.  

Simple arguments demonstrate that the mean age of h and $\chi$ Persei is most likely between 
$\approx$ 10 Myr and $\approx$ 30 Myr.  The field includes only one O star (HD 14434), which is 
likely an interloping field star \citep{Sl02}.  Otherwise, the earliest stars in the cluster 
are later than B0.  Between B0 and B2.5, stars are predominately giants and supergiants: dwarfs only 
clearly dominate the stellar population later than B3--B4.  Because the main 
sequence lifetime of early B stars is at least $\sim$ 10 Myr \citep[cf.][]{Bertelli1994,Schaller1992}, 
the clusters are likely at least 10 Myr old.  The main sequence lifetime of B4--B5 stars is 
$\lesssim$ 30 Myr; thus, the lack of giants among mid-B stars implies an upper age limit of 30 Myr.

Figure \ref{isozero} illustrates the age independence of the luminosity and temperature of B5--A5 stars for 
over the 10--30 Myr age range.   We plot the predicted V vs. V-J loci for 10 and 20 Myr-old 
pre-main sequence stars from the Siess evolutionary tracks \citep{Siess2000} and D'Antona and 
Mazzitelli tracks \citep{Dm94}\footnote{This is the only place in the paper where we use the \citet{Siess2000} 
and \citet{Dm94} tracks.  We do not use the \citet{Ba98} tracks here because they do not extend to 
sufficiently high masses.}, using our conversions from T$_{e}$ to colors.  The leftmost 
point for each pre-main sequence track corresponds to the highest mass, 
bluest star yet to reach the main sequence.  We also overplot the 
10--30 Myr Padova post-main sequence evolution tracks \citep{Marigo2008} and zero-age main sequence.
  Over a significant range in M$_{V}$ and V-J color, there is substantial agreement between all pre-main sequence and 
post-main sequence tracks regardless of age.  In particular, stars with M$_{V}$ = 0.5--2, 
V-J = -0.3--0.15 ($\sim$ B5 -- A5 stars) are all on the main sequence.   
If we consider only the \citet{Dm94} pre-main sequence tracks, agreement expands to 
M$_{V}$ = 2.5, V-J = 0.5 ($\sim$ A9 stars). 

Motivated by these comparisons, we measure the distance modulus primarily by identifying where the zero-age main sequence 
lines up with B5--A5 stars (i.e. where the density of stars on the ZAMS is highest).  Additionally, because stars contract \textit{onto} the main sequence from 
the pre-main sequence (losing luminosity), the zero-age main sequence cannot lie above the locus of later-type stars.
We divide stars on the field into three groups: those within 10' of the $\chi$ Persei center, those within 10' of the h Persei center, 
and those in the low-density halo region.

\subsubsection{The Distance to $\chi$ Persei}
The results of our main sequence fitting are listed in Table \ref{msfits} and illustrated in Figures \ref{msfitchi} and \ref{msfith}.  
Figure \ref{msfitchi} reveals that main sequence fitting of $\chi$ Persei stars can determine the cluster's distance modulus with high precision.  In particular, 
the V vs. spectral type distribution (top panel) defines a sharp locus for nearly all stars earlier than G0, including 
 main sequence B5--A5 stars.  
While slightly more dispersed, the positions of stars in V/V-J (bottom panel) as well as V vs. log(T$_{e}$), V/V-H, and V/V-K (not shown) 
 also define very narrow distributions.  Distance moduli measured from spectral types, T$_{e}$, and the three infrared colors are 
 essentially identical.

From main sequence fitting, we derive a median distance modulus to $\chi$ Persei (thick line) 
 dM$_{\chi Per}$ = 11.85, which we calculate by taking the median value for the four individual estimates.  
The main sequence locus for dM = 11.85 clearly runs through the middle of the main distribution of B5--A5
$\chi$ Per stars.  By comparison, loci for dM = 11.77 and 11.93 (thin dashed lines) lie above and below the 
main distribution of B5--A5 stars in most cases.  We cannot clearly identify disagreement between the observed distribution and main sequence loci for 
 distance moduli between these extrema (e.g., 11.82, 11.88).  Therefore, we consider our mean distance modulus determination 
to be accurate within 0.08 mag.  Thus, 
\begin{equation}
dM_{\chi Per} = 11.85 \pm 0.08 \;(d = 2344^{+88}_{-85}\;pc). 
\end{equation}

\subsubsection{The Distance to h Persei}
Though the locus of h Persei stars shows a wider dispersion as a function of spectral type and color, main sequence fitting 
yields a precise estimate for its median distance modulus, 
\begin{equation}
dM_{h Per} = 11.80 \pm 0.08 \; (d = 2290^{+87}_{-82}\;pc).
\end{equation} 
Thus, the cluster is slightly foreground to $\chi$ Persei.  While the uncertainty in distance modulus is sufficiently large 
that formally h and $\chi$ Persei are at the same distance (within errors), h Persei is systematically foreground by the same 
amount as inferred from each of the four estimates by identical amounts (see Table \ref{msfits} and Figure \ref{msfith}).  
Moreover, because the distance modulus is derived from analyzing main sequence stars, not post-main sequence stars, 
h Persei's offset cannot be due to subtle differences in its evolved star population.

The difference in distance modulus is plausibly real because the difference is systematic and also because it is consistent 
with some previous independent analysis performed using completely separate methods.  Using the Q method\footnote{The Q 
method determines the mean cluster extinction iteratively by using relationships between optical reddening laws and the 
intrinsic colors of O and B stars \citep[see][]{Johnson1953}.}, \citet{Mayne2008} also found a small but systematic difference in 
distance modulus nearly identical to ours.  The distance moduli derived 
by \citet{Mayne2008} -- 11.78 for h Persei and 11.82 for $\chi$ Persei -- are within 0.02--0.03 mags of 
our estimates.  Our dM estimates also agree with those from \citet[][dM = 11.85 for both clusters]{Sl02}.  Because their 
spectroscopic survey was limited to early-to-mid B stars, they were unable to measure a clear difference 
in dM for the two clusters.  The main sequence steeply rises in M$_{V}$ for 
a given spectral type for B stars; the main sequence is more horizontal for 
late B stars/early A stars, which makes differences in distance modulus stand out more.

\subsubsection{Distance to the halo population of h and $\chi$ Persei}
Our most unique contribution regarding the distance to h and $\chi$ Persei stars is a first precise estimate for 
stars in the halo population.  Figure \ref{msfithalo} plots our fits.  The huge density change between the population of 
stars earlier and later than B5 is solely a reflection of our survey bias: the earlier stars were too bright to be selected for 
Hectospec observations, were derived solely from FAST observations which were fewer in number, while Hectospec data covers later 
stars.  To more clearly present the empirical locus of main sequence stars, we shrink the symbol sizes by 30 \%.
The upper main sequence of halo stars (B5--A5) is separable from the field star population despite the higher level of contamination.

Interestingly, we find that stars in the halo population have a well defined locus implying a distance modulus essentially identical 
to that of the core region stars, especially $\chi$ Persei's:
\begin{equation}
dM_{halo} = 11.85 \pm 0.08 (d = 2344 pc^{+88 pc}_{-85 pc}).
\end{equation}
The halo stars with spectra are drawn from throughout the $\sim$ 1 square degree region surrounding both clusters.  At the 
distance of h and $\chi$ Persei, the projected size of the halo is $\approx$ 41 pc, which is comparable to the difference 
in distance modulus between the cluster cores.  If the halo region has a roughly spherical shape, it may surround 
both cluster cores.  Previous estimates for the halo population's distance modulus 
place it within 30\% of the core regions \citep{Sl02}.  Our analysis indicates that stars in the low-density regions 
surrounding both cores \textit{that are most plausibly associated with the cores} are at distances within $\sim$ 5\% 
of the cores' distances.

\subsubsection{Uncertainties in Distance Modulus}
To evaluate the origin of the uncertainties in our estimates for the distance modulus, we 
now consider each point in the process.
Table \ref{msfits} indicates that uncertainties in dereddening infrared photometry from filter to filter contribute a negligible 
level of uncertainty in distance modulus estimates.  A more important source of uncertainty is the assumed ratio of optical 
extinction to reddening in B-V.  We adopt the standard value of R$_{V}$ = 3.12, appropriate for reddening due primarily to 
interstellar dust.  Diffuse mid-IR nebular emission, expected in dark cloud regions where R$_{V}$ is large, is minimal
in the Spitzer IRAC and MIPS mosaics of the Double Cluster \citep{Currie2010}; \citet{Sl02} also argue that 
reddening is likely entirely due to line-of-sight extinction from the ISM.  Our assumed R$_{V}$ brackets previous 
estimates of R$_{V}$ = 3.0 $\pm$ 0.1 and R$_{V}$ = 3.2 $\pm$ 0.04 by \citet{Johnson1965} and \citet{Uribe2002} for 
cluster stars.  Over this range in R$_{V}$, the distance moduli formally have a systematic uncertainty of $\approx$ 0.06 mags.  However, 
considering data from \cite{Sl02} and \citet{Currie2010}, the evidence for deviations from the standard R$_{V}$ is 
far from convincing.  

Slightly more important is the error in determining the dereddened V band magnitude by uncertainties in E(B-V).  As noted in 
\S 2, the dispersion in E(B-V) derived in each optical/IR color is $\lesssim$ 0.03 mags.  This uncertianty results in a maximum dereddening 
error of $\delta$V$_{o}$ $\approx$ 0.09.  E(B-V) derived for most sources show even better internal consistency; the distributions of 
$\bar{E(B-V)}$ - E(B-V)$_{J,H,K}$ show a gaussian distribution centered on 0.00-0.01.  Therefore, uncertainties in E(B-V) simply widen 
the dereddened V vs. spectral type/log(T$_{e}$) and V vs. J/H/K loci, and do not systematically shift the position of the locus.

The primary source of uncertainty in distance modulus (and other parameters derived later) is metallicity, which affects the 
luminosity of stars at a given age.  Throughout the paper, we assume that h and $\chi$ Persei has an approximately solar metallicity.  
While the research literature shows significant disagreement over the Double Cluster's metallicity, we explain why 
a solar metallicity is more likely in Appendix B.

\subsection{The Post-Main Sequence Ages of h and $\chi$ Persei}

We now estimate the post-main sequence ages of h and $\chi$ Persei stars via two methods.
Identifying the location of the main sequence turn off, where stars of increasing luminosity 
become cooler as they evolve to become giants, provides a robust age estimate for massive clusters.  We measure 
the location of the main sequence turnoff by comparing the spectral types and colors of stars of a given V magnitude to 
predictions from the Padova stellar evolution models \citep{Girardi2002,Marigo2008}.  Isochrones whose spectral types/colors 
bisect the stellar population at a given V magnitude at the turnoff best reflect the clusters' main sequence turnoff age.  
Figures \ref{msfitchi} and \ref{msfith} imply that the turnoff occurs at V $\approx$ 8 for both clusters.  The main sequence 
turnoff age is the source of nearly all previous age estimates for h and $\chi$ Persei \citep[e.g.][]{Ke01, Sl02, 
Bragg2004}.  With our large spectroscopic survey, we can strengthen previous constraints on the Double Cluster's main sequence 
turnoff age(s).

The luminosities and colors of M supergiants also constrain the ages of h and $\chi$ Persei.  
Between 10 and 20 Myr, stars with masses $\approx$ 10--15 M$_{\odot}$ reach the 
M supergiant phase and rapidly evolve in luminosity; M supergiants dim by nearly 
an order of magnitude over this age range \citep[e.g.][]{Bertelli1994, Schaller1992}.  
Because the temperature of the reddest supergiants also varies with age \textit{and} metallicity \citep{Bertelli1994, Marigo2008}, 
combining V-band luminosities and long-baseline colors (e.g., V-J, V-H, and V-K) provides a sensitive 
probe of the stars' ages and may constrain the clusters' metallicities (see Appendix B).  
Because isochrones did not extend to very cool temperatures thought to characterize 
M supergiants \citep[e.g.][]{Sl02},  most previous investigations did not infer cluster ages from 
these stars.  However, the recent recalibration of the M supergiant 
T$_{e}$ scale to systematically higher temperatures by \citet{Levesque2005} makes age estimates 
possible.
Table \ref{hxPerage} summarizes our results, which are described in more detail below.

\subsubsection{Ages from the Main Sequence Turnoff}
Figure \ref{msturnoffchi} and \ref{msturnoffh} plot V vs. spectral type and V vs. 
log(T$_{e}$) for stars within 10' of the cluster cores against the Padova
post main sequence isochrones for 10 Myr and 20 Myr and one intermediate age, which varies from panel to panel.
In each figure, we assume the distance moduli derived in previous sections.  If the cluster's age and 
distance spread is minimal, nearly all stars should deredden to a single, well-defined locus.  However, 
Be stars, which are abundant in h and $\chi$ Persei \citep[e.g.,][]{Br02,Cu08a}, can have intrinsically red 
near-IR colors \citep{Do91, Do94} due to circumstellar gas/dust shells.  Therefore, we do not consider these 
stars in locating the main sequence turnoff.

Adopting a solar metallicity and the Padova stellar evolution models, the main sequence 
turnoff age of $\chi$ Persei is about 14 Myr.  In Figure \ref{msturnoffchi}, 
the agreement between the 14 Myr isochrone (solid line, top panels) and the observed turnoff 
is unambiguous.  The curve in the isochrone at V$_{o}$ $\sim$ 7.5--9.5 clearly rules out a 10 Myr age because almost all 
sources lie to the right in both spectral type and log(T$_{e}$) space.  Similarly, a 20 Myr age is ruled out because 
the vast majority of stars in both panels lie to the left of its isochrone.
The 14 Myr isochrone bisects the stars' positions in both top panels, especially in V$_{o}$ vs. spectral type space.  

The HR diagram in V$_{o}$--log(T$_{e}$) space allows a firmer constraint on the age of $\chi$ Persei.
The 12 Myr isochrone (lower left panel of Figure \ref{msturnoffchi})
clearly overestimates the temperatures of the earliest B stars at the turnoff by $\sim$ 0.05 dex and overpredicts 
the temperatures at a given M$_{V}$ for all bright cluster stars except for the Be stars.  Conversely, the 16 Myr isochrone 
plotted in the lower right panel generally predicts too cool temperatures, including at the turnoff ($\sim$ 0.05 dex too cool).
Though we consider the 14 Myr isochrone to provide the best visual fit, isochrones for 13 Myr and 
15 Myr also correctly predict the temperature at the MS turnoff.  Therefore, $\chi$ Persei has a 
formal MS turnoff age of
\begin{equation}
t_{\chi Per, MS turnoff} = 14 \pm 1 {\rm Myr}.
\end{equation}

Figure \ref{msturnoffh} reveals that the main sequence turnoff for h Persei is nearly indistinguishable.  For h Persei, 
both the 13 Myr (shown) and 14 Myr (not shown) isochrones correctly predict the turnoff temperature and follow the distribution of 
cluster stars from V$_{o}$ = 6.5 to 10.  Given the larger photometric and spectroscopic scatter for h Persei, 
 it is impossible to choose confidently between the two isochrones.  Thus, we 
average them for our best-fit value.  Compared to $\chi$ Persei, h Persei's range of possible turnoff ages 
may be very slightly shifted towards younger ones.  Comparing the bottom panels of Figures \ref{msturnoffchi} 
and \ref{msturnoffh}, the 12 Myr isochrone does not overpredict the turnoff temperature as badly for h Persei 
as it does for $\chi$ Persei.  However, because the intrinsic dispersion of V$_{o}$ vs. log(T$_{e}$) is much 
larger for h Persei, we interpret this observation to mean that the age uncertainty is larger.  Our adopted 
MS turnoff age for h Persei is:
\begin{equation}
t_{h Per, MS turnoff} = 13.5 \pm 1.5 {\rm Myr}.
\end{equation}

Because our spectroscopic sample is very incomplete for B stars in the low-density halo regions surrounding 
the cluster cores, our turnoff age estimate for the halo population is formally more uncertain.  
However, as Figure \ref{msturnoffhalo} suggests, the likely turnoff age for h and $\chi$ Persei halo stars 
is about the same as the core dominated regions.  The shape of the observed turnoff clearly rules out 
both the 10 Myr and 16 Myr isochrones; isochrones with intermediate ages are plausibly consistent.  Therefore, 
our derived turnoff age for the halo population is:
\begin{equation}
t_{halo, MS turnoff} = 13 \pm 2 {\rm Myr}
\end{equation}

Thus, within our measurement uncertainties the turnoff ages for all regions of h and $\chi$ Persei are 
$\sim$ 14 Myr determined from V$_{o}$ vs. spectral type and log(T$_{e}$).  While differences in 
cluster loci for different ages are more pronounced for V$_{o}$ vs. log(T$_{e}$), the loci are 
also more prone to random errors.  Specifically, in Figures \ref{msturnoffchi} through \ref{msturnoffhalo} 
the distribution of V$_{o}$ vs. log(T$_{e}$) for h and $\chi$ Per stars exhibits a larger 
dispersion about the 14 Myr locus than they do in V$_{o}$ vs. spectral type space.  
Taken at face value, the V band magnitudes and temperatures for some bright, hot stars 
are more consistent with the 20 Myr isochrone.
We cannot definitively rule out the existence of a small population of 20 Myr-old stars.  
However, the tight clustering in V$_{o}$ vs. spectral type indicates that uncertainties 
in log(T$_{e}$) provide a more simple explanation for this larger dispersion than 
a true age spread. 

\subsubsection{Ages Determined from M supergiants}

The red supergiants provide additional evidence for an age of $\approx$ 13--14 Myr. 
In each panel of Figure \ref{msupergiants}, we plot the 
predictions for 10 Myr and 16 Myr isochrones as dashed lines; the solid line represents the intermediate age isochrone, which 
is varied between 12 Myr and 14 Myr.  Because there are no M supergiants within 10' of h Persei, we can only estimate 
ages for $\chi$ Persei and the halo population.

Inspection of the panels clearly rules out 10 Myr and 16 Myr as the ages for $\chi$ Per and the halo region.  From
the positions of M supergiants relative to the isochrones,
 13 Myr and 14 Myr are equally plausible ages for both the $\chi$ Per core and halo regions.  The 
12 Myr isochrone slightly overpredicts the maximum luminosity of M supergiants and underpredicts the maximum V-J color.  
The 15 Myr isochrone (not shown) slightly underpredicts the typical luminosities and overpredicts the maximum V-J color.
Therefore, we consider the best-estimate ages for $\chi$ Persei and the halo population as:
\begin{eqnarray}
t_{\chi Per, M supergiants} = 13.5 \pm 1.5 Myr\\
t_{halo, M supergiants} = 13.5 \pm 1.5 Myr
\end{eqnarray}

\subsection{The Pre-Main Sequence Ages of h and $\chi$ Persei}
\begin{figure}
\centering
\plotone{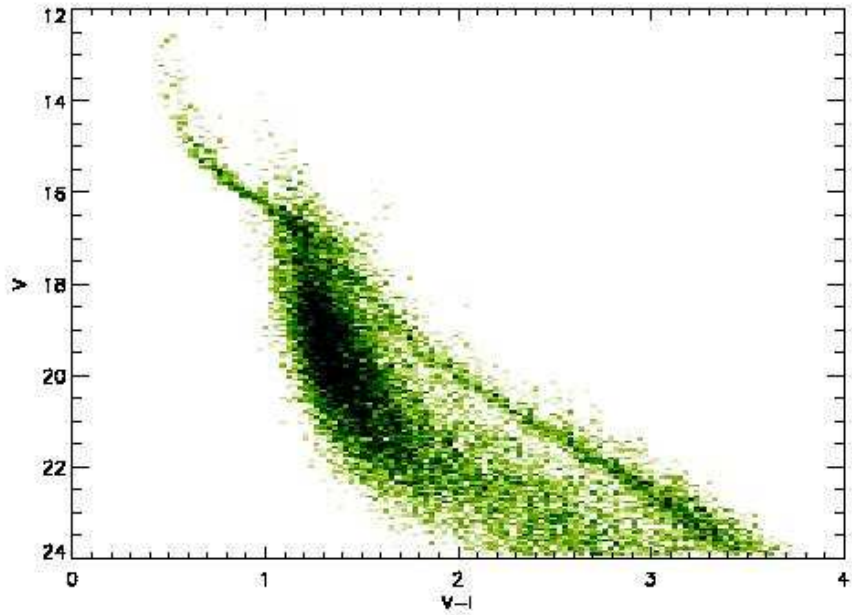}
\caption{V/V-I color-magnitude Hess diagram for all stars detected with our photometry.  
The positions of h and $\chi$ Persei stars are clearly identifiable as a narrow locus stretching 
from the top-left to lower-right region of the plot.  With the possible exception of 
a region with V-I $\sim$ 1.1--1.6, the locus of cluster stars is easily distinguishable from 
the background field star population.}
\label{Hess}
\end{figure}

\begin{figure}
\centering
\epsscale{0.85}
\plotone{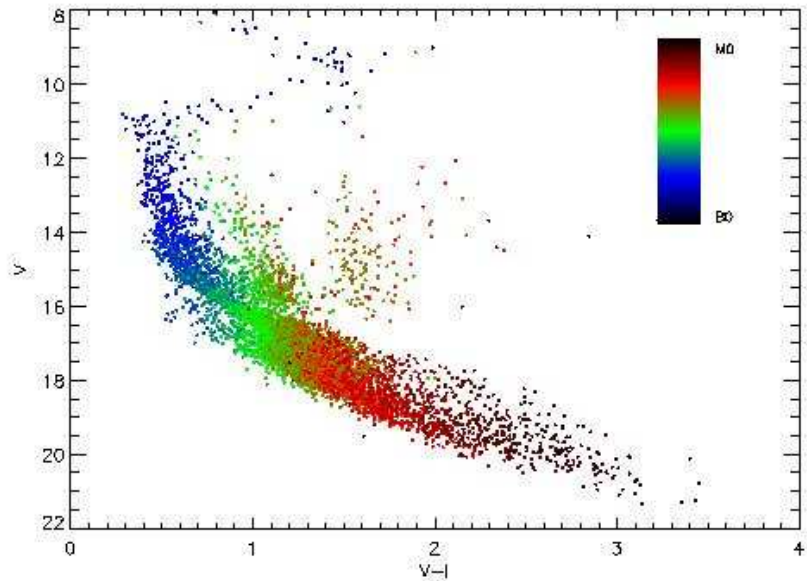}
\plotone{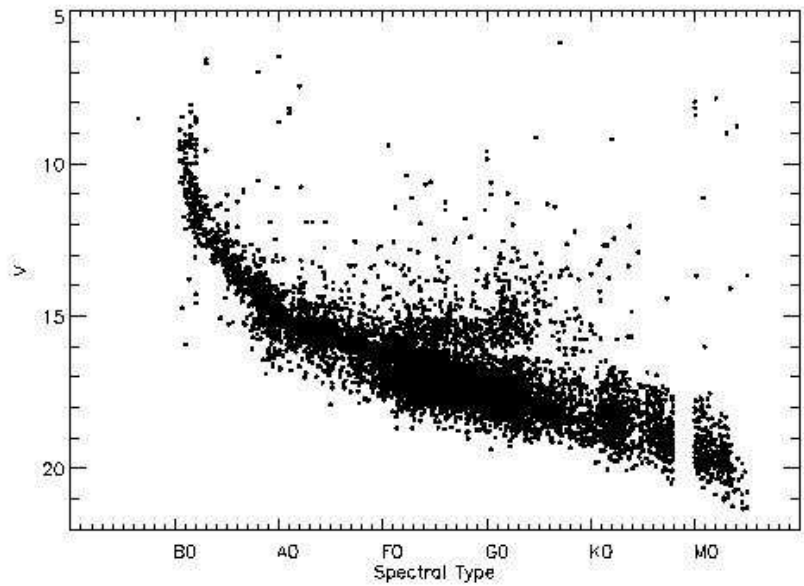}
\caption{(Top) V/V-I color-magnitude diagrams of stars with optical spectra 
color-coded by spectral type.  (Bottom)  Observational Hertzsprung-Russell diagram 
for the same population. }
\label{HSdiagram}
\end{figure}

\begin{figure}
\epsscale{1}
\centering
\plottwo{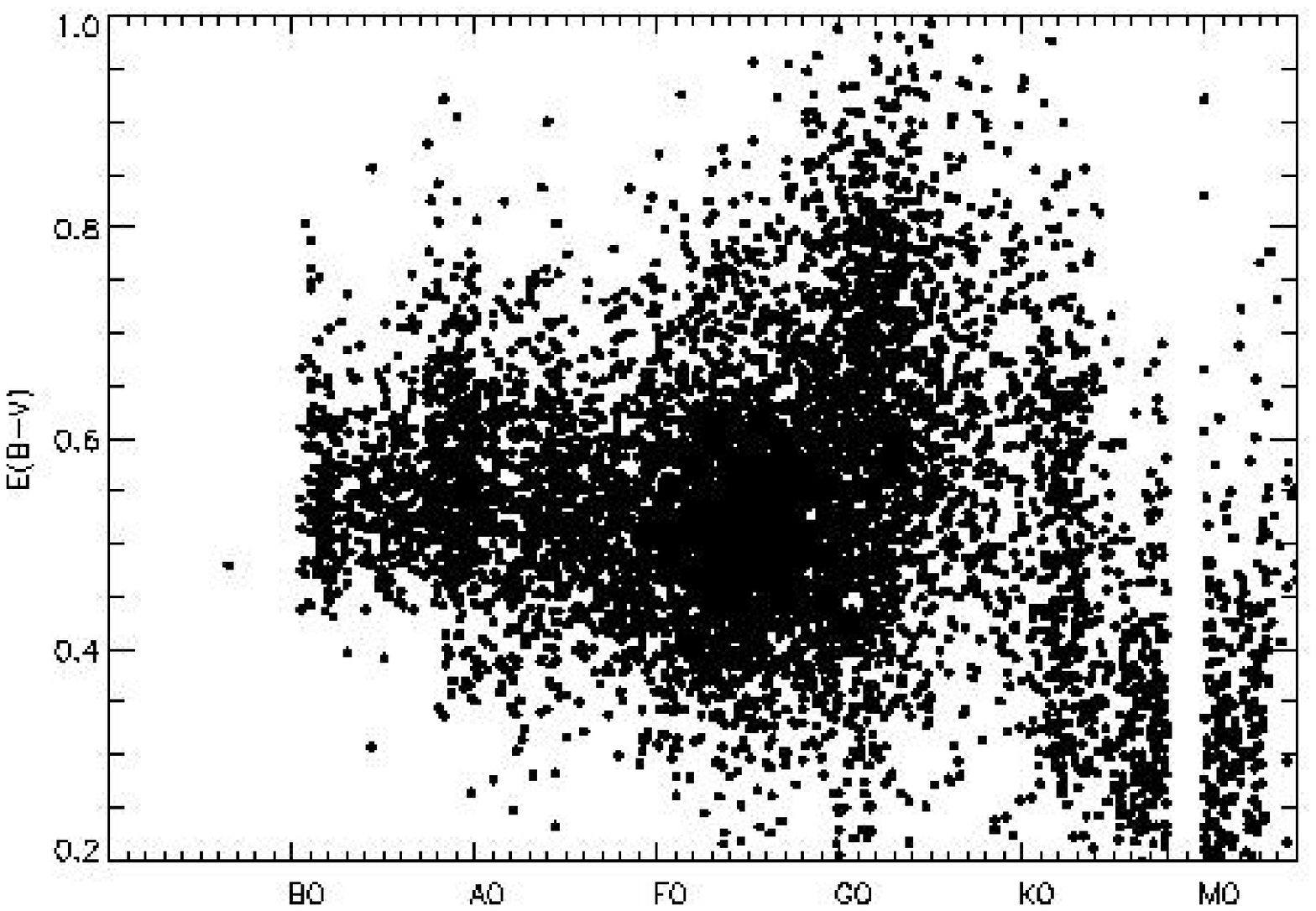}{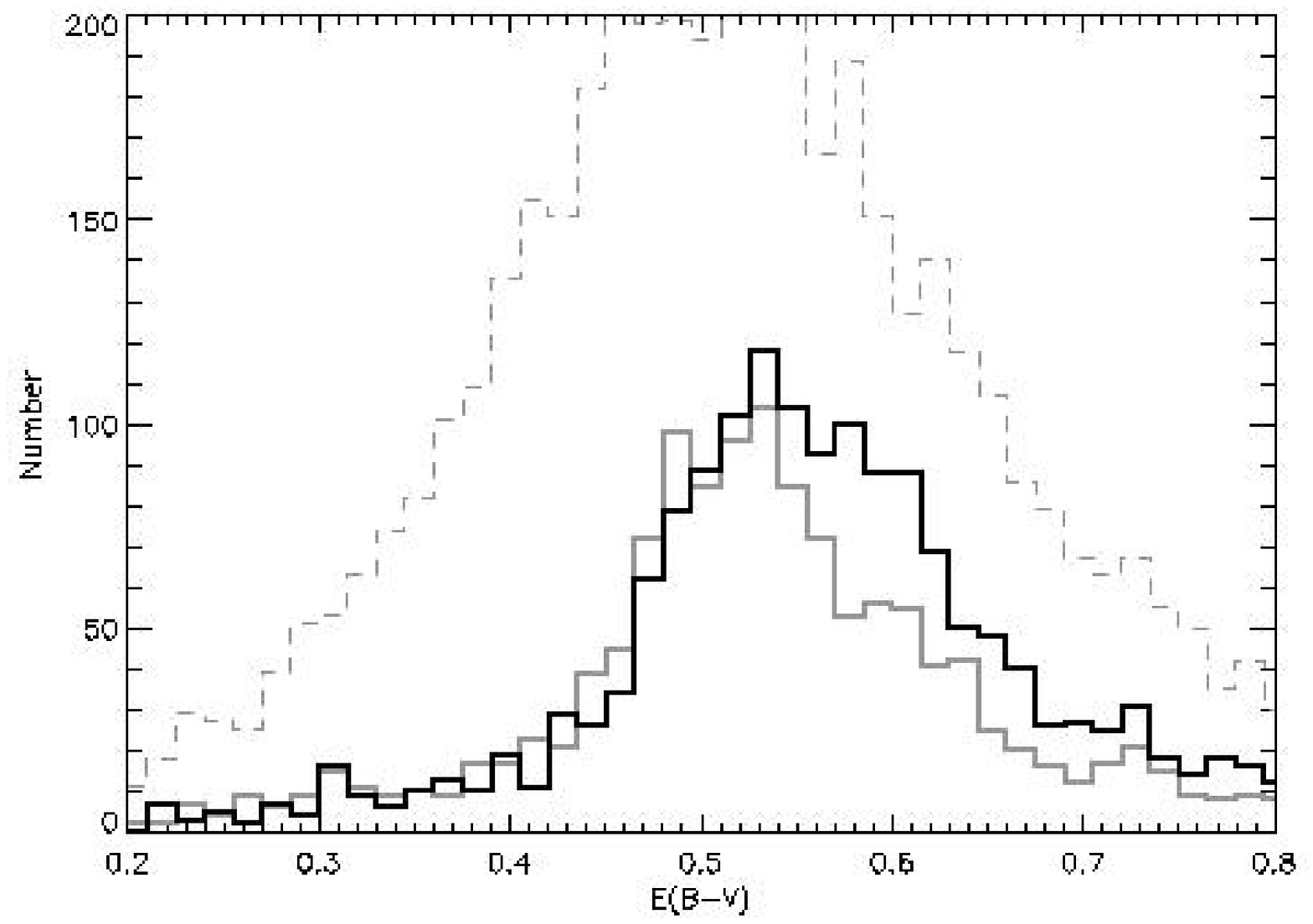}
\plottwo{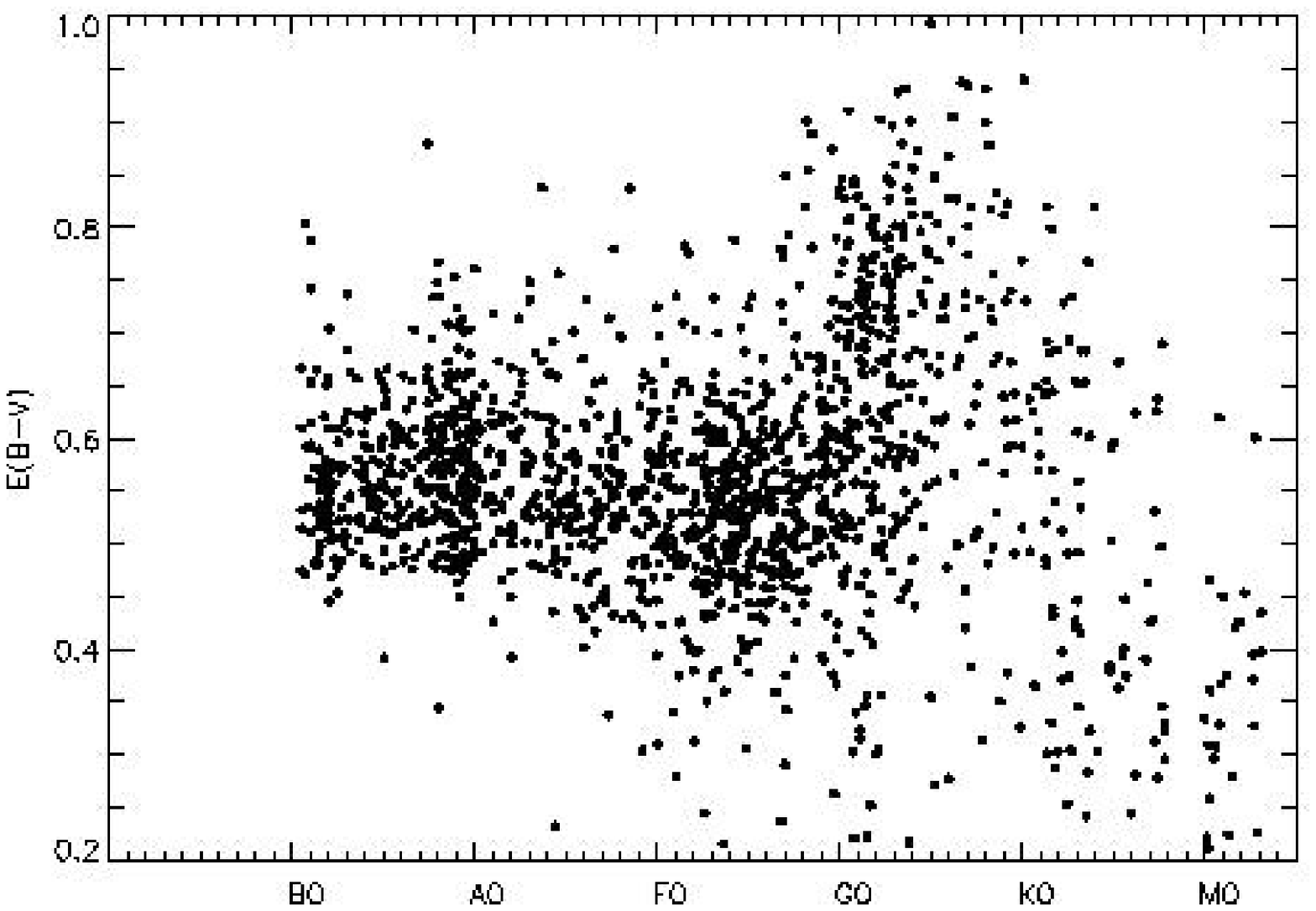}{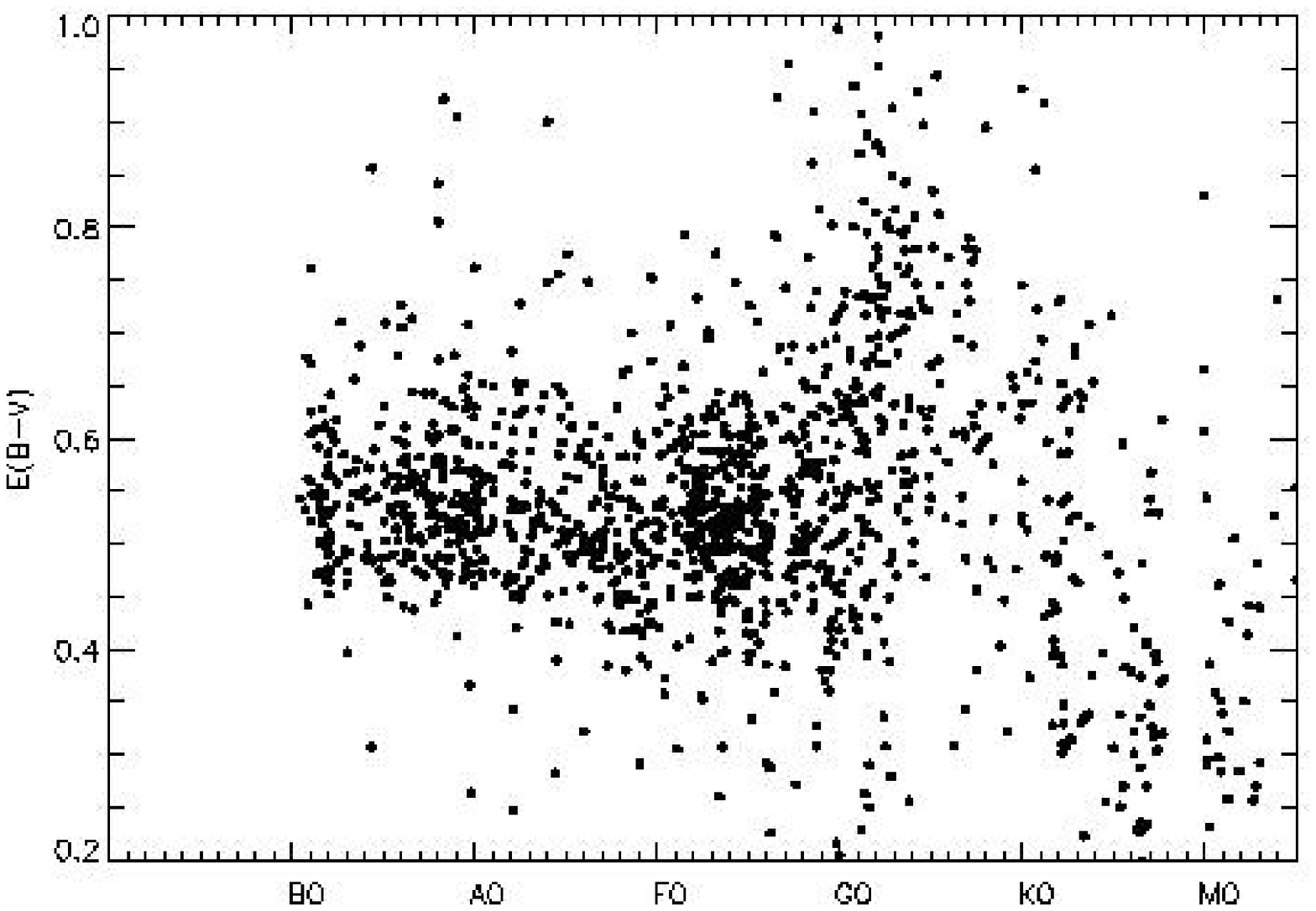}
\caption{Reddening distribution for spectroscopically observed stars.  (Top-left) 
The distribution of E(B-V) vs. spectral type for all stars on the field.  (Top-right) 
Histogram plot of E(B-V) for stars in the h Per core-dominated region ($<$ 10'; thick black line), 
$\chi$ Per core-dominated region ($<$ 10'; thick grey line), and low-density halo 
region (dashed grey line).  (Bottom panels) The reddening distributions for the 
h Per core region (left) and $\chi$ Per core region (right).
}
\label{ebmvdist}
\end{figure}

\begin{figure}
\centering
\epsscale{0.8}
\plotone{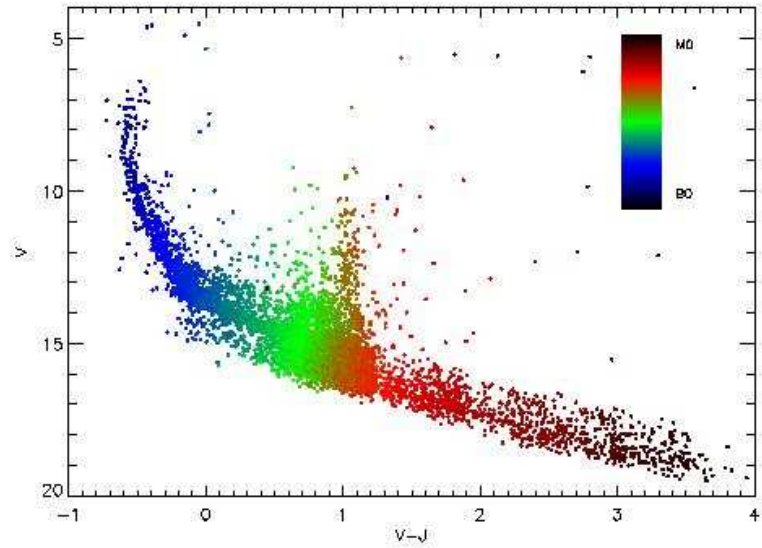}
\plotone{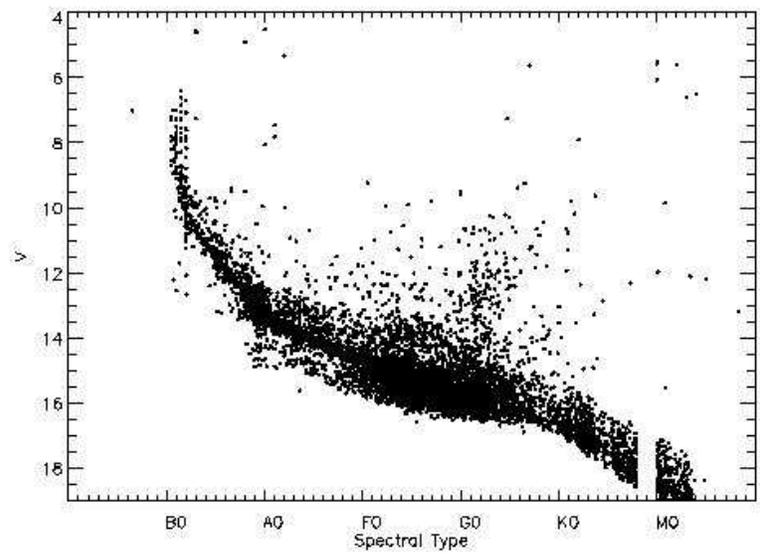}
\caption{The dereddening V/V-J color-magnitude diagram (top) and the 
dereddened V vs. spectral type HR diagram (bottom).}
\label{dredall}
\end{figure}

\begin{figure}
\centering
\epsscale{1}
\plotone{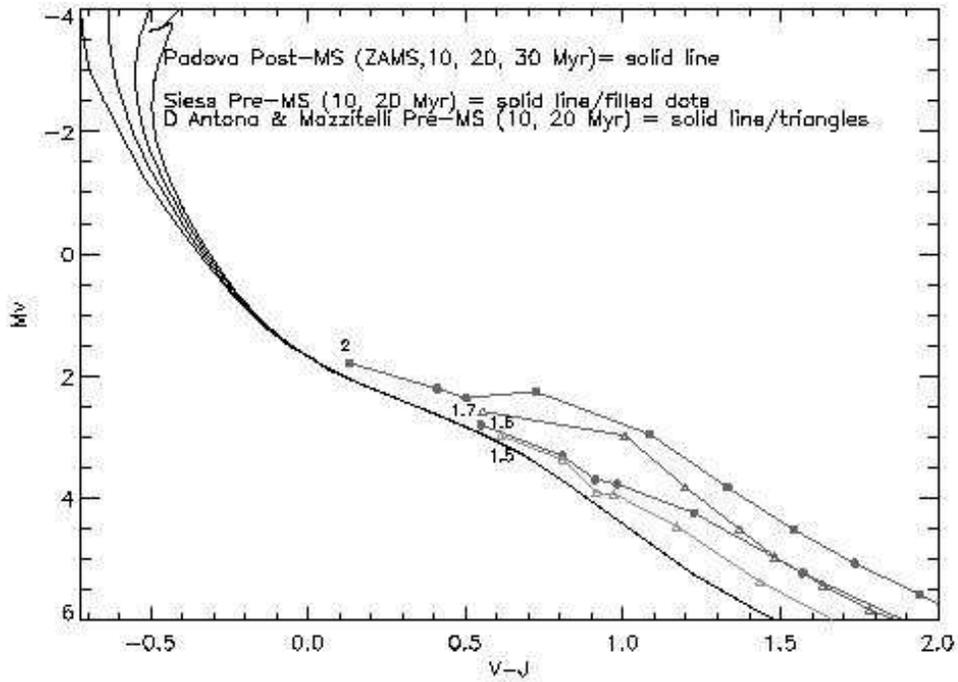}
\caption{Comparing isochrones from \citet{Marigo2008}, \citet{Dm94}, and 
\citet{Siess2000} to illustrate the age independence of the luminosity of late B to early A stars.  
The numbers identify the most massive star yet to reach the main sequence for each pre-main sequence isochrone. }
\label{isozero}
\end{figure}

\begin{figure}
\centering
\epsscale{0.75}
\plotone{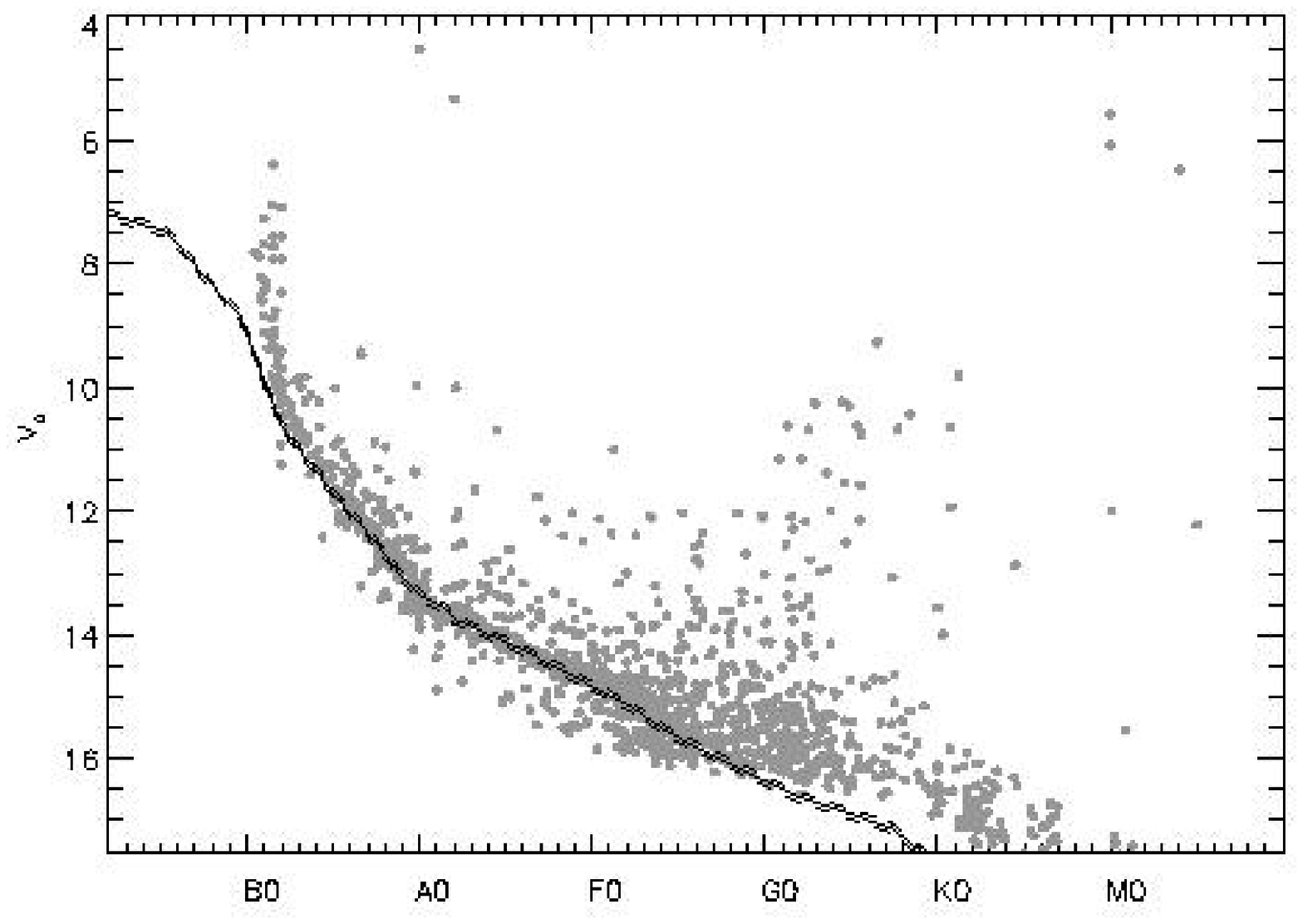}
\plotone{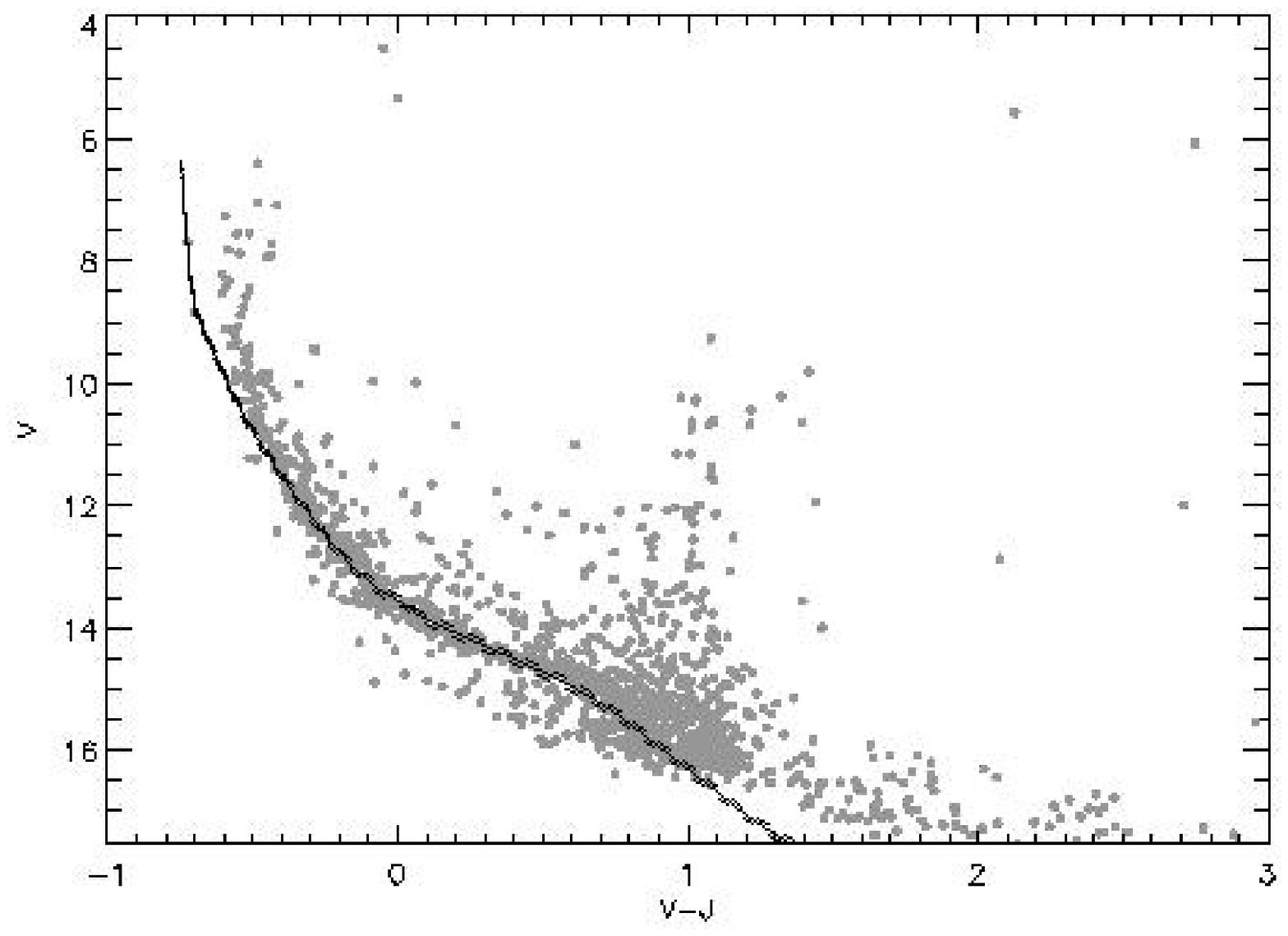}
\caption{(Top) V vs. spectral type plot of stars within 10' of the $\chi$ Persei center.  (Bottom) V vs. V-J color-magnitude 
diagram of the same stars.  The solid line in both plots corresponds to the zero-age main sequence with a distance modulus of 
dM = 11.85.  Dashed lines show the zero-age main sequence for dM = 11.77 and 11.93.  Main sequence fitting is performed 
between B5 and A5 spectral types and V- J = -0.3--0.5.}. 
\label{msfitchi}
\end{figure}

\begin{figure}
\centering
\plotone{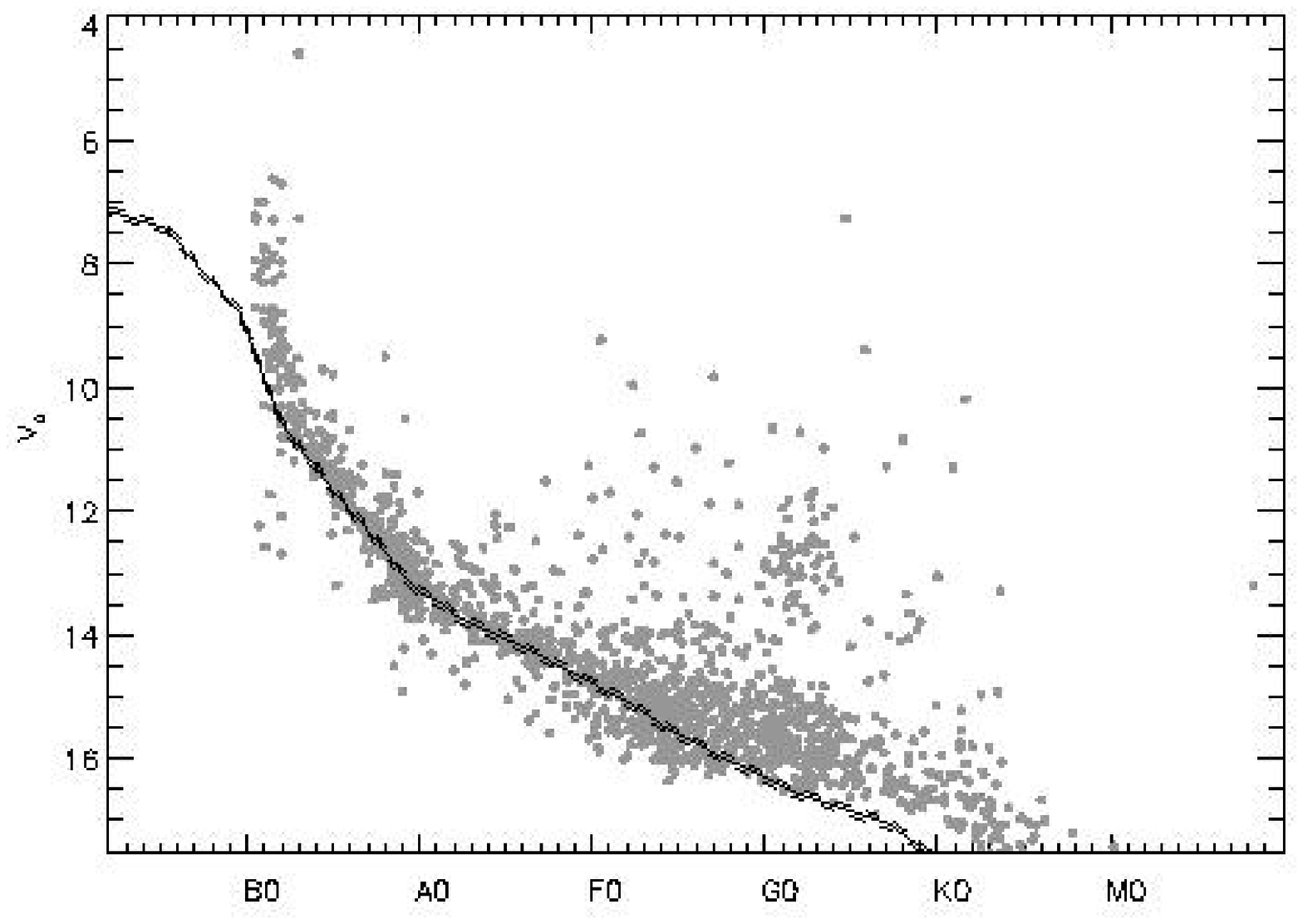}
\plotone{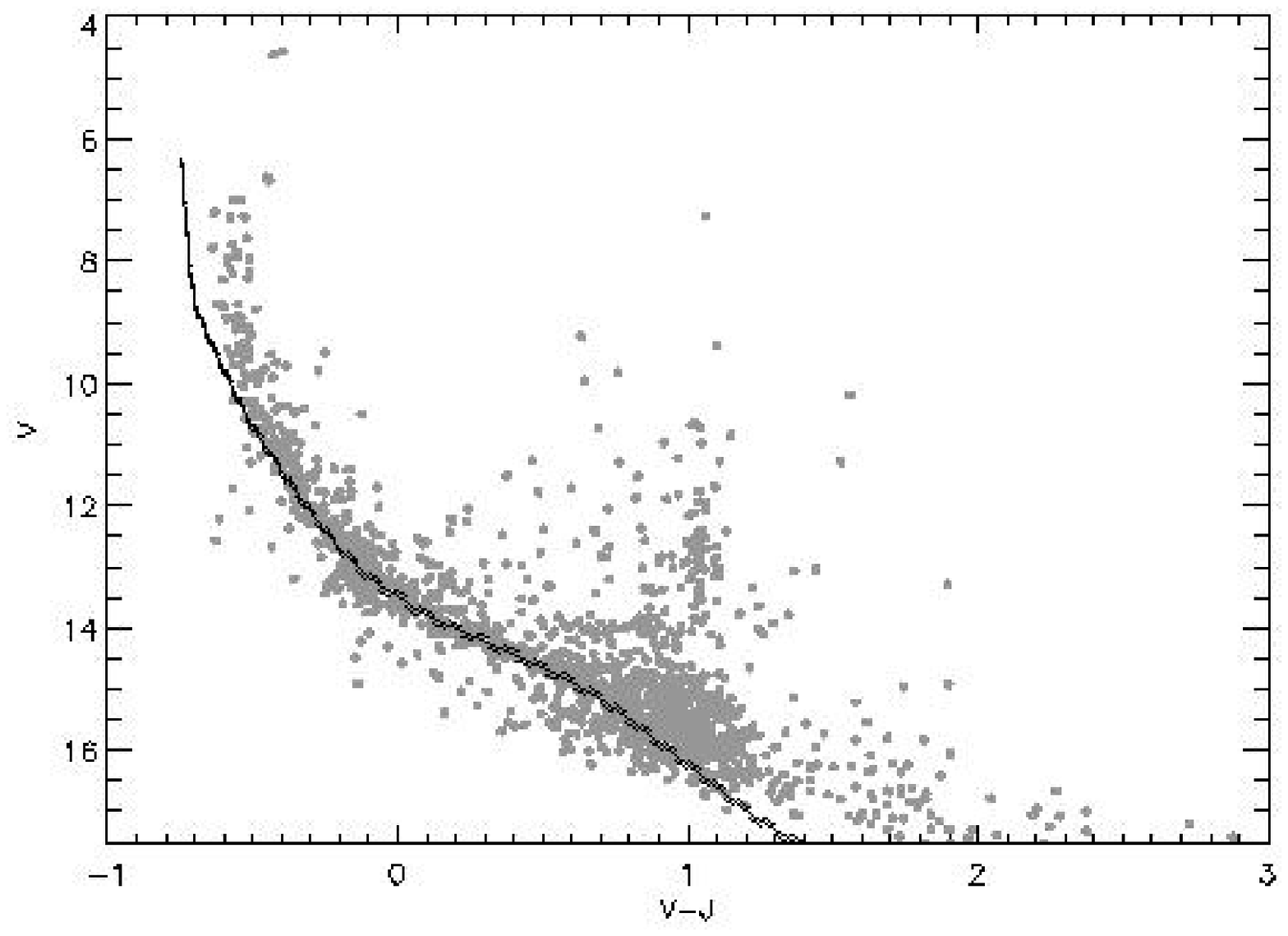}
\caption{Same as previous plot except for stars within 10' of the h Persei center with a distance modulus of dM = 11.8 (thick line) and 
extrema of dM = 11.72 and 11.88.}
\label{msfith}
\end{figure}

\begin{figure}
\centering
\plotone{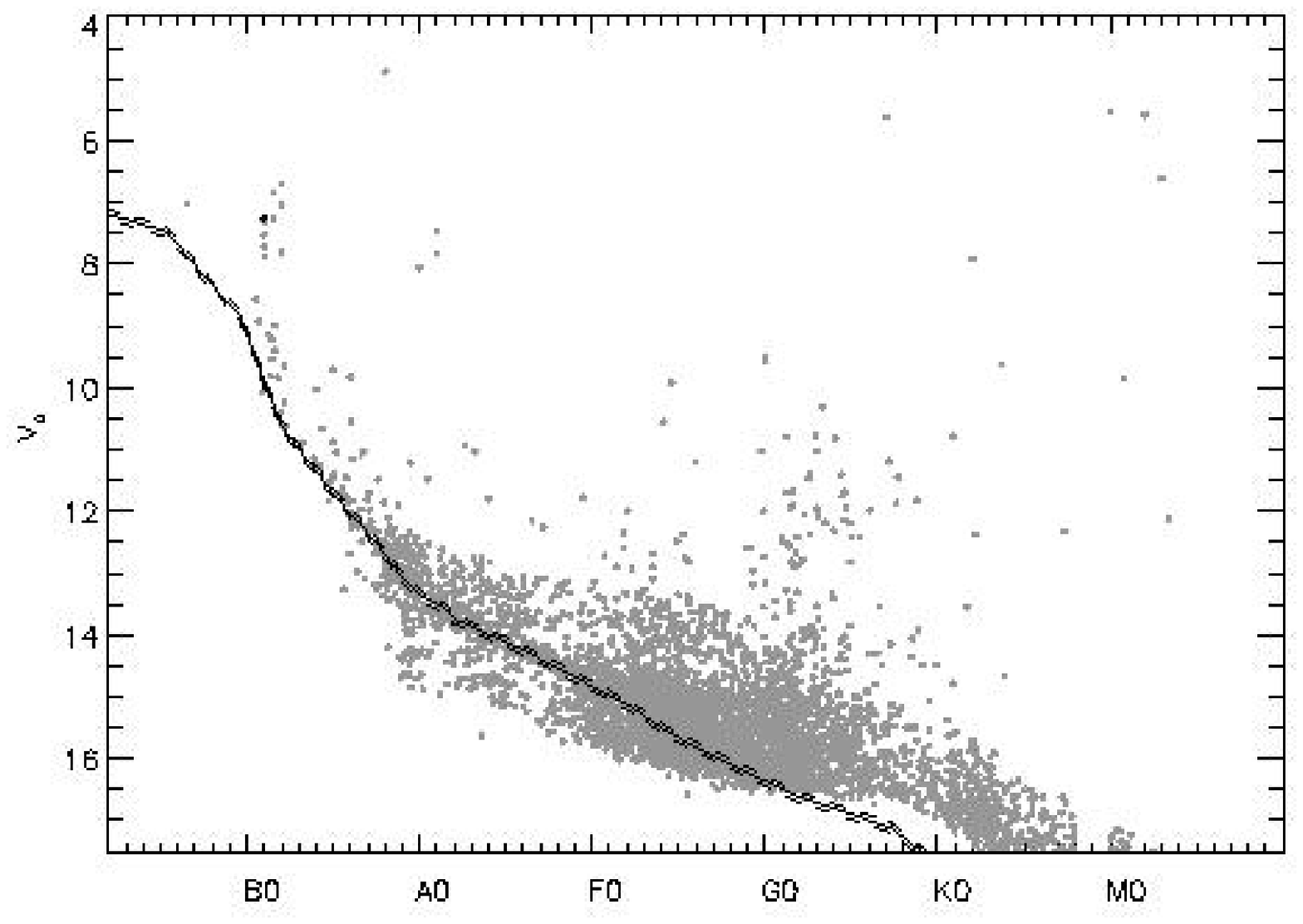}
\plotone{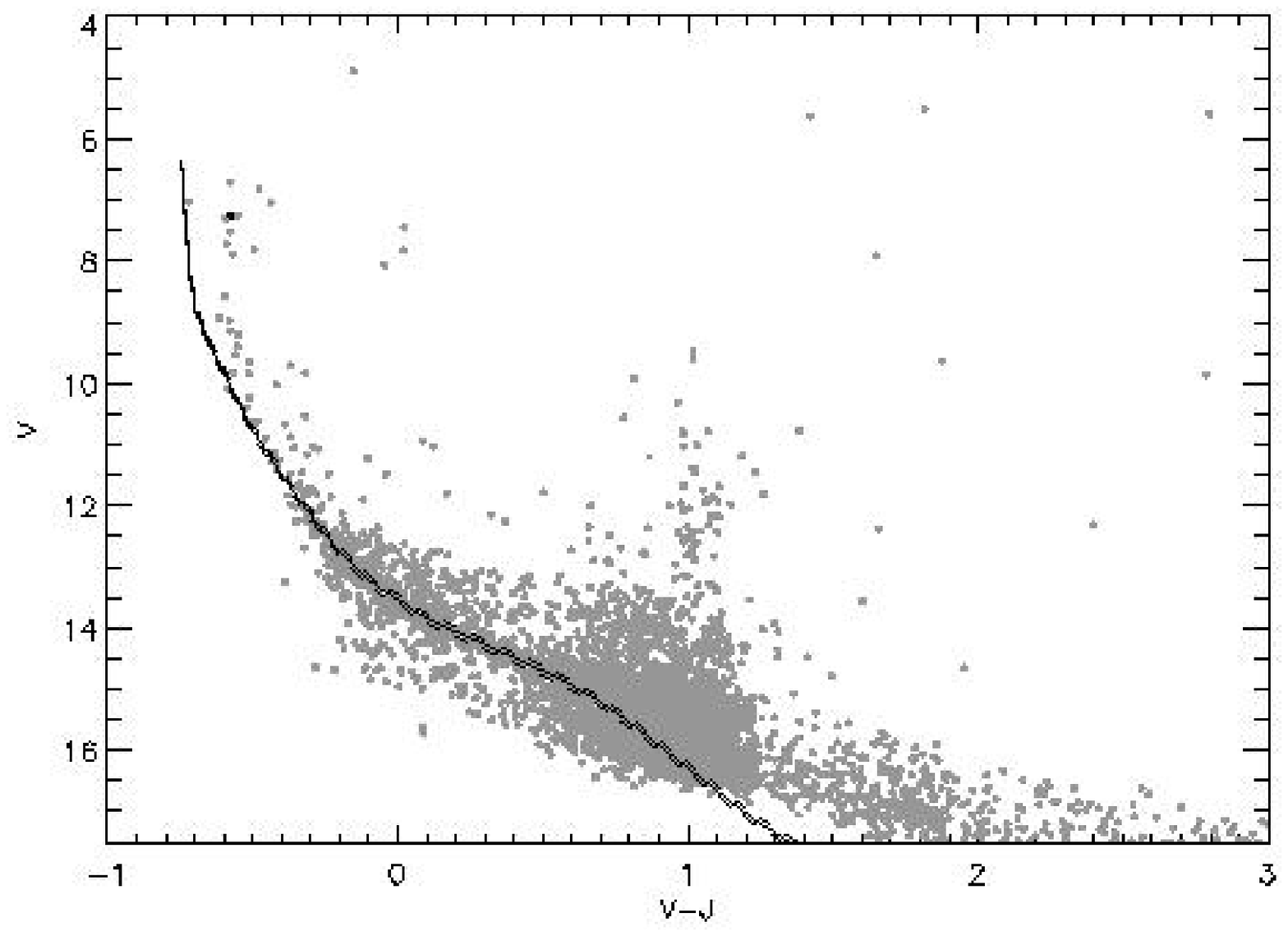}
\caption{Same as previous plot except for stars beyond 10' from the h and $\chi$ Persei centers, the halo population,
 with a distance modulus of dM = 11.85 (thick line) and extrema of dM = 11.77 and 11.93.}
\label{msfithalo}
\end{figure}

As shown by Figure \ref{Hess}, our new optical photometry 
reveals faint, low-mass stars plausibly associated with h and $\chi$ Persei 
whose V band luminosity at a given color is significantly higher than that 
of the background field star population.  The photometry extends over 14 magnitudes in 
V: over this range the positions of solar/subsolar-mass stars in V vs. V-I color-magnitude diagrams are 
a sensitive function of stellar age \citep[e.g.][]{Ba98}.  We now 
derive the pre-main sequence age of h and $\chi$ Persei stars by comparing their 
color-magnitude diagram positions to predictions from the \citet{Ba98} isochrones.  

Even though a plausible cluster locus is clearly observable and can be fit 
from visual inspection of Figure \ref{Hess}, we follow a slightly more statistical approach to obtain 
a more robust pre-main sequence age.  Field star contamination is larger for fainter stars.
A Hess diagram of all stars on the field will have an increased density in regions occupied by 
fainter cluster stars (at a given color), which may artificially dim the apparent cluster locus.  
Thus, to be conservative, we must subtract out the background field star population near the cluster locus.

Our method is as follows.  We construct Hess diagrams of stars within the h and $\chi$ Persei cores (r $<$ 7')\footnote{This 
radius is adopted instead of 10' because it yields a cleaner background subtraction.  We obtain identical results for the pre-MS age 
with a 10' radius.} and halo region (10'--20' away from both cluster centers).  Next, we construct a Hess diagram of the 
'background', which we obtain from stars greater than 21' distant from both cluster centers.  The cluster and halo Hess diagrams are 
 box-car smoothed by two resolution elements; the background diagram is smoothed by four resolution elements, essentially yielding 
an 'unsharp mask' of the background V/V-I distribution.  Finally, we scale the background diagram to the densities of the core/halo diagrams 
and subtract the background from the core/halo diagrams.  We determine the pre-main sequence age by overplotting solar metallicity 
\citet{Ba98} isochrones, assuming a mixing length parameter of 1.9 H$_{p}$ and distance moduli/reddenings equal to those found in previous sections.

Figures \ref{chiPerPreMS}, \ref{hPerPreMS}, \ref{haloPreMS} illustrate our results.  In each figure, the black line (dash-three dots) 
identifies the main sequence/post-main sequence locus from the Padova isochrones.  The solid dark grey line identifies the 
\citet{Ba98} isochrone position for 0.6 M$_{\odot}$ (lower right end) to 1.4 M$_{\odot}$ (upper left end) stars.   In Figures 
\ref{chiPerPreMS} and \ref{hPerPreMS}, the entire cluster locus is plainly visible.  The background has been almost completely 
subtracted out.   The halo population is more poorly defined (Figure \ref{haloPreMS}): the locus of cluster stars is entirely absent in some regions 
(V-I $\sim$ 1.5--1.6).  This poorer definition probably occurs because even regions $>$ 20' 
away have some small population of halo stars, so the halo population is 
partially subtracted out.  However, both the upper main sequence and the pre-main sequence redder than V $\sim$ 1.7--2 are clearly visible 
and well separated from the background field star population.  Comparing both panels of Figure \ref{haloPreMS} demonstrates that the 
apparent cluster locus in the subtracted diagram is real.

 Assuming that the faintest h and $\chi$ Per stars (V $\sim$ 23.5--24) are as reddened as the bright, spectroscopically observed 
stars analyzed in the previous section, they typically have intrinsic V-I $\sim$ 3.  According to Table \ref{dwarfcolors} and 
\citet{Ba98}, these stars likely have T$_{e}$ $\approx$ 3000 K and masses $\sim$ 0.1 M$_{\odot}$.  Although these determinations are highly uncertain, 
\textit{it is possible that the faintest cluster stars detected in our survey have masses approaching the hydrogen burning limit}.
At the very least, cluster stars with V $\sim$ 23--24 likely probe well into the M dwarf spectral type range.

Comparing the distribution of h and $\chi$ Per stars to the \citet{Ba98} isochrones clearly shows that the pre-main sequence ages 
for all three components of the Double Cluster are nearly identical to the post-main sequence ages:
\begin{eqnarray}
t_{pre-MS, \chi Per} = 14^{+1.9}_{-1.4} {\rm Myr} \\
t_{pre-MS, h Per} = 14^{+1.9}_{-1.4} {\rm Myr}\\
t_{pre-MS, halo} = 14^{+1.9}_{-1.4} {\rm Myr}.
\end{eqnarray}
In all cases, the 14 Myr isochrone best bisects the densest regions of the cluster loci: isochrones for 12.6 Myr overpredict the V band luminosity 
at a given color, while the 15.9 Myr isochrone underpredicts the luminosity.  Moreover, the slope of the 14 Myr \citet{Ba98} isochrone accurately 
tracks the observed locus.  This agreement is remarkable considering that 
the important processes driving post-main sequence evolution for high-mass stars -- changes in internal structure fundamentally due to nuclear reaction rates and 
the lack of hydrogen and helium in the stellar core-- are not the same as those driving pre-main sequence evolution for low-mass stars: Kelvin-Helmholtz contraction and 
the onset of hydrogen burning as a star follows a radiative or convective track onto the main sequence.

\subsection{Pre-Main Sequence Age Estimates Based Off of the \citet{Dm94} and \citet{Siess2000} isochrones}
The \citet{Ba98} isochrones are only one of many that are often used to determine cluster ages from the observed locii of pre main sequence stars.  
To compare the \citeauthor{Ba98} isochrones with others, we overplot 10--20 Myr isochrones from \citet{Dm94,Dm97} and \citet{Siess2000} on the color-magnitude 
diagram for $\chi$ Persei in Figures \ref{premsdm97} and \ref{premssiess}.  For both sets of isochrones, we transform the computed luminosities and 
effective temperatures into V band magnitudes and V-I color using our effective temperatures and intrinsic colors.   We display isochronal positions 
for stars more massive than 0.5 M$_{\odot}$.

While neither set of isochrones yields ages less than 10 Myr or greater than 20 Myr, they show a clear disagreement with 14 Myr \citet{Ba98} locus
and a failure to accurately reproduce the shape of the observed pre-main sequence locus.  The slope of both sets of isochrones are systematically steeper 
than the \citet{Ba98} isochrones.  The mismatch between the \citet{Dm94,Dm97} isochrones 
and the observed locus is particularly striking.  The 10 Myr isochrone overpredicts the luminosities of 1.5--2 M$_{\odot}$ stars (V-I $\sim$ 1--1.5) 
but underpredicts the luminosities of slightly redder stars (V-I $\sim$ 1.5--2).  The 20 Myr isochrone underpredicts the luminosity of cluster 
stars redder than V-I $\sim$ 1.  The \citet{Dm94,Dm97} grid lacks published entries at intermediate ages.  However, unless only the 1.5--2 M$_{\odot}$ stars
fade in V while at a constant temperature (unlikely), isochrones at intermediate ages would not clearly provide a better match to the observed locus.
Thus, we cannot derive a best-fit pre-main sequence age based off of these isochrones.

Compared to the \citet{Dm94,Dm97} isochrones, the \citet{Siess2000} isochrones show slightly better agreement with the observed locus, especially for stars 
redder than V-I $\sim$ 1.5.  In particular, the 14 Myr and 16 Myr isochrones (dashed lines) show reasonably good agreement for V-I $\sim$ 1.75--2.5.  In contrast, 
the 12 Myr and 18 Myr isochrones systematically overpredict and underpredict the luminosities of cluster stars over this range in V-I.  
Thus, based on the \citet{Siess2000} isochrones for stars with V-I $\sim$ 1.75-2.5, $\chi$ Persei has an age of $\approx$ 14--16 Myr, though this estimate 
is somewhat uncertain given the poorer fidelity that the isochrones have to the slope of the locus for V-I = 1.5--2.5.

For bluer, earlier stars (V-I $\sim$ 0.75--1), all \citet{Siess2000} isochrones systematically overpredicts luminosities.  Based solely on the apparent 
position of the pre-main sequence "turn on" -- where the observed locus first becomes brighter than the main sequence -- the \citeauthor{Siess2000} 
isochrones would yield an age greater than 20 Myr, which is inconsistent with all other age estimates.  Moreover, \citeauthor{Siess2000} predicts 
that over a narrow color range the V band luminosities of cluster stars \textit{increase} with redder color and thus with decreasing mass.
We find no clear evidence for such a trend in the data.  Likewise, none of the \citet{Ba98} isochrones for 0.6--1.4 M$_{\odot}$ stars 
and ages of 1--20 Myr have this trend: it is likely a unique feature of the \citet{Siess2000} models.

In summary, comparing the photometric data with isochrones from \citet{Dm94,Dm97} and \citet{Siess2000} constrains the age of h and $\chi$ Persei to be between 
10 Myr and 20 Myr.  However, the shape of these isochrones in V vs. V-I for any age provide a significantly poorer match to the observed locus.
These comparisons justify our choice of the \citet{Ba98} isochrones to determine pre-MS ages, given their far stronger agreement with the observed 
cluster locus shape for a wide range of color.  
Adopting the \citeauthor{Ba98} tracks as our benchmark, the \citeauthor{Dm97} tracks are inaccurate for redder pre-MS 
stars.  They are potentially accurate for bluer, higher-mass pre-MS stars, though this is uncertain given their sparse age sampling.  The \citet{Siess2000} 
tracks yield reasonable age estimates for redder pre-MS stars but are inaccurate for bluer, higher-mass pre-MS stars.
If these trends are indicative of the isochrones for all ages, we caution against estimating cluster ages from the main sequence turn-on from the \citet{Siess2000} 
tracks and against deriving ages from the observed locus of red pre-MS stars from the \citet{Dm94,Dm97} tracks.

\section{Membership in h and $\chi$ Persei}
Despite the huge volume of research devoted to studying the Double Cluster, essentially no studies formally 
list h and $\chi$ Persei's main sequence/pre main sequence members.  Most membership lists consist of 
post-main sequence B stars and red supergiants on the cluster field identified by visual inspection, spectroscopy, and proper motions 
\citep[e.g.][]{Sl02, Uribe2002}.  
There has been far less progress in identifying (pre) main sequence cluster members.  Nearly all work 
focused on such stars simply identifies candidate cluster members in different color-color diagrams, without listing these members, 
or it focuses on particularly x-ray luminous members \citep{Cu07a, CurrieEvans2009}.

As demonstrated in previous sections, our spectroscopic and photometric data reveal loci of post-main sequence, main 
sequence and pre-main sequence stars that are clearly separated from the population of older field stars.  
These data allow us to identify likely members from optical spectroscopy
and VI photometry, greatly expanding upon previous studies.  In this section, we 
determine the first catalog of h and $\chi$ Persei members that includes main sequence and pre-main sequence stars.

\subsection{Method of Membership Determination}
Using the dereddened V vs. spectral type HR diagram, we first establish a spectroscopically determined list of h and $\chi$ Persei members.  
Because each component of h and $\chi$ Per has an identical post main sequence and pre-main sequence age, we 
define the locus of cluster members by joining together the Padova 14 Myr post main sequence isochrone and the Baraffe 14 Myr 
pre-main sequence isochrone.  The width of the cluster locus is determined by a) the physical extent of the cluster, b) binarity, 
and c) uncertainties in spectral types.  Including the halo population, the h and $\chi$ Per region extends to at least 
$\approx$ 20--25' away from either cluster center.  Assuming a distance modulus of $\approx$ 11.8--11.85 and spherical distribution of cluster 
and halo stars, the h and $\chi$ Persei region is $\approx$ 52.5 pc in diameter, which translates into $\delta$(dM)= $\pm$ 0.05.  
The typical dispersion in E(B-V) as determined from the V-J, V-H, and V-K colors is $\lesssim$ 0.03.  This uncertainty in 
E(B-V) leads to a formal error in the dereddened V magnitude of $\approx$ 0.09 ($\approx$ 0.1) for each source, which is larger than the $\delta$(dM) from 
the clusters' physical sizes and slightly widens the V$_{o}$ vs. spectral type locus.  The uncertainty in the distance modulus is $\approx$ 0.08 (see \S 3).
Equal mass binaries in h and $\chi$ Per may be up to 0.75 mags more luminous than the cluster locus for single stars.
Finally, we compute the median formal spectral type uncertainties as a function of V magnitude and define the width of the locus from left to right 
by these uncertainties.  The uncertainties are $\approx$ 2 subclasses for most spectral types.  Therefore, we identify 
cluster stars in a locus centered on the 14 Myr Padova-Baraffe isochrone with boundaries of -0.75 and +0.2 in V magnitude above and below the 
isochrone and typically $\pm$ 2 subclasses in spectral type.  Figure \ref{specmemill} illustrates.

Second, we identify probable photometric members from the V vs. V-I color-magnitude diagram again using the 14 Myr Padova-Baraffe isochrone 
reddened by the median average cluster reddenings.  The upper bound defining the cluster locus 
is the same as before; for the lower bound, we choose the larger of 0.05 (the physical size of the clusters in mags) and the V band uncertainty.
  We choose the boundaries in color assuming that the dispersion in reddening ($\approx$ 0.1 mag) and photometric errors 
in V-I color (($\sigma_{V}$$^{2}$+$\sigma_{I}$$^{2}$)$^{0.5}$) widen the locus.  The dispersion in E(B-V) translates into a dispersion of 0.125 mags in 
V-I.  The median errors in color range from $\lesssim$ 0.01 mags at V $\le$ 10--12 to 0.08 mags at V = 23.  Therefore the widening of the left/right boundaries 
of the locus from these two sources of error bound of the locus increases from $\sim$ 0.125 mags for bright stars to $\sim$ 0.2 mags for faint stars.
  Figure \ref{photmemill} illustrates the region within with we identify probable members.

\subsection{Membership Estimates}
\textbf{Spectroscopic Membership} --Based on our selection criteria, we identify a total of 4,702 stars as spectroscopic members: 
$\sim$ 63\% of the 7,465 spectroscopically observed stars with V-band photometry.  The frequency of members is highest for the 
cluster-dominated regions.  Within 10' of the h Persei center, 71\% (1138/1606) of stars are members.  Within 10' of the $\chi$ Persei center,
 68\% (906/1330) of stars are members.  The slightly lower membership frequency of 
59\% (2676/4529) in the low-density halo regions is consistent with the region's greater level of field star contamination.

\textbf{Photometric Membership} -- The number of photometric candidate members is substantial: 14,307 stars, or 
$\sim$ 30\% of the 47,060 stars on the field.  Over half of these stars reside in the low-density regions surrounding the 
cluster cores (8,817, or 62\% of the total population).  The other candidates are almost evenly split between those within 10' of the 
h Persei center (2,724) and 10' of the $\chi$ Persei center (2,766).  Within both core-dominated regions, about 40\% of 
stars are probable members: 2724/6519 for h Persei and 2766/6674 for $\chi$ Persei.  As with the spectroscopic 
sample, field contamination is far heavier in the halo region: $\sim$ 26\% of halo-region stars are consistent with membership (8817/33867).

Taken at face value, these results seem to contradict \citet{Cu07a} and \citet{Currie2008} who argue that the core-dominated regions 
have as many stars as the halo.  Furthermore, they conflict with previous studies showing that h Persei is $\sim$ 30\% more 
massive/populous than $\chi$ Persei \citep[e.g.][]{Sl02, Bragg2004, Bk05}.  The first difference is one of semantics: both papers identify the halo region with stars 
at distances greater than 15' away from either cluster center and extending to 25' distant, not 10' as in this paper.  
Adopting the previous definition of the core and halo regions yields much more similar results.  For example, the number of photometric members 
would be 4808 (out of 14567) for the h Persei core, 4993 (out of 15,026), for the $\chi$ Persei core, and 4283
(out of 17,467) for the h and $\chi$ Per halo population.  We primarily adopt 10' as the core boundary because 
the density of stars is low \citep{Cu07a} and our 
spectroscopic survey is unbiased within this radius, though we consider either definition to be reasonable.

The conflicting results for cluster richness are probably due to survey incompleteness near the h Persei core.  
The h Persei core has a higher density of early, bright stars \citep[e.g.][]{Cu07a, Bk05, Sl02}.  The scattered light 
from these stars reduces the contrast between faint, low-mass stars and the background, which restrict source detection 
to a brighter limit than for the $\chi$ Persei core.  Restricting our analysis to brighter stars recovers previous 
results.  For example, we identify 607 stars brighter than V = 16 -- likely earlier than A2 (see Figure \ref{HSdiagram})-- 
within the h Persei core as photometric members but only 469 within the $\chi$ Persei core as members.  Adopting these 
criteria implies that h Persei is $\sim$ 30\% more massive, in complete agreement with previous results.  Assuming the standard h Per to $\chi$ Per mass 
ratio, there are 700 more lower-mass h Per stars unaccounted for by our survey, bringing the total number to $\sim$ 3424.

\subsection{Membership Estimates Corrected for Field Star Contamination}
Some stars identified as members are likely interlopers.
In the HR diagrams and color-magnitude diagrams presented here, 
the locus of cluster stars intersects the main locus of field stars around G spectral types.
Thus, we expect some field star contamination among G stars. 
The increasing scatter of the colors of field stars with increasing V probably 
also leads to some contamination along the lower boundary of the pre-main sequence locus in V/V-I.
  We cannot subtract a field star population from 
the cluster population and estimate the number of interlopers, 
because the halo region fills the entire optical survey. 

Comparing the membership results for stars based on photometry and spectroscopy yields an estimate of the field 
star contamination in the photometric member list.  There are 3,984 stars identified as candidate photometric members 
\textit{and} spectroscopic members; 763 stars identified as candidate members from photometry are rejected as members 
based on spectroscopy.  Therefore, about $\sim$ 16\% of the photometric member list is comprised of interloping field stars, 
not bona fide cluster members.  Even though field contamination is greater for fainter magnitudes, analysis in \S 3.3 
demonstrates that the region defining the locus of low-mass h and $\chi$ Per stars (V $\sim$ 18--24, V-I $\sim$ 1.5--4) 
is less contaminated than the region (V $\sim$ 16.5--17.5, V-I $\sim$ 1--1.5) comprising most of our spectroscopically observed members.

The number of B stars can also be used to estimate contamination. Because early to mid B dwarf stars identified 
as members above cannot be old field stars, they are highly likely to be young cluster members.  
Assuming a cluster mass function (e.g. Miller-Scalo), the number of B stars yields an expected 
number of total cluster stars.  By comparing the ratio of the expected 
number of members to the derived number of members, we estimate the number of 
field stars misclassified as members/candidate members.  We choose the B0--B6 star population 
in the halo as our reference, assume that the B0--B6 star population extends to V $\sim$ 13.45 (cf. 
Figure \ref{HSdiagram}) and assume that the field star density is independent of 
position.  Main sequence B6 stars should 
have masses of $\approx$ 4 M$_{\odot}$ (or log(M$_{\star}$/M$_{\odot}$) $\approx$ 0.6)\footnote{The 
Padova isochrones list T$_{e}$ $\sim$ 14,600 K for 4 M$_{\odot}$ main sequence stars, which places at 
a spectral type of $\approx$ B5.75 on our T$_{e}$ scale.}.  We assume that our survey reaches 
stars with masses $\approx$ 0.2 M$_{\odot}$ (log(M$_{\star}$/M$_{\odot}$ $\sim$ -0.7): this lower 
limit is higher than that determined in \S 3.3 ($\sim$ 0.1--0.15 M$_{\odot}$) but is chosen to be conservative.  We derive the expected 
number of cluster stars using Table 9 in \citet{MillerScalo1979}.

The halo population includes 217 candidate members brighter than V = 13.45, so it should contain 
about 7,380 stars total.  Compared to the 8,817 stars listed as candidate members, the percentage 
of interloping field stars is also $\sim$ 16\%, which is essentially identical to the estimate derived by comparing 
membership lists.  Because the area of the photometric survey consisting 
of the halo region is $\approx$ 0.43 square degrees -- 0.6 sq. deg - 2$\times$$\pi$10'$^{2}$ -- 
the density of interloping field stars is $\sim$ 3,381 deg$^{-2}$, which yields $\sim$ 295 interlopers 
in each core region.  Thus, within our 0.6 square-degree field there are probably 
3,129 cluster members for h Persei, 2,471 for $\chi$ Persei, and 7,380 for the low-density halo, yielding a total of 12,980
stars in h and $\chi$ Persei.  If lower-mass halo stars associated with the halo cover one square degree on the sky, 
consistent with our spectroscopic membership results for higher-mass stars, \textit{\textbf{then h and $\chi$ Persei may contain up 
to $\approx$ 20,000 members}}.  

Several other factors indicate that our membership estimate is reasonable and conservative.  First, the color-magnitude diagram 
boundaries identifying photometric members assume that all h and $\chi$ Per stars have E(B-V) within 0.1 mag 
of the derived median values for each core region and the halo.  However, our spectroscopic membership list 
clearly identifies some members with reddening outside this range.  It is plausible that other members 
 lacking spectroscopic data lie outside the photometric membership boundaries but would deredden to the cluster locus. 
Second, analysis in \S 3.3 indicates that our survey may have reached close to the hydrogen burning limit ($\approx$ 0.1 M$_{\odot}$).  
If so, then the expected number of cluster stars determined from extrapolating the cluster IMF from the B star population 
will be slightly greater than derived in this section.  Because more cluster stars are expected, the number of interlopers drops.
Third, because the regions identifying members are defined purely by analysis uncertainties, we implicitly 
assume that all 13,000--20,000 stars in h and $\chi$ Persei formed simultaneously.  Analysis in \S 3 indicates that the median 
age of h and $\chi$ Per stars is 14 Myr with no evidence for a substantial age spread.  
However, our analysis cannot preclude the existence of smaller, 1--2 Myr age spreads \citep[see 
also discussion in][]{Sl02}.  Even 1--2 Myr age spreads would widen the cluster loci and cause us to 
underestimate the true number of members.

\subsection{Combined Membership List}
After removing the 763 candidate photometric members rejected by spectroscopy, we compile 
a final membership list of 14,160 stars including both types of members.  For each entry, we list ID numbers, 
the position, optical photometry, spectral types, and reddening.  Table \ref{memberlist} lists 
the members of h and $\chi$ Persei.

\section{The Mass and Structure of h and $\chi$ Persei}
Using our membership lists, we can explore other bulk properties of the Double Cluster: 
its mass and spatial distribution.  Following previous investigations \citep{Sl02, Bk05}, 
we estimate the total cluster mass by determining the total mass in stars above some 
mass limit and extrapolating to lower masses by assuming a cluster mass function.
The spatial distribution of photometric members reveals complex structure in the 
regions with a high density of cluster stars.  By comparing the number of stars with 
different masses, we probe the clusters' mass segregation, which yields information 
in their formation histories.

\subsection{Cluster Mass}
With a list of spectroscopically confirmed members, we estimate the 
total mass of h and $\chi$ Persei and probe the spatial distribution of cluster
 stars in core-dominated regions.  To determine the mass, we add up the 
derived masses of stars earlier than B6 (4 M$_{\odot}$), assume a Miller-Scalo IMF, calculate 
the masses of the core regions, and use the ratio of members in the 
core and halo to determine the total mass.   We use the spatial density 
of cluster stars to investigate mass segregation.

From these assumptions, we derive a mass of 4,704 M$_{\odot}$ for the h Persei core
and 3,699 M$_{\odot}$ for the $\chi$ Persei core.  Assuming the ratio of the core 
populations to the halo (5600/7380), the halo population has a total mass 
of 11,074 M$_{\odot}$.   Not including the halo regions beyond our photometric coverage, 
the total mass of h and $\chi$ Persei is $\approx$ 19,477 M$_{\odot}$: an 
order-of-magnitude larger than the Orion Nebula Cluster and the Pleiades.

Our deep spectroscopic survey supports the conclusion of \citet{Sl02} that the 
h and $\chi$ Persei region is unique in mass and structure among nearby (d $<$ 3 kpc) regions and 
much more similar to other massive 'double clusters' such as NGC 1818 in the Large Magellanic Cloud \citep{Ja01} 
and massive single clusters such as the Arches cluster near the galactic center \citep{Fi99}.
\subsection{Mass Segregation}
Mass segregation also probes cluster structure.
Dynamical mass segregation occurs because cluster members gravitationally 
attract one another, exchanging kinetic energy and momemtum, which tends to 
drive the system into energy equipartition.  As the cluster ages, this process leaves more massive cluster stars more 
concentrated in the cluster core and less massive stars concentrated in lower-density regions.  
Alternatively, clusters may exhibit \textit{primordial} mass segregation, which occurs 
as an outcome of the cluster formation process.

To probe mass segregation, we compute the relative number of B star members to A and F star members as a function 
of distance from the cluster centers.  If the clusters lack mass segregation, then the ratio of 
B stars to A/F members should be nearly constant with distance.  Clusters with mass segregation 
should have a decreasing ratio with cluster-centered distance.  

As shown in Figure \ref{massseg}, both clusters exhibit some evidence for mass segregation.  The strongest 
evidence is for h Persei stars (top) within $\approx$ 3--4' of the cluster center.  Within this region, 
the relative number of B stars to A stars drops from $\sim$ 4.5 within 1' to $\sim$ 2 for larger distances. 
The ratio of A stars to F stars is also highest within 2' and reaches a constant level ($\sim$ 0.5--1) by $\sim$ 3'.  
The ratio of the number of B stars to F stars, comparing a wider range in masses ($\ge$ 2.2 M$_{\odot}$ to 1.35--1.5 M$_{\odot}$), 
varies the most, exceeding 22 within 1' and reaching constant levels at distances greater than 4'.  In all cases, 
the high number of B stars to later stars implies that higher-mass stars are preferentially concentrated in 
the cluster centers, consistent with the cluster being mass segregated.

For $\chi$ Persei, the extent of mass segregation and the physical scale over which it occurs may be comparatively smaller.  
The ratio of B stars to F stars is high ($\sim$ 4--14) within 2' of the cluster center and drops to a constant 
level by 3'.  The ratios of B stars to A stars and A stars to F stars is high for stars within 2' of the cluster center after which they reach constant levels.  
These trends are qualitatively similar to those found for h Persei.  However, the smaller B/F star ratio in 
the cluster center and slightly smaller radius beyond which the ratio is constant (3' vs. 4' for h Per) indicates 
that mass segregation in $\chi$ Persei is not as strong as it is in h Persei \citep[see also][]{Sl02, Bk05}.

Comparing the dynamical mass segregation timescales with cluster ages determines whether 
mass segregation must be primordial or dynamical.
Dynamical mass segregation occurs on relaxation timescales \citep{Bt87}, which for h and $\chi$ Persei is:
\begin{equation}
t_{relax} \sim \frac{1.8\times10^{10} yr}{ln \Lambda(r)}[\frac{\sigma(r)}{10 km s^{-1}}]^{3}
(\frac{1 M_{\odot}}{m})[\frac{10^{3}M_{\odot} pc^{-3}}{\rho(r)}],
\end{equation}
where $\Lambda$ is the number of stars interior to radius r, $\sigma$ is the velocity dispersion, 
m is the typical stellar mass (0.63 M$_{\odot}$), and $\rho$(r) is the stellar density.  \citet{Bk05} compute a
 relaxation timescale of $\sim$ 15 Myr for h Persei at a core radius of 2.8' and $\sim$ 20 Myr 
for $\chi$ Persei at a core radius of 3.5'.  The age of h Persei is almost identical to its relaxation 
timescale at 2.8' from the cluster center.  While the age of $\chi$ Persei is 5 Myr less than its relaxation timescale, 
the cluster only clearly exhibits mass segregation at distances within 2'--3' from the cluster center.  Because 
the relaxation timescale is inversely proportional to stellar density and the natural log of the number of stars, 
the relaxation timescale for regions interior to 2'--3' is much less than 20 Myr.  Conversely, even though 
the relaxation timescale rises well above 15--20 Myr for distances greater than 3'--4', there is little evidence 
for mass segregation at these separations.  Thus, it is not clear whether mass segregation in h and $\chi$ Persei 
must be primordial.   Similarly, for the Arches cluster it is not clear whether mass segregation is primordial 
or dynamical.

\section{Summary and Future Work}
\subsection{Summary of Analysis and Major Findings}
This paper describes the first extensive photometric and spectroscopic survey of 
main sequence and pre-main sequence stars in h and $\chi$ Persei.
  By analyzing optical photometry for 47,000 stars and spectroscopy of 11,000 stars, 
we derived the median reddening, distance modulus, and age for each component 
of h and $\chi$ Per.  We then constructed the first extensive membership list of h and $\chi$ Per stars, 
ranging from massive B--M supergiants to pre-main sequence stars whose masses are likely 
comparable to the hydrogen burning limit.  Our study yields the following major results:

\begin{itemize}
\item The median reddening values for the h Persei core, $\chi$ Persei core, and halo region are 
 comparable: E(B-V) $\sim$ 0.55, E(B-V) $\sim$ 0.52, and E(B-V) $\sim$ 0.52, respectively.

\item The distance moduli for h Per, $\chi$ Per, and the halo region are also nearly identical: 
dM$_{h Per}$ = 11.8, dM$_{\chi Per}$ = 11.85, and dM$_{halo}$ = 11.85.

\item Ages for all three components of h and $\chi$ Persei are identical: $\sim$ 14 Myr.  
Moreover, post-main sequence ages and pre-main sequence ages for each component are identical.
  Thus, the properties of h and $\chi$ Persei are consistent with the coeval cores and the low-density 
halo population emerging from a single, explosive star-forming event.

\item Within 10' of the two cluster centers, the Double Cluster contains at least $\sim$ 5,000 stars; h Persei is 
about 30\% more populous than $\chi$ Persei.  The halo region contains at least $\sim$ 7,000 stars 
and as many as 15,000 stars, bringing the total number of stars in h and $\chi$ Persei to 
$\sim$ 20,000.  The estimated masses for h Persei, $\chi$ Persei, and the 
halo region are $\sim$ 4,700 M$_{\odot}$, $\sim$ 3,700 M$_{\odot}$, and $\sim$ 11,000 M$_{\odot}$.

\item Both clusters show clear evidence of mass segregation within 3' of the cluster centers, though 
it is stronger for h Persei.  The relaxation timescales for both clusters are comparable to or less than their 
ages within 3'.  Therefore, mass segregation may either be primordial or dynamical. 
\end{itemize}

These results support and extend recent studies of h and $\chi$ Per conducted by \citet{CurrieEvans2009}, \citet{Mayne07}, 
\citet{Cu07a}, \citet{Bk05}, \citet{Sl02}, and \citet{Ke01}.  Remarkably, four mutually exclusive sets of authors using 
different techniques converge on essentially the same properties for the clusters' reddening, distance, and age.
Combined with previous work, our results clearly refute earlier claims that the Double Cluster and its environs have 
a substantial age spread \citep{Wildey1964, Marco2001}, have a substantially different age \citep{Schild1965, Schild1967, Marco2001}, or are 
located at substantially different distances \citep{Schild1967, Kharchenko2005}.   All indications are 
that the different components of h and $\chi$ Persei substantially differ only by spatial distribution and mass.

\subsection{Future Research on the Stellar Population of h and $\chi$ Persei}
\begin{figure}
\centering
\epsscale{1.}
\plottwo{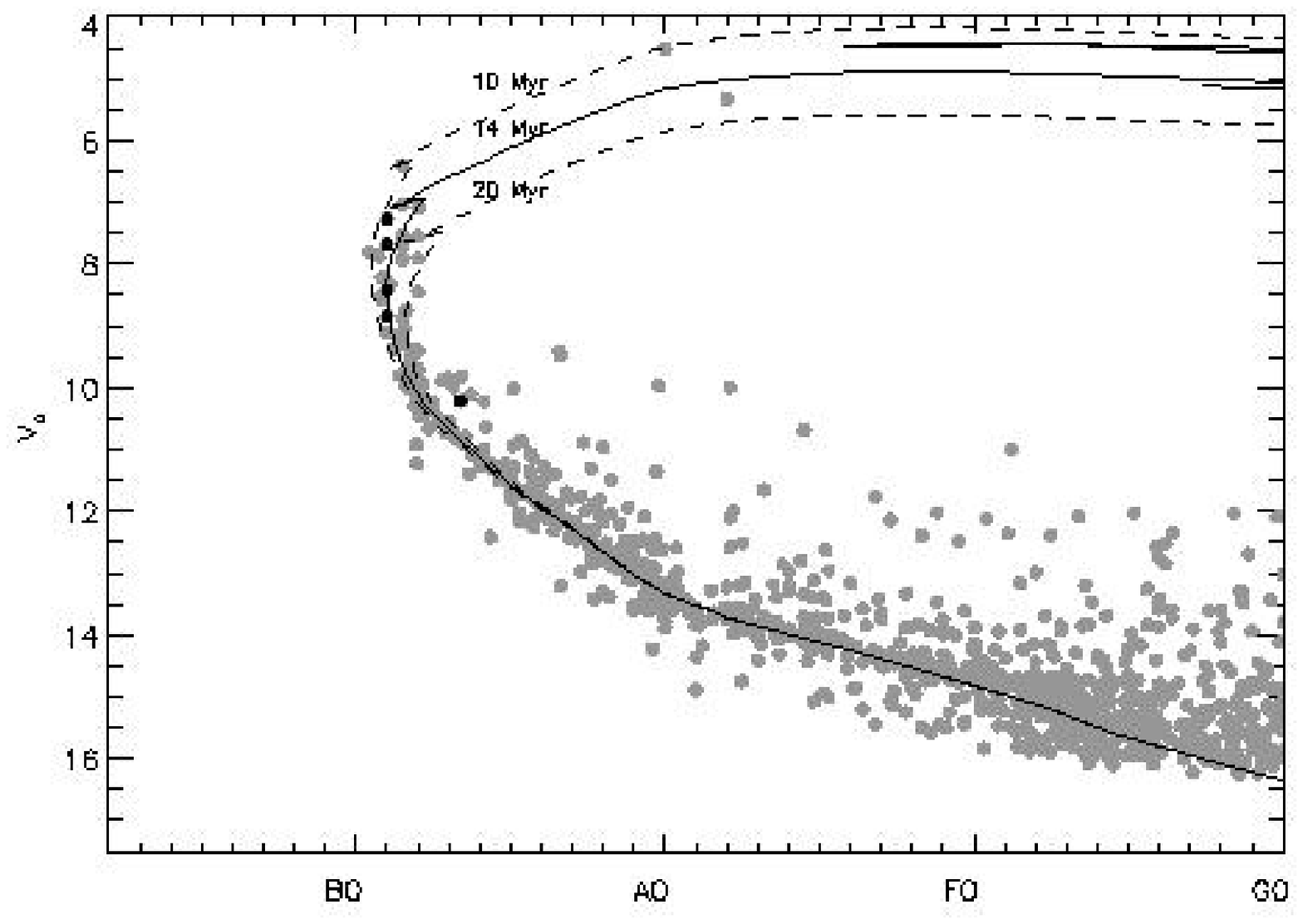}{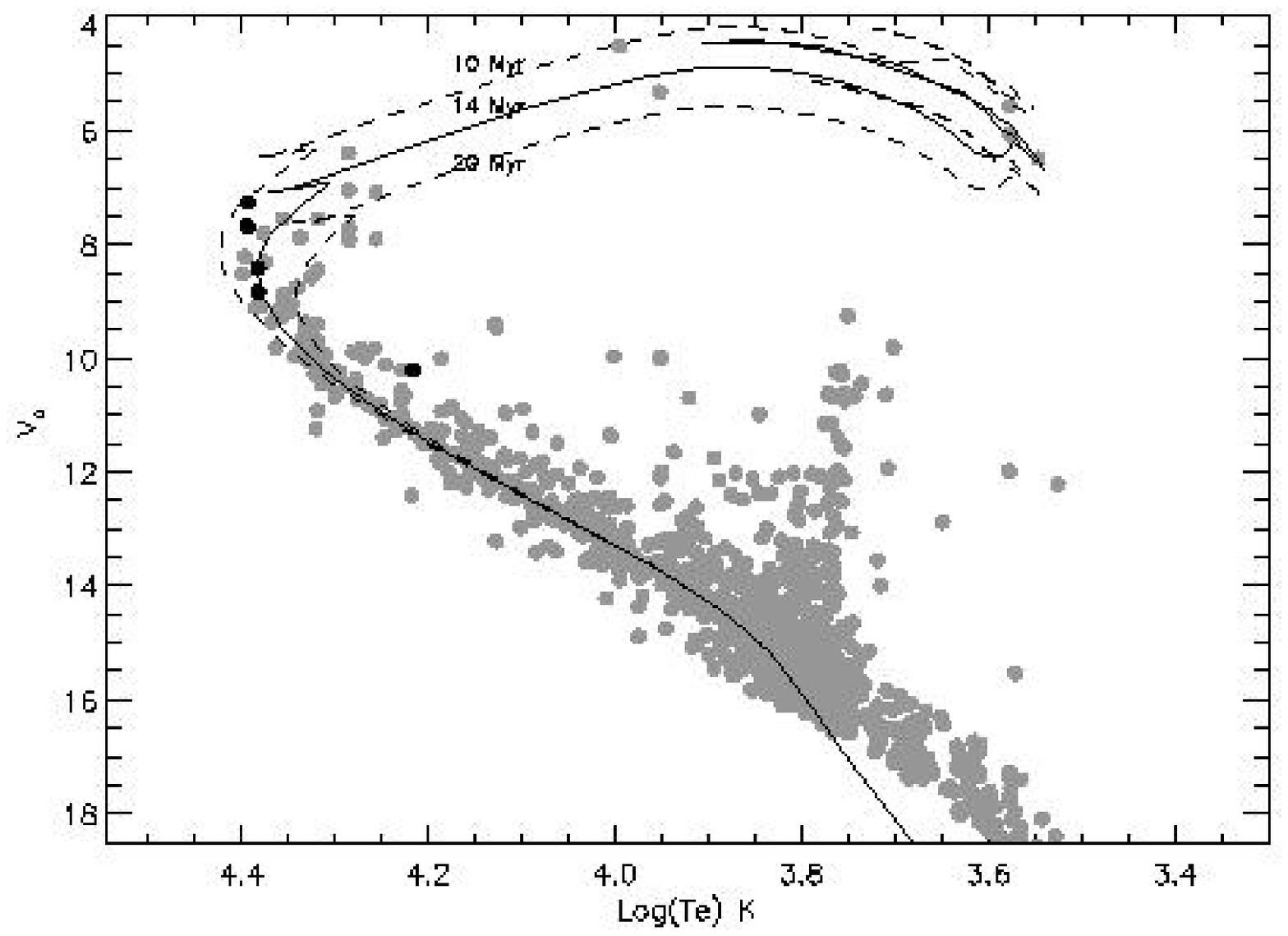}
\plottwo{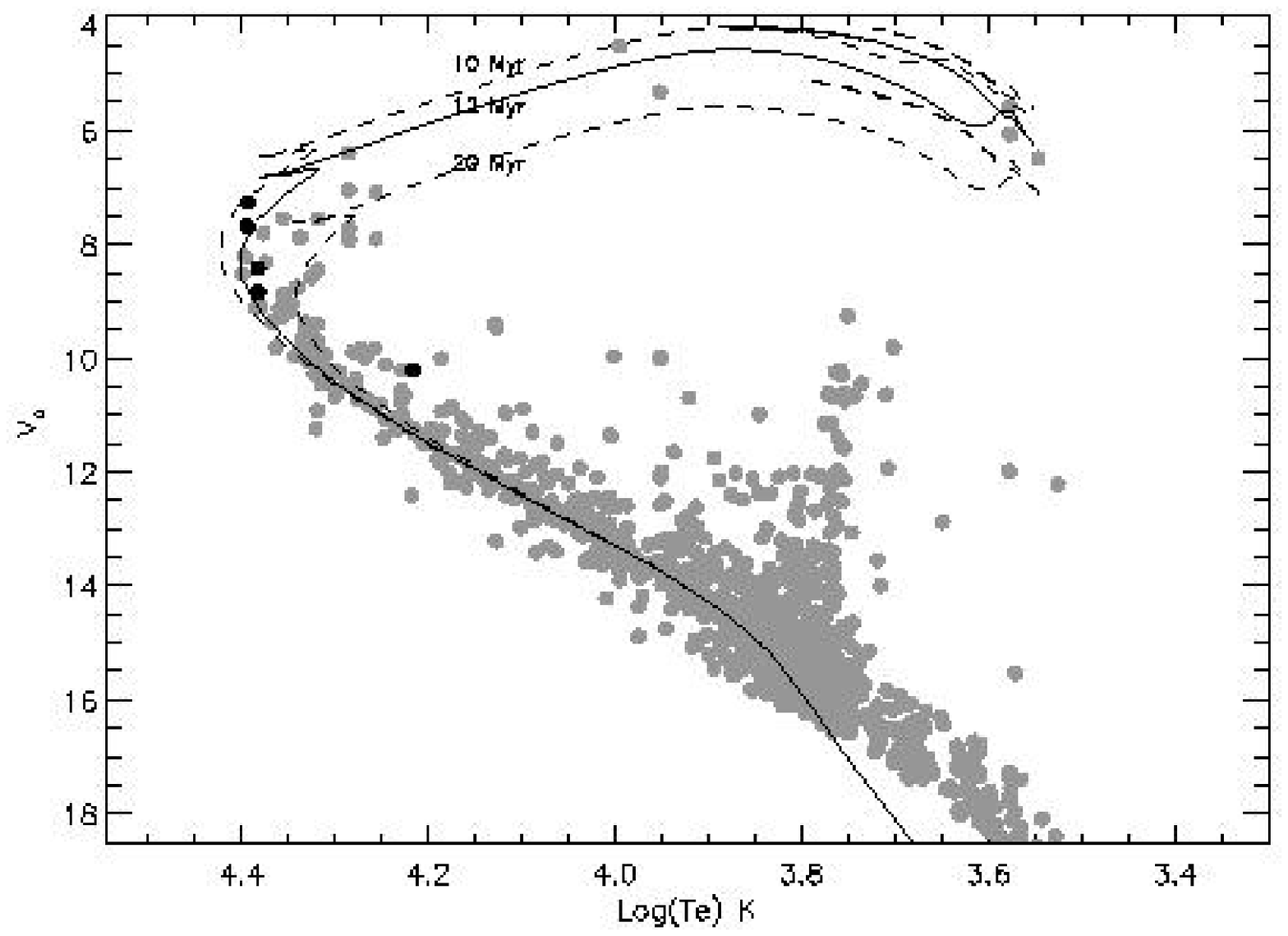}{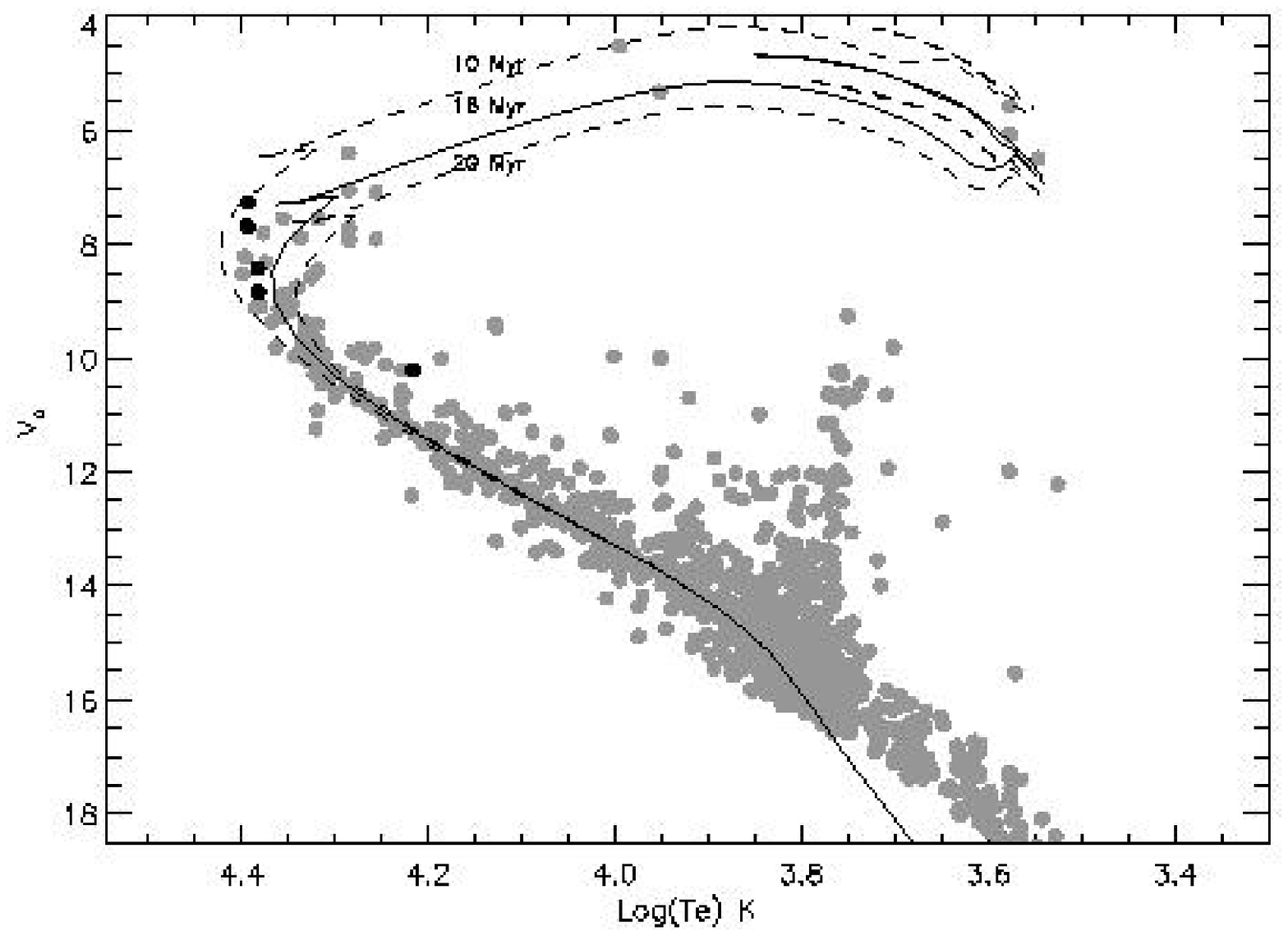}
\caption{(Top Panels) Dereddened V vs. spectral type (left) and dereddened V vs. log(T$_{e}$) (right) HR diagrams showing 
the main sequence turnoff for $\chi$ Persei.  Black dots indicate the positions of Be stars.  
Overplotted are the Padova post-main sequence isochrones corresponding to 
ages of 10 and 20 Myr (dashed lines) and 14 Myr (solid line).  (Bottom Panels) The dereddened 
V vs. log(T$_{e}$) diagrams replacing the 14 Myr isochrone with 12 Myr and 16 Myr isochrones.}
\label{msturnoffchi}
\end{figure}

\begin{figure}
\centering
\plottwo{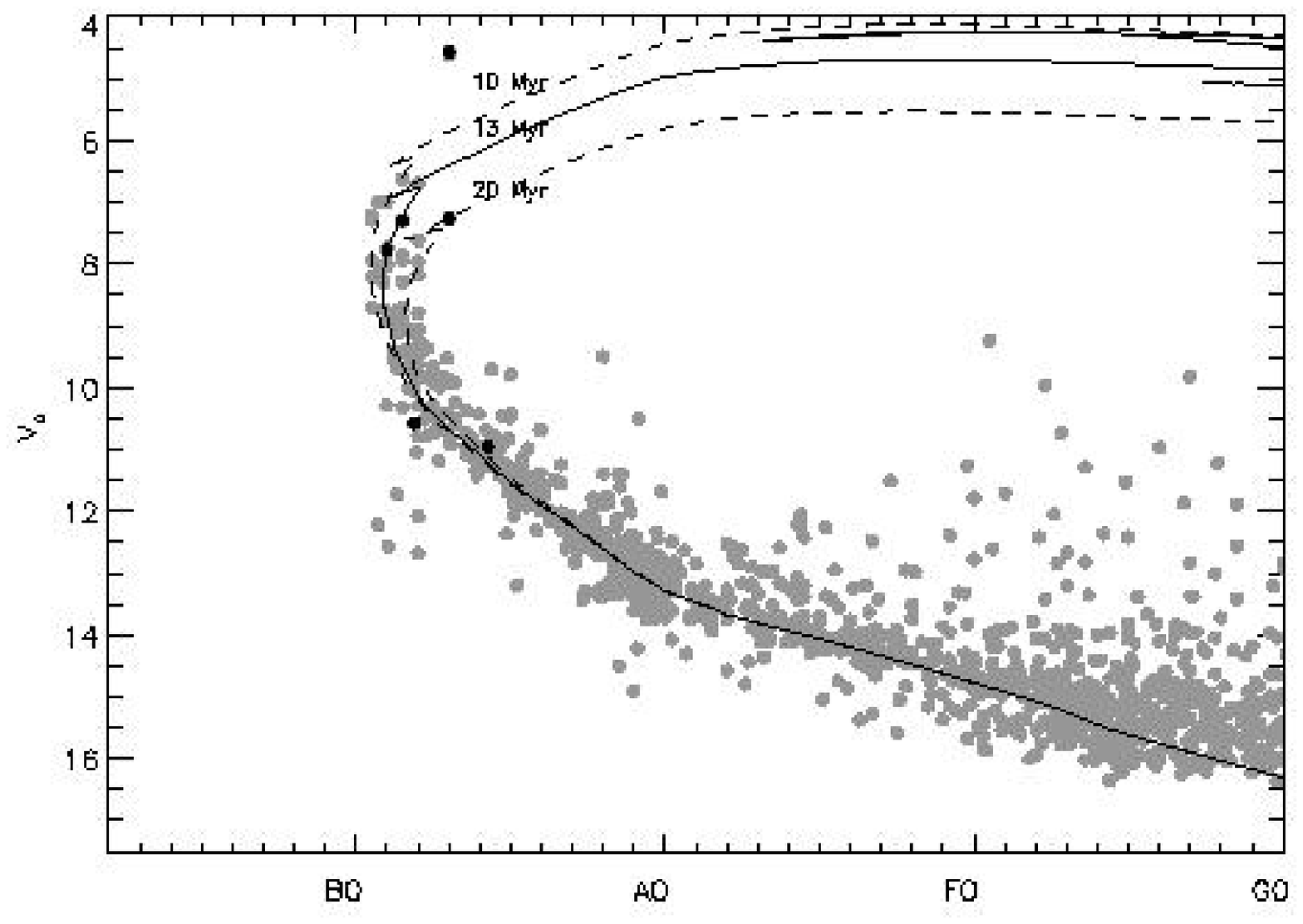}{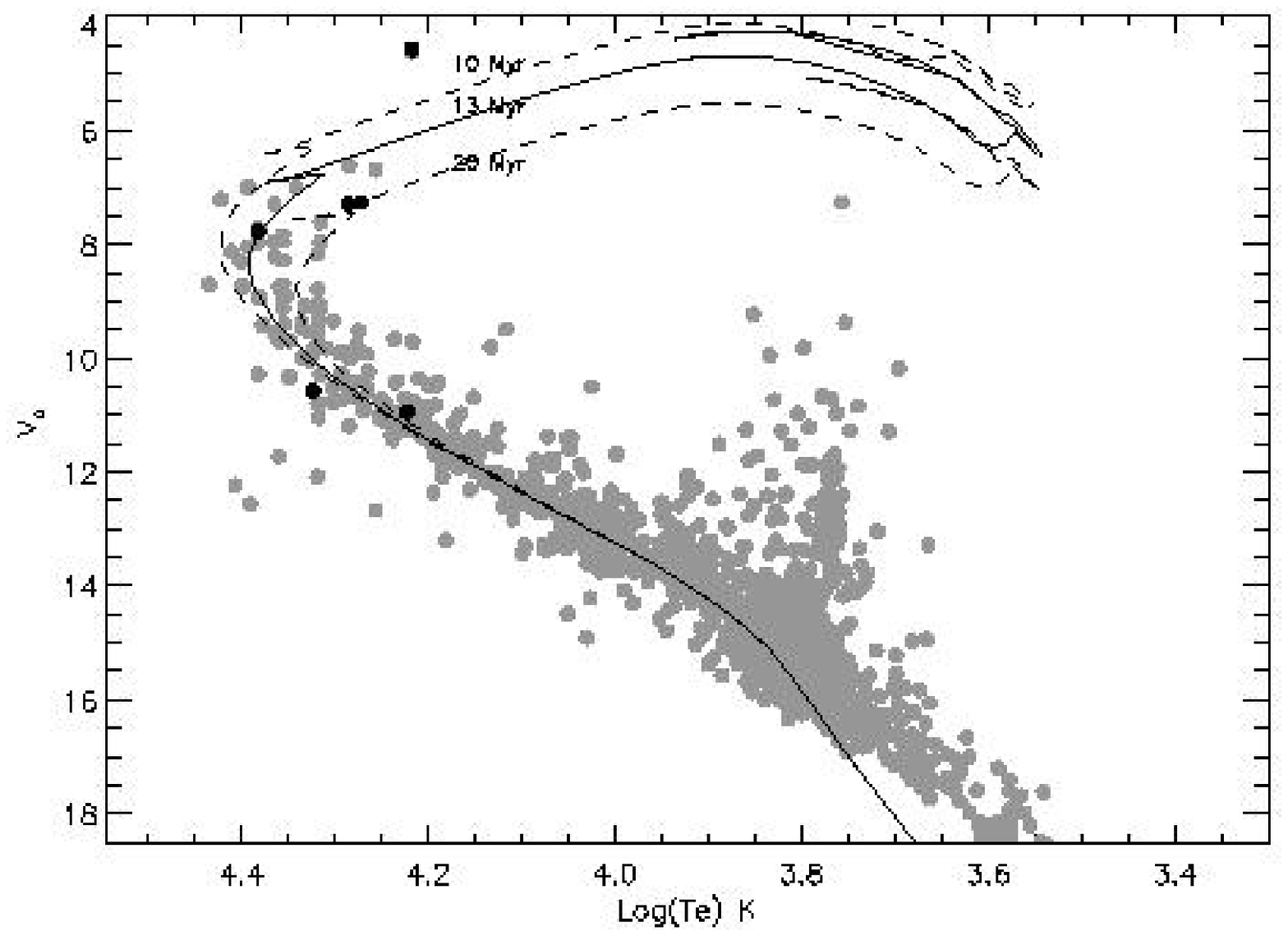}
\plottwo{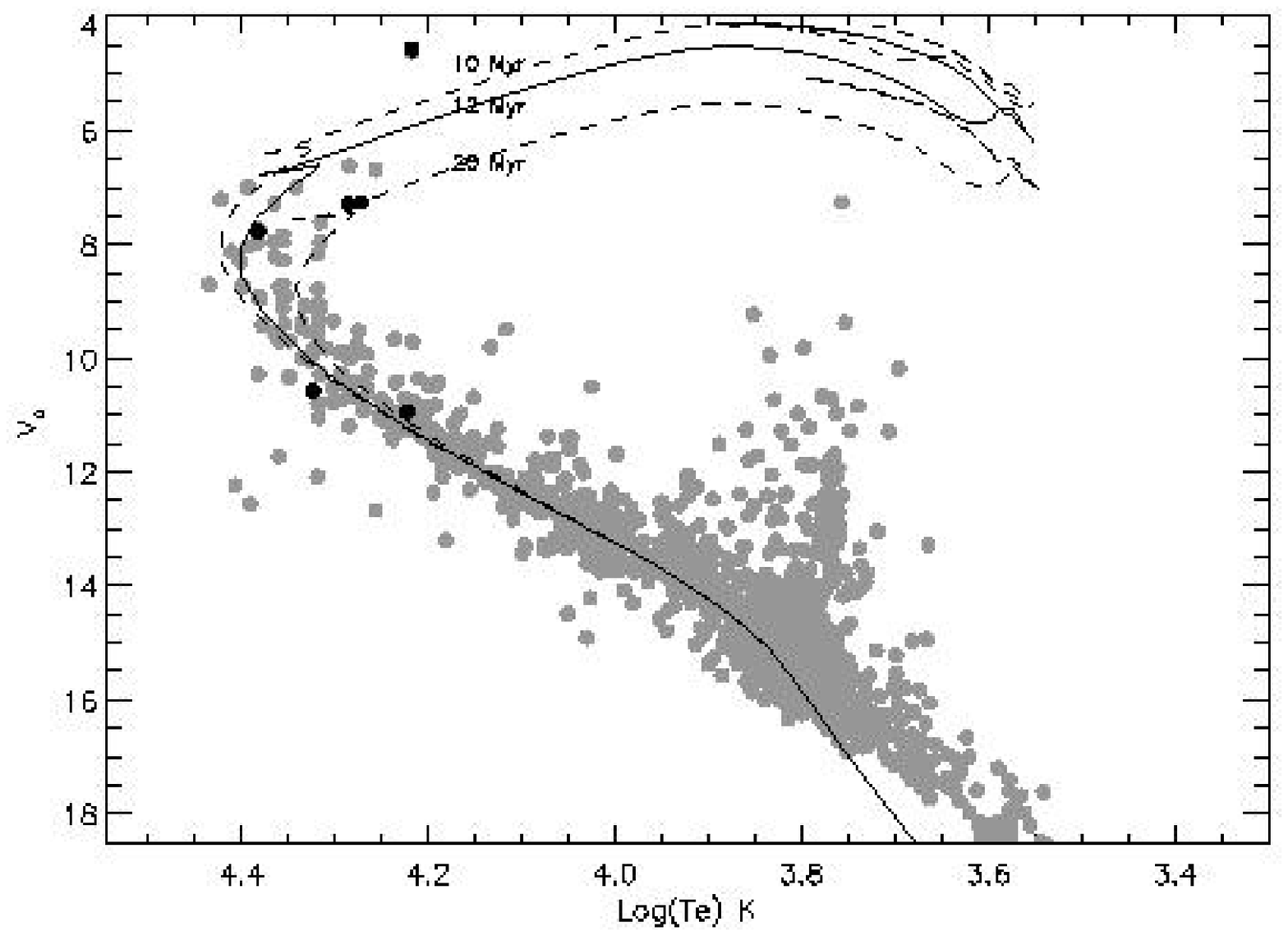}{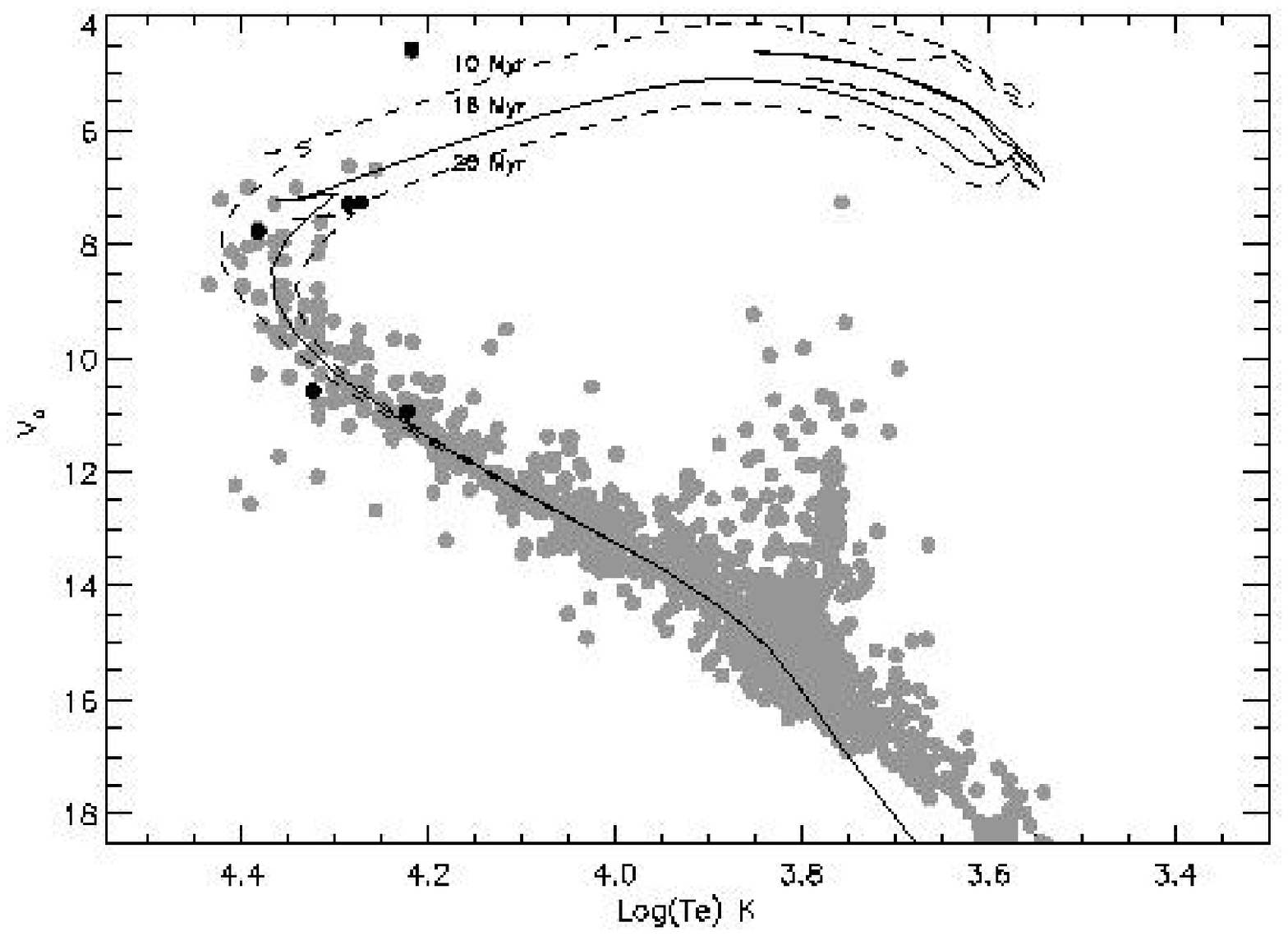}
\caption{(Top Panels) Dereddened V vs. spectral type (left) and dereddened V vs. log(T$_{e}$) (right) HR diagrams showing 
the main sequence turnoff for h Persei.  Black dots indicate the positions of Be stars.  
Overplotted are the Padova post-main sequence isochrones corresponding to 
ages of 10 and 20 Myr (dashed lines) and 13 Myr (solid line).  (Bottom Panels) 
The dereddened V vs. log(T$_{e}$) diagrams replacing the 13 Myr isochrone with 12 Myr and 16 Myr isochrones.}
\label{msturnoffh}
\end{figure}

\begin{figure}
\centering
\plottwo{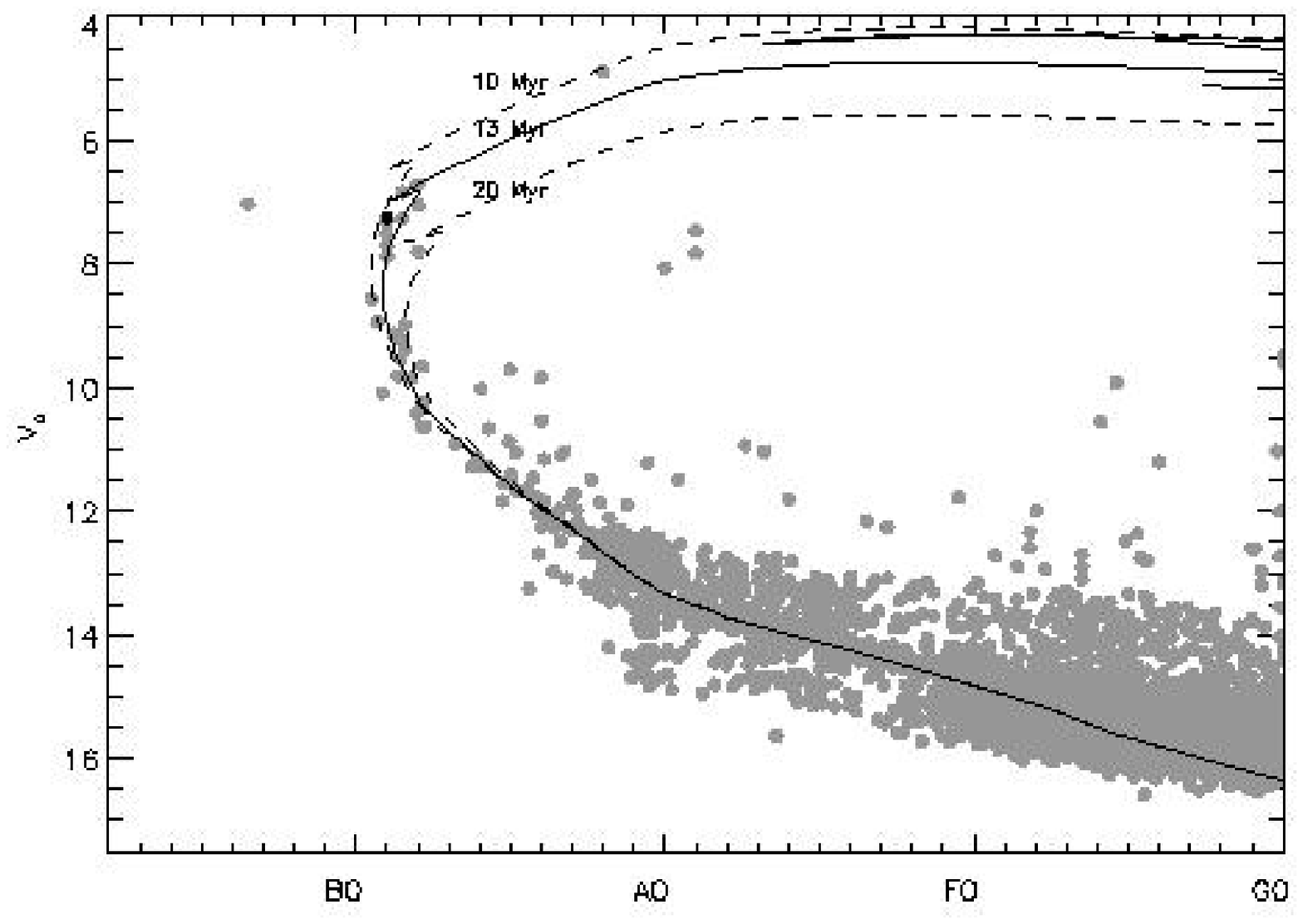}{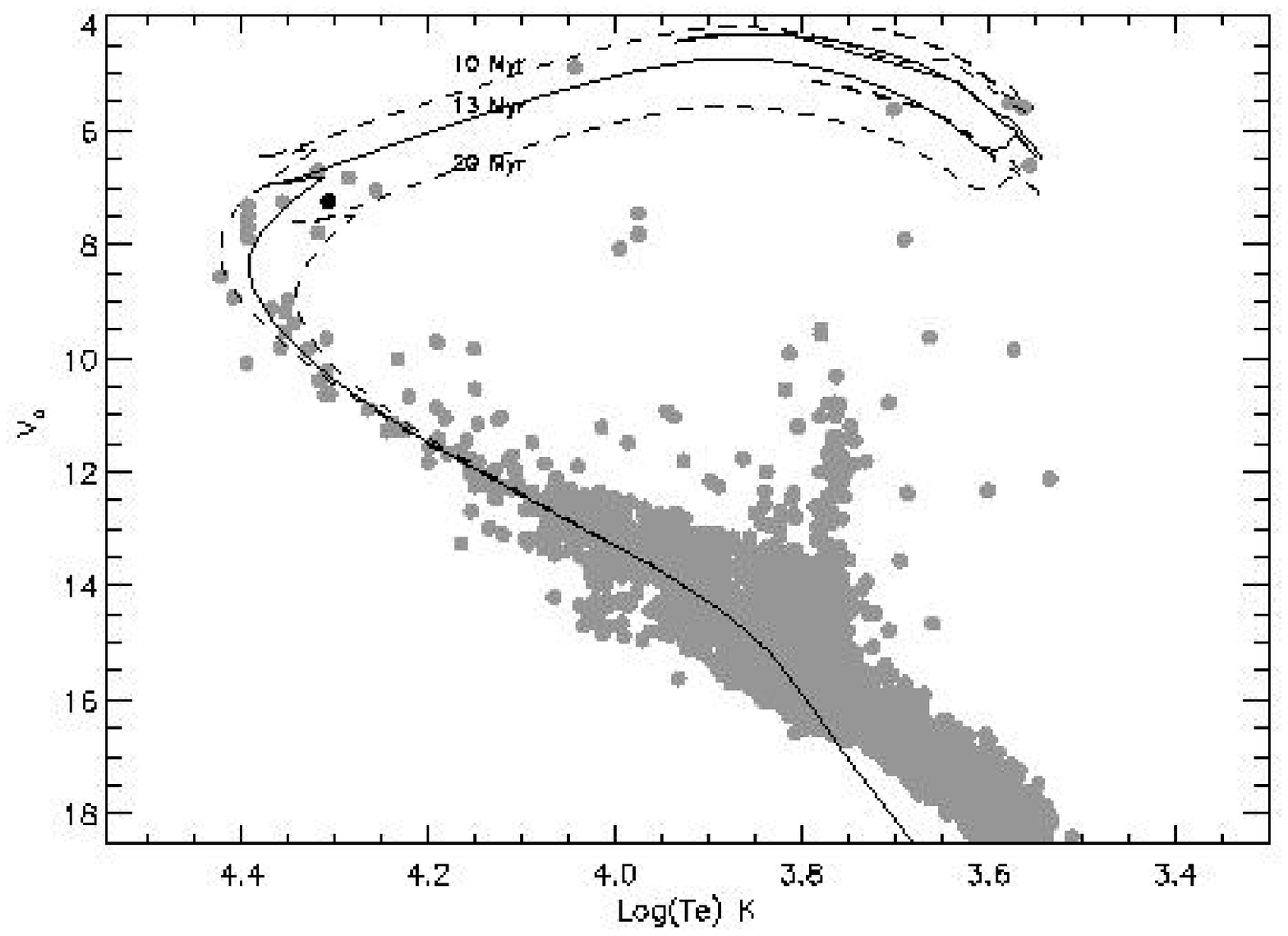}
\plottwo{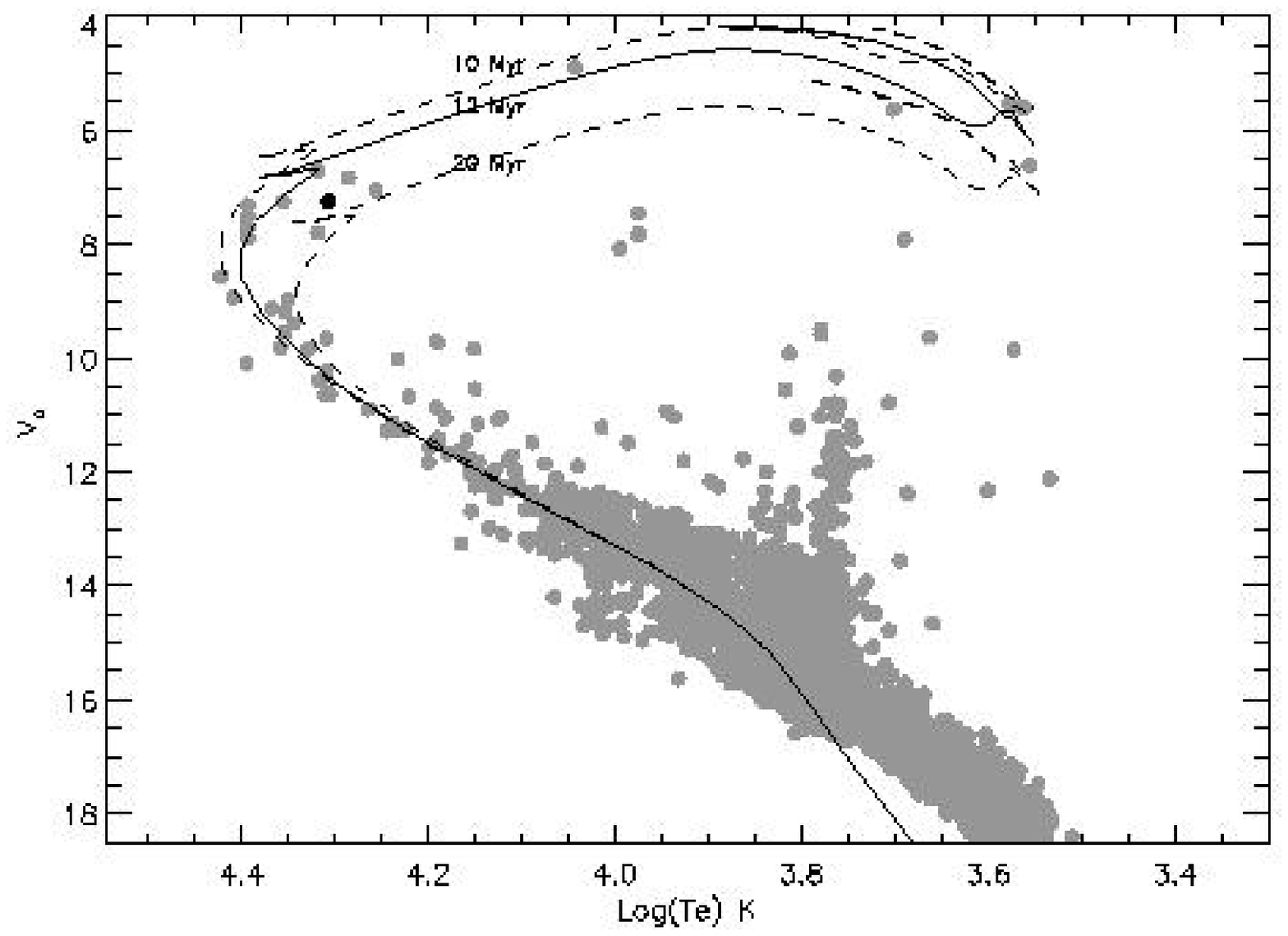}{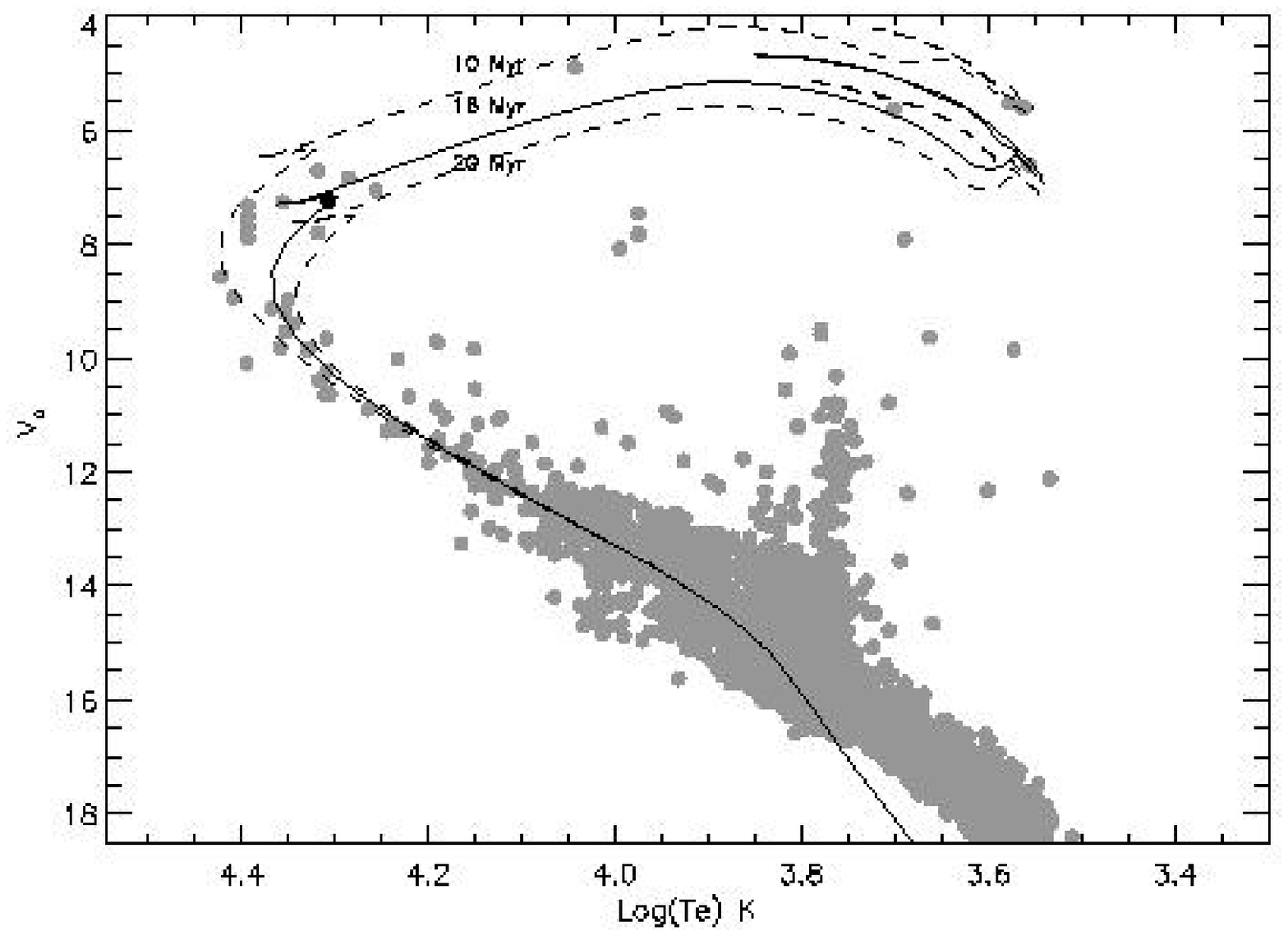}
\caption{Same as Figure \ref{msturnoffh} except for stars in the low-density regions surrounding the 
h Persei and $\chi$ Persei cores ($>$ 10' away from the cluster centers).}
\label{msturnoffhalo}
\end{figure}

\begin{figure}
\centering
\epsscale{1}
\plottwo{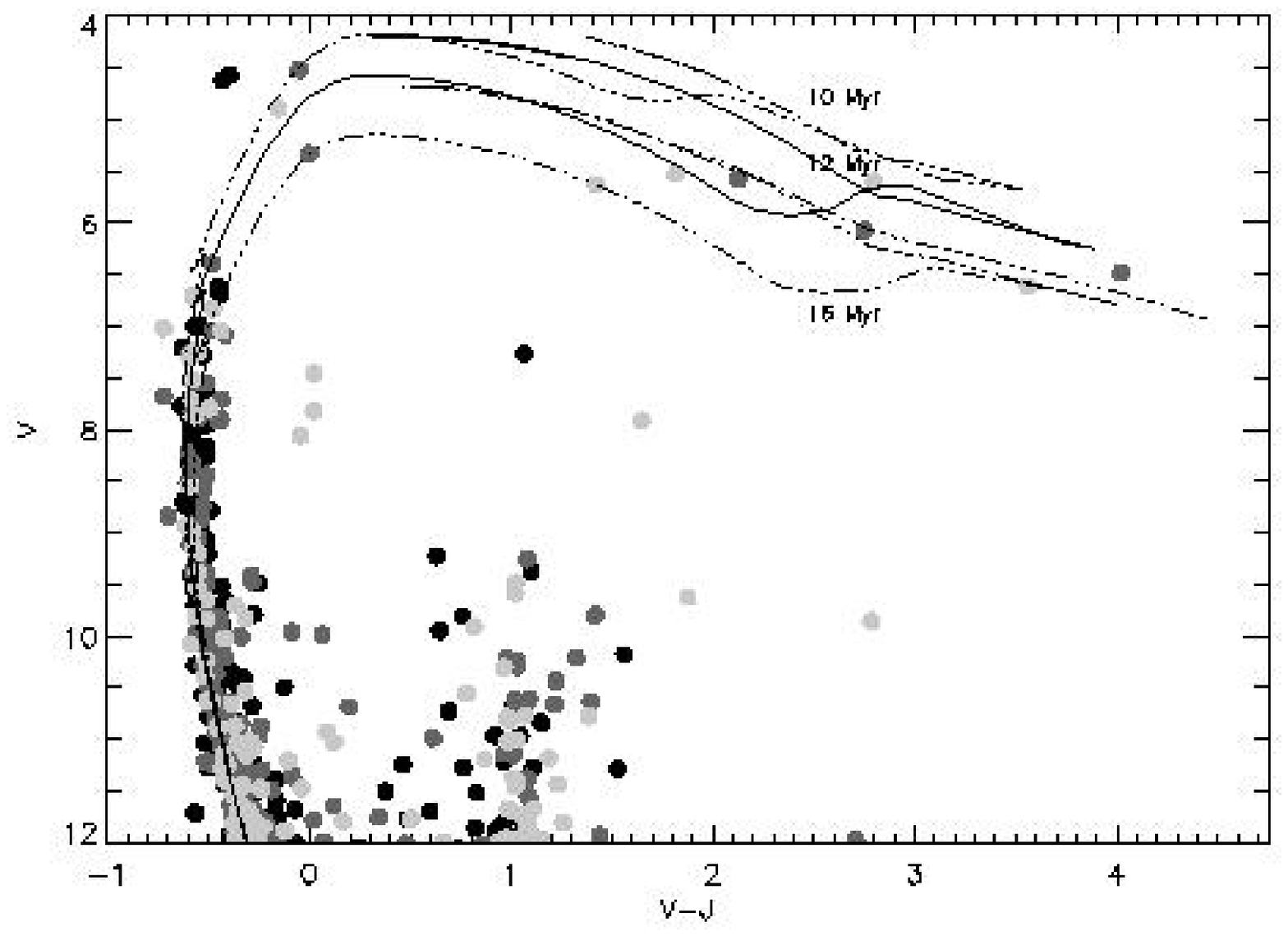}{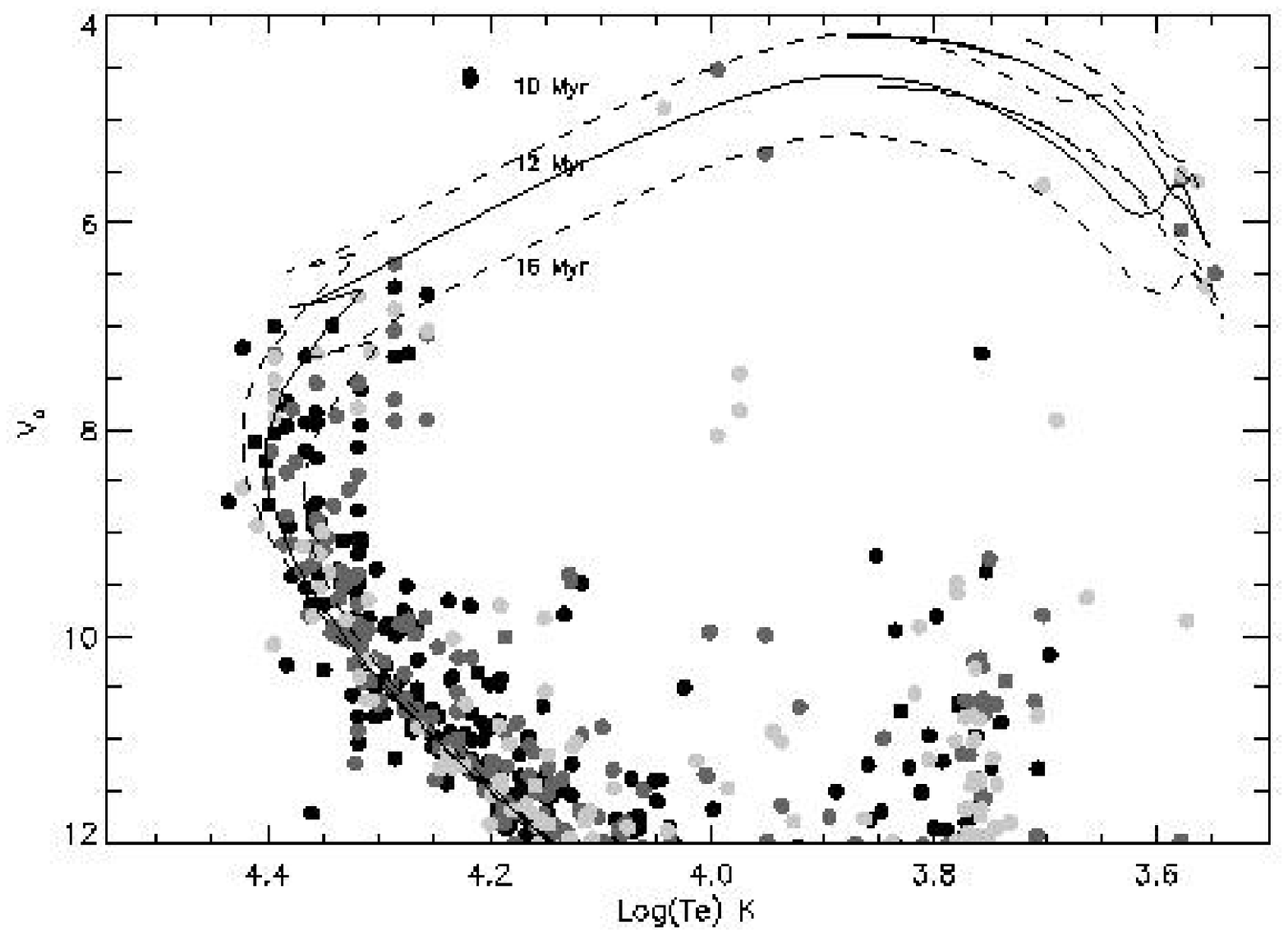}
\plottwo{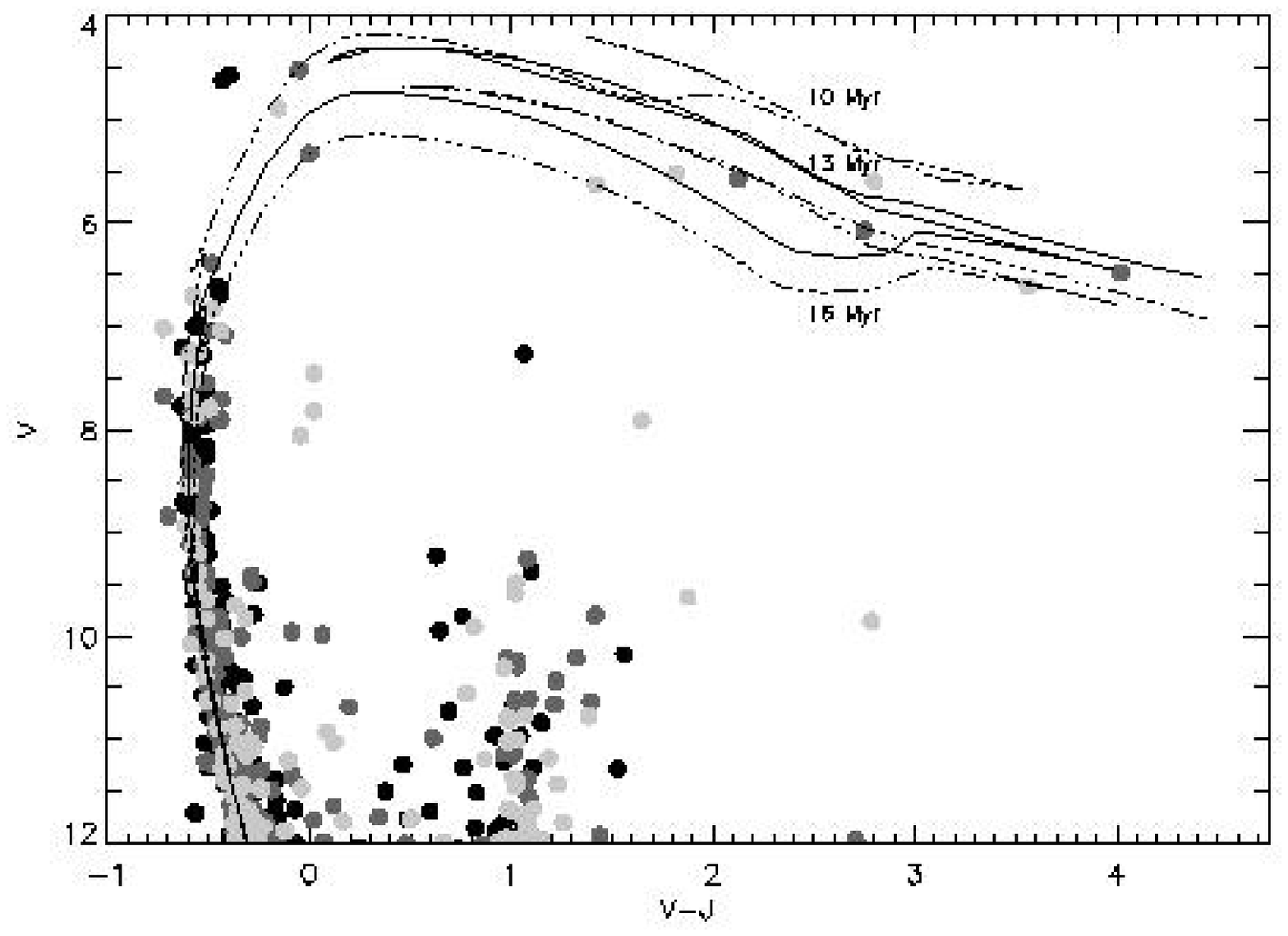}{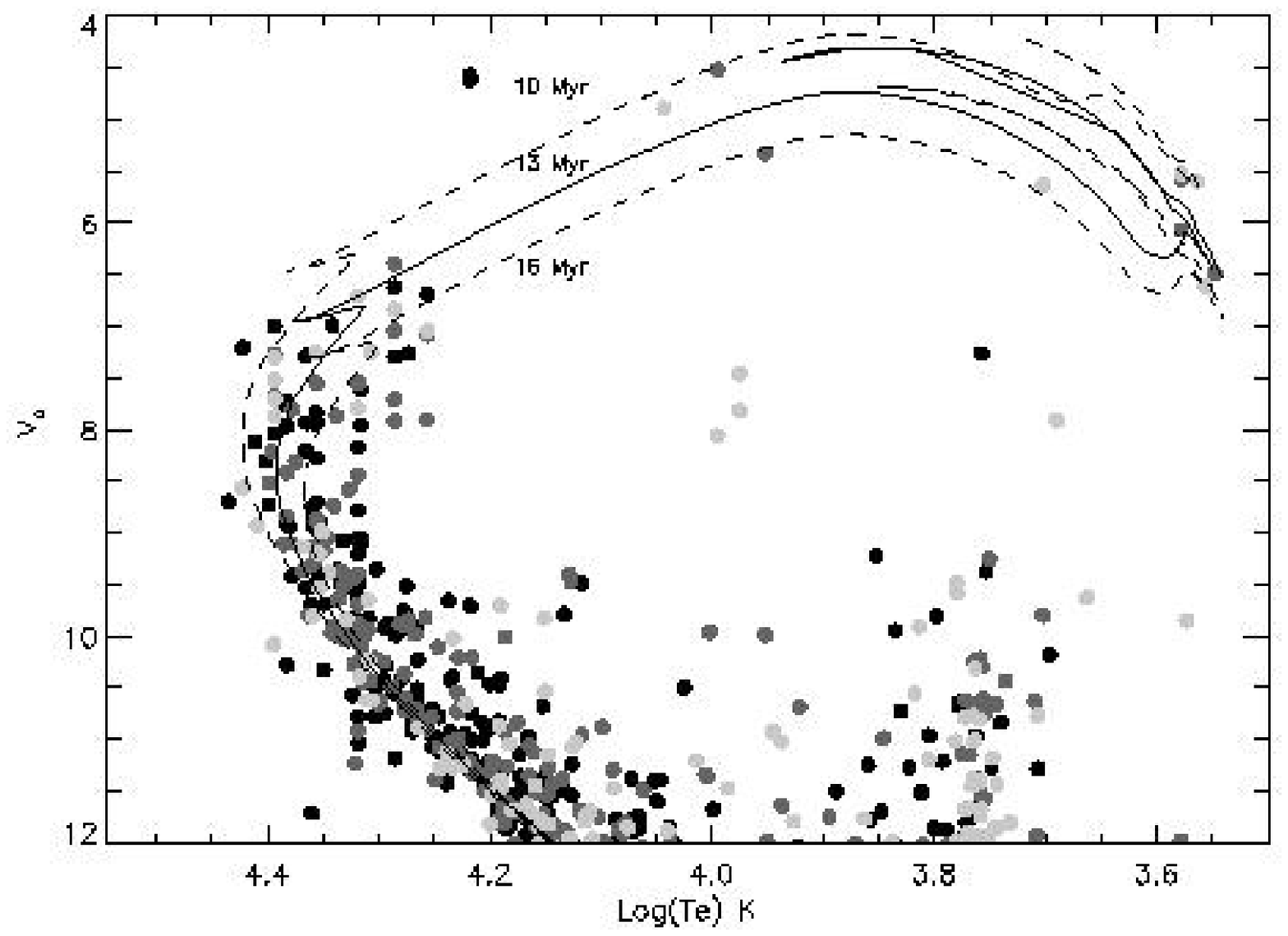}
\plottwo{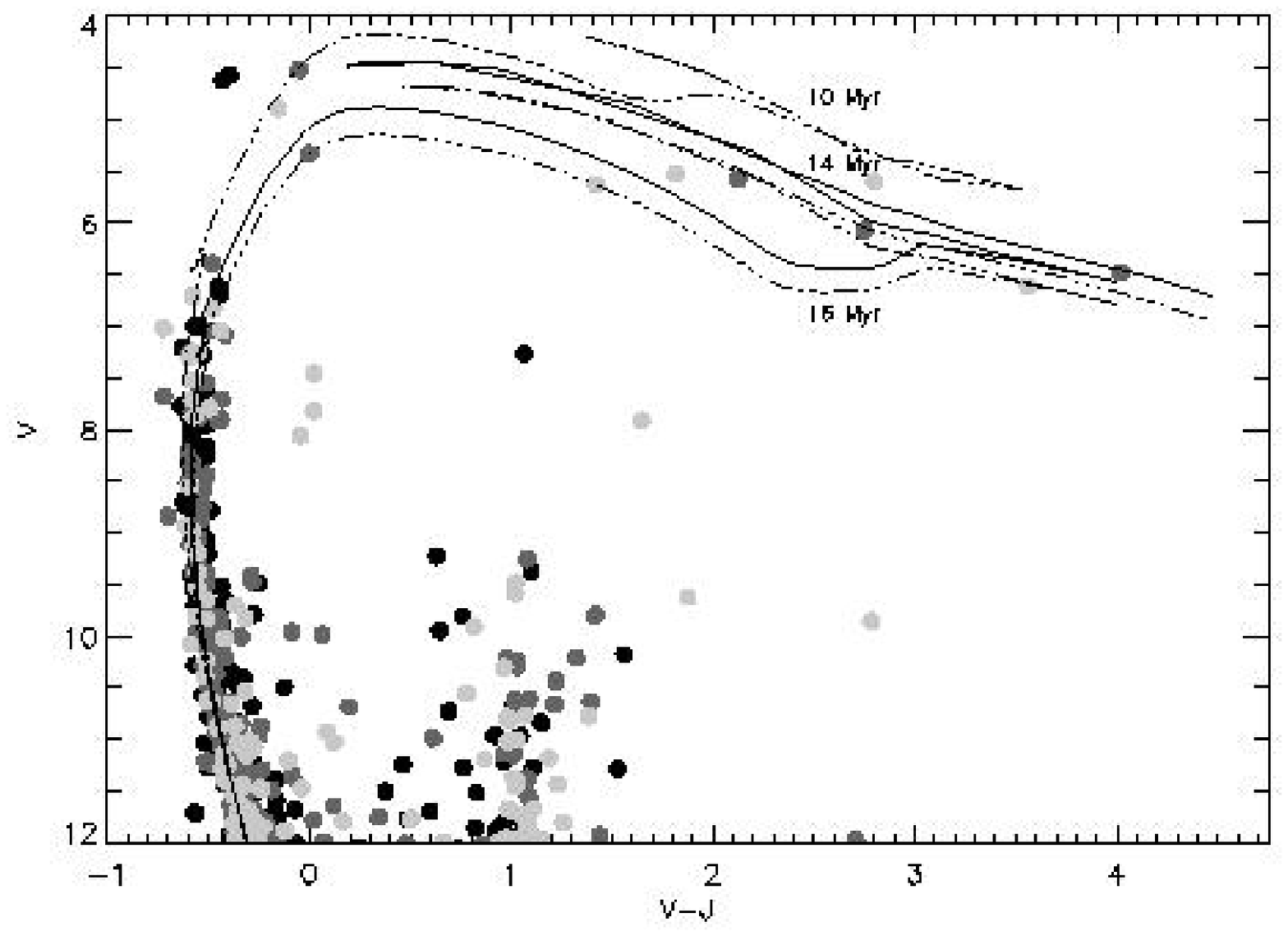}{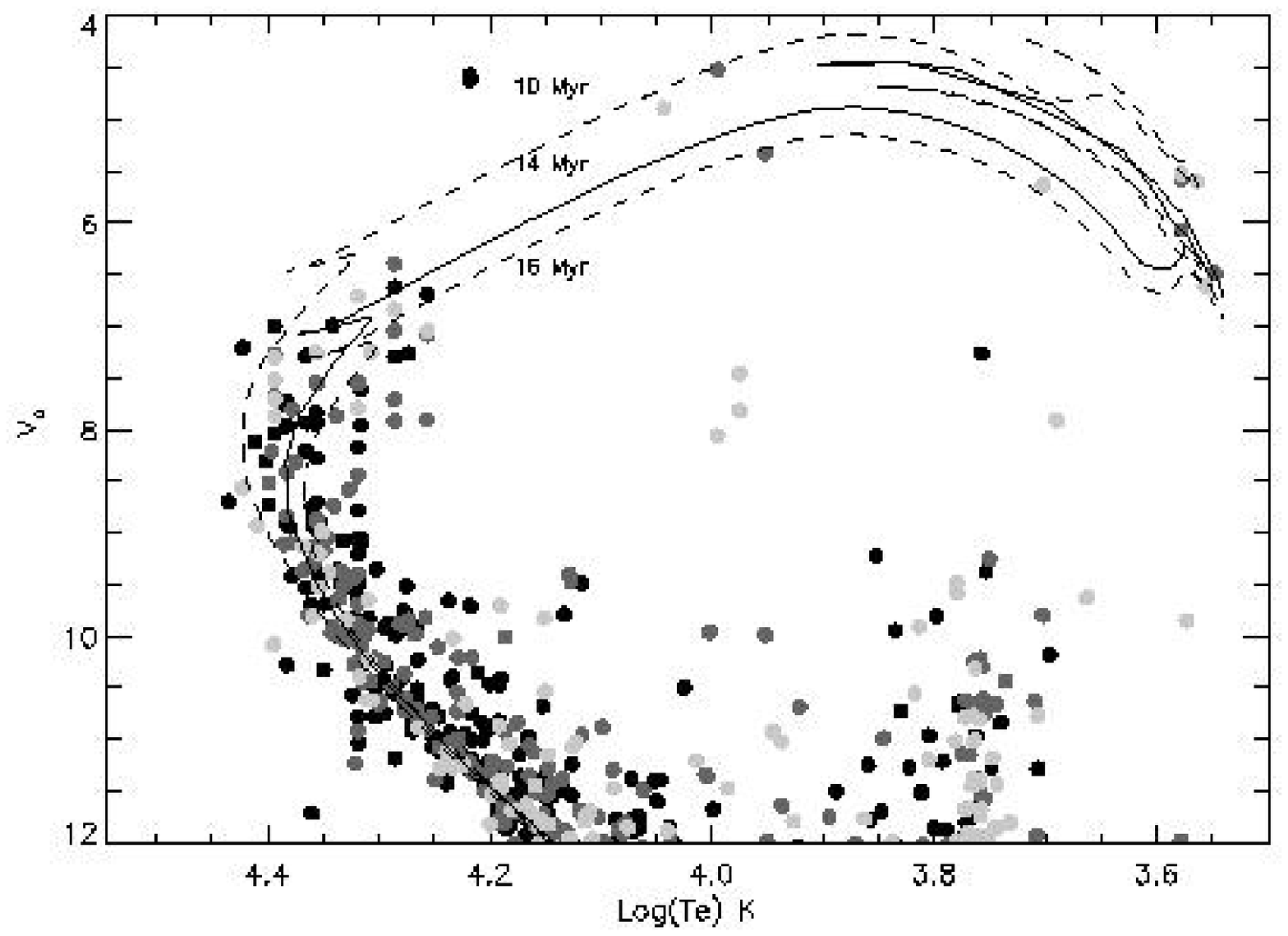}
\caption{(Left Panels) dereddened V vs. V-J and (Right Panels) dereddened V vs. log(T$_{e}$) 
comparing the Padova isochrones of 12 Myr, 13 Myr, and 14 Myr (organized from top to bottom; solid lines in all cases)
to the positions of red supergiants.  Black dots indicate stars in the h Persei core, grey dots indicate stars 
in the $\chi$ Persei core, and light-grey indicates halo region stars.}
\label{msupergiants}
\end{figure}

\begin{figure}
\centering
\plotone{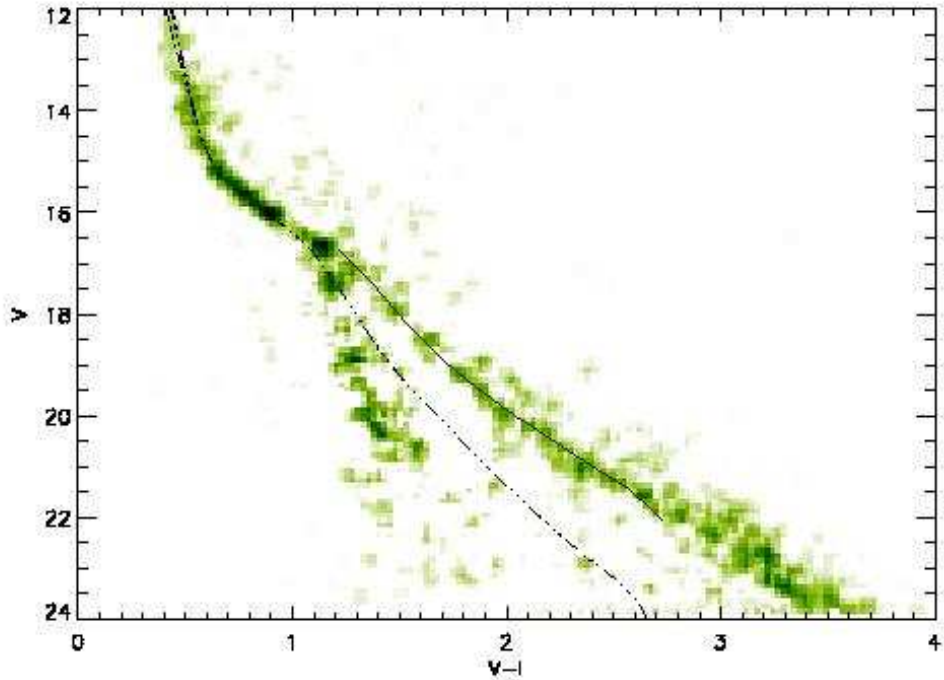}
\caption{The pre-main sequence age determination for $\chi$ Persei from the V/V-I color-magnitude diagram.  The data for 
stars within 7' of the cluster center are represented by a 
box-car smoothed Hess diagram with the background field star population statistically subtracted (green regions).  Darker regions correspond to 
higher density.  The black line (dash-three dots) represents (from left to right) the zero-age main sequence and the 
zero-age main sequence + 14 Myr post-main sequence Padova isochrone.  The solid dark grey line represents the 14 Myr pre-main sequence 
isochrone from \citet{Ba98}.}
\label{chiPerPreMS}
\end{figure}

\begin{figure}
\centering
\plotone{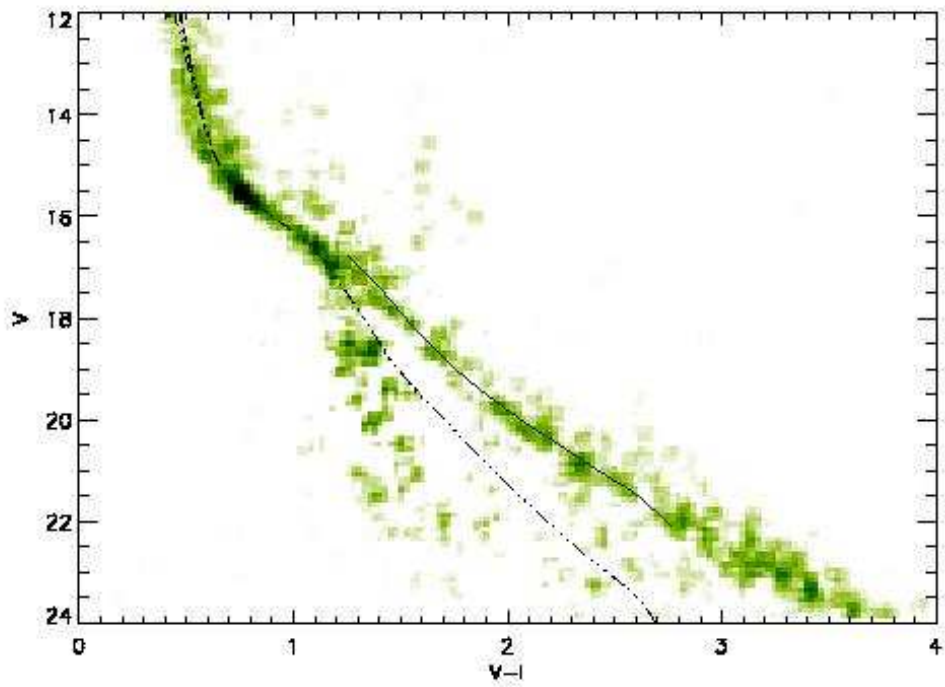}
\caption{Same as previous figure except for stars within 7' of the h Persei center.}
\label{hPerPreMS}
\end{figure}

\begin{figure}
\centering
\epsscale{0.8}
\plotone{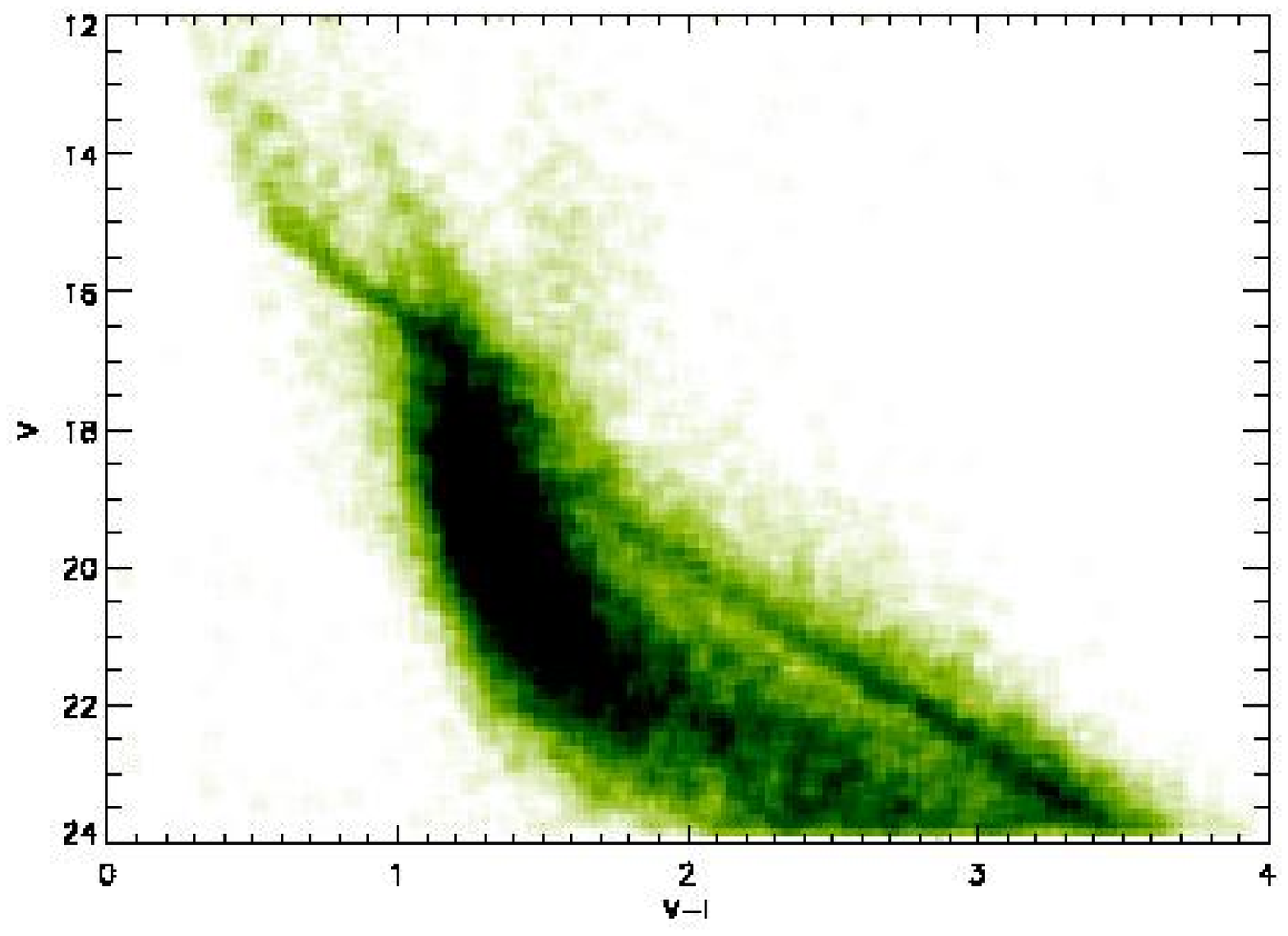}
\plotone{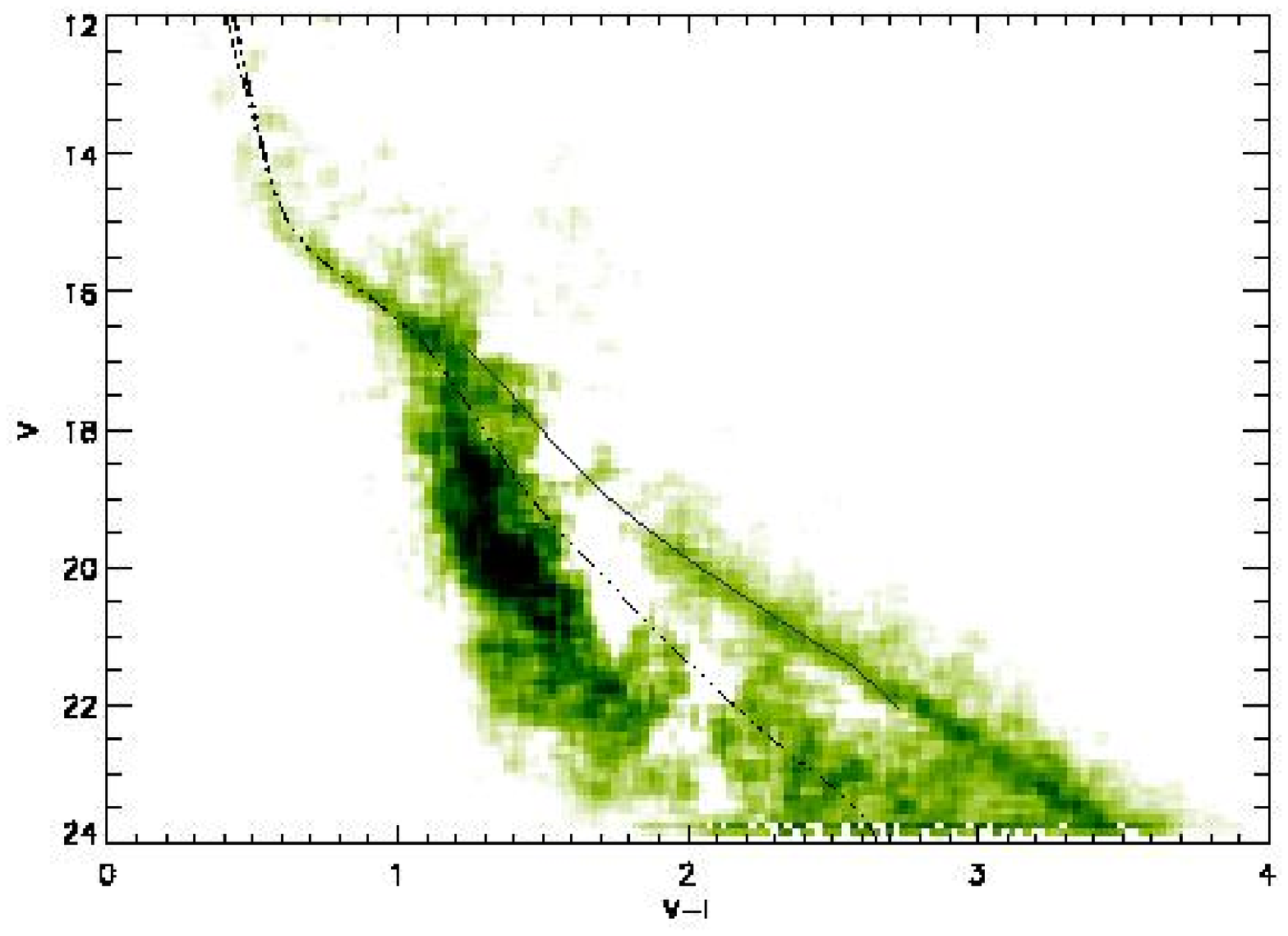}
\caption{(Top panel) The raw box-car smoothed Hess diagram for stars between 10' and 20' from the cluster centers.  (Bottom panel)  
Same as previous figure except for stars between 10' and 20' from the cluster centers.}
\label{haloPreMS}
\end{figure}

\begin{figure}
\epsscale{1}
\plotone{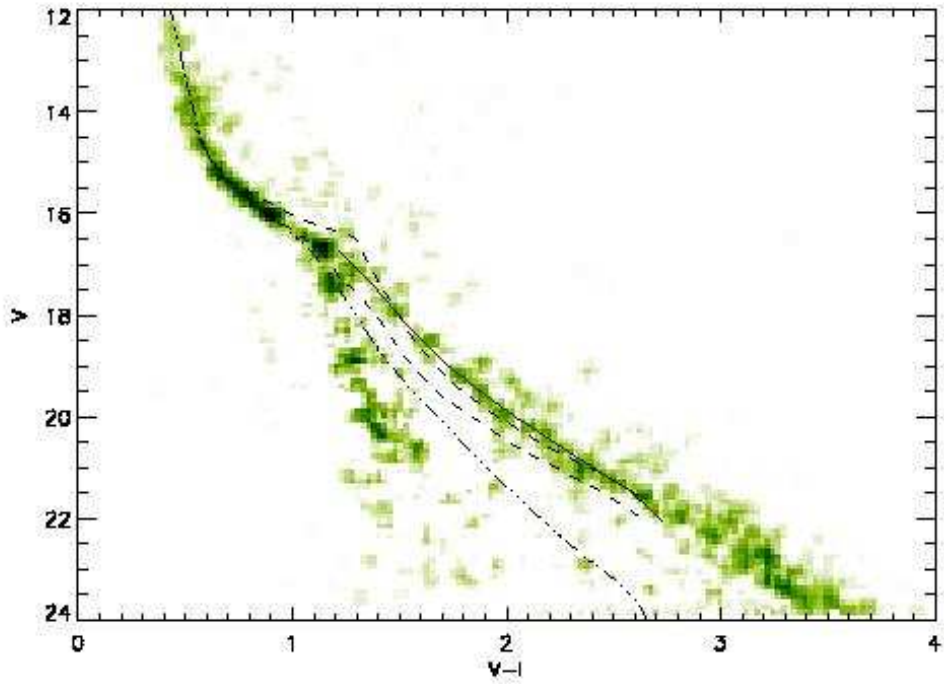}
\caption{Same as previous figures except with the \citet{Dm94,Dm97} isochrones 
for 10 Myr and 20 Myr overplotted.  The \citet{Dm94,Dm97} tracks clearly fail to 
reproduce the observed shape of the pre-main sequence locus.}
\label{premsdm97}
\end{figure}

\begin{figure}
\plotone{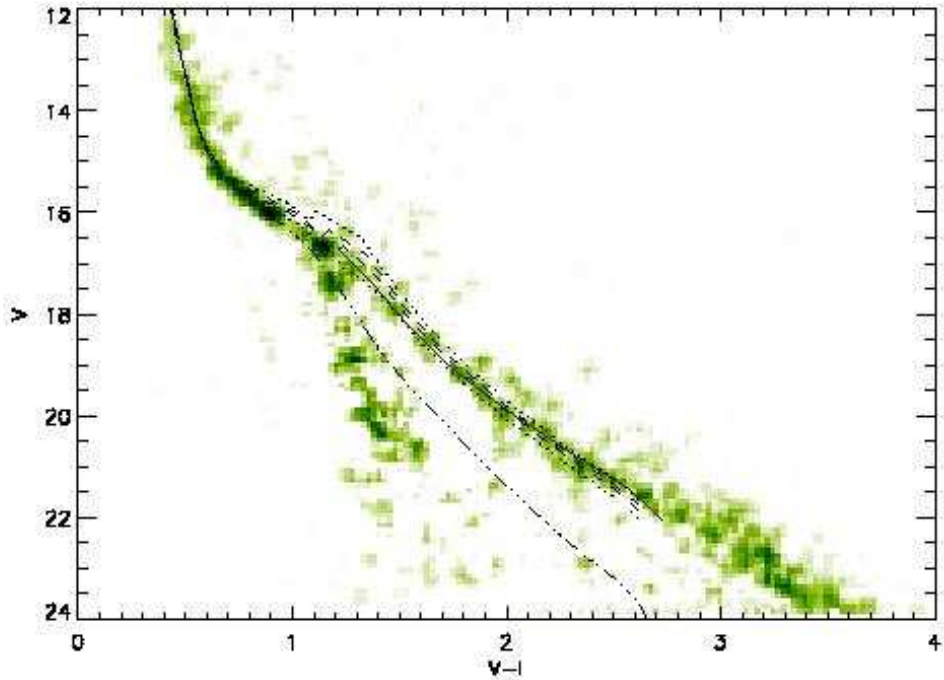}
\caption{Same as previous figure except with the \citet{Siess2000}
isochrones for 12 Myr (top dotted line), 14 Myr (top dashed line), 16 Myr (bottom dashed line), 
and 20 Myr (bottom dotted line) overplotted.  The \citet{Siess2000} do better than 
\citet{Dm97} in reproducing the observed shape of the cluster locus but generaly overestimate 
the luminosity of $\sim$ 1.5--2 M$_{\odot}$ stars (V-I $\sim$ 1--1.5).}
\label{premssiess}
\end{figure}

\begin{figure}
\centering
\plotone{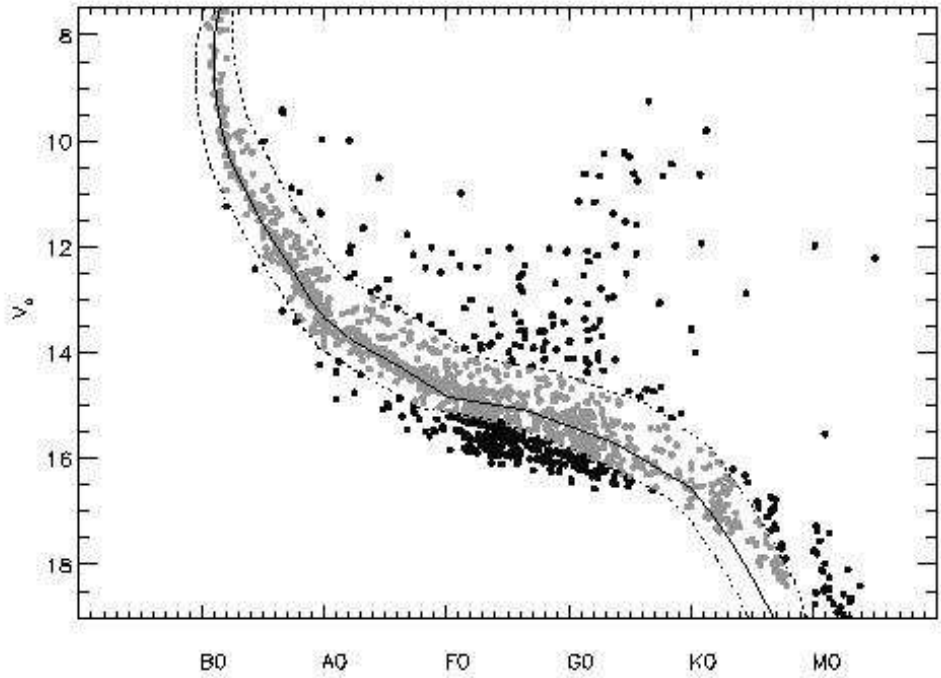}
\caption{The V$_{o}$ vs. spectral type HR diagram for $\chi$ Persei illustrating the region within which 
we identify spectroscopic members.  As described in the text, the boundaries of the locus identifying members (dashed lines) 
is determined from the dispersion in E(B-V), the physical size of the h and $\chi$ Per region, binarity, and uncertainties 
in spectral types.}
\label{specmemill}
\end{figure}

\begin{figure}
\centering
\plotone{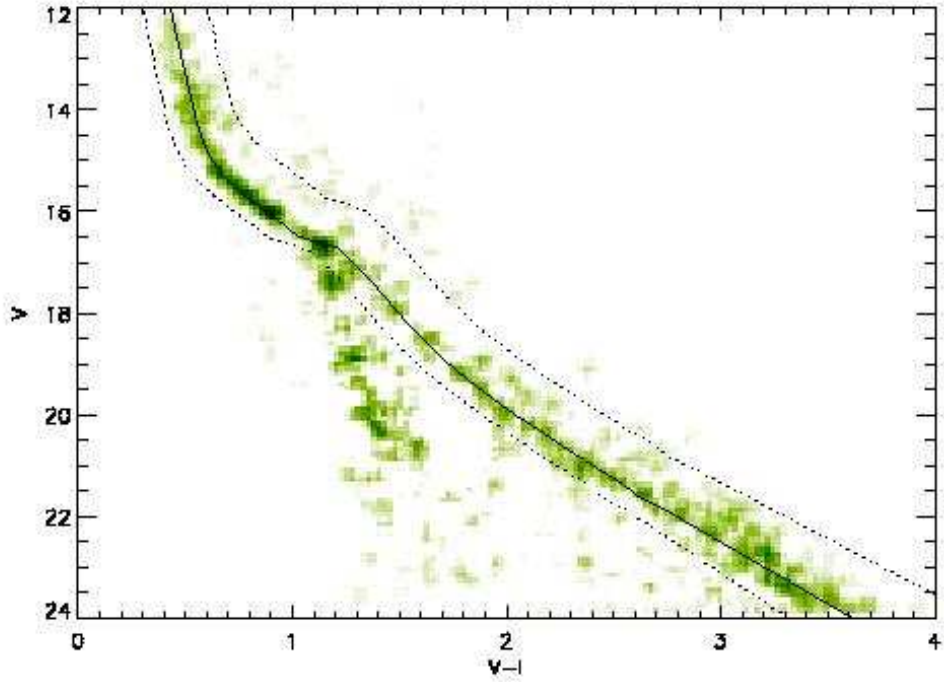}
\caption{Plot of the Hess diagram for $\chi$ Persei from Figure \ref{chiPerPreMS} now illustrating the range of colors 
and magnitudes identifying candidate members.  As described in the text, the boundaries of the locus identifying members (dashed lines) 
is determined from the dispersion in reddening, binarity, and photometric errors.}
\label{photmemill}
\end{figure}

\begin{figure}
\centering
\epsscale{0.8}
\plotone{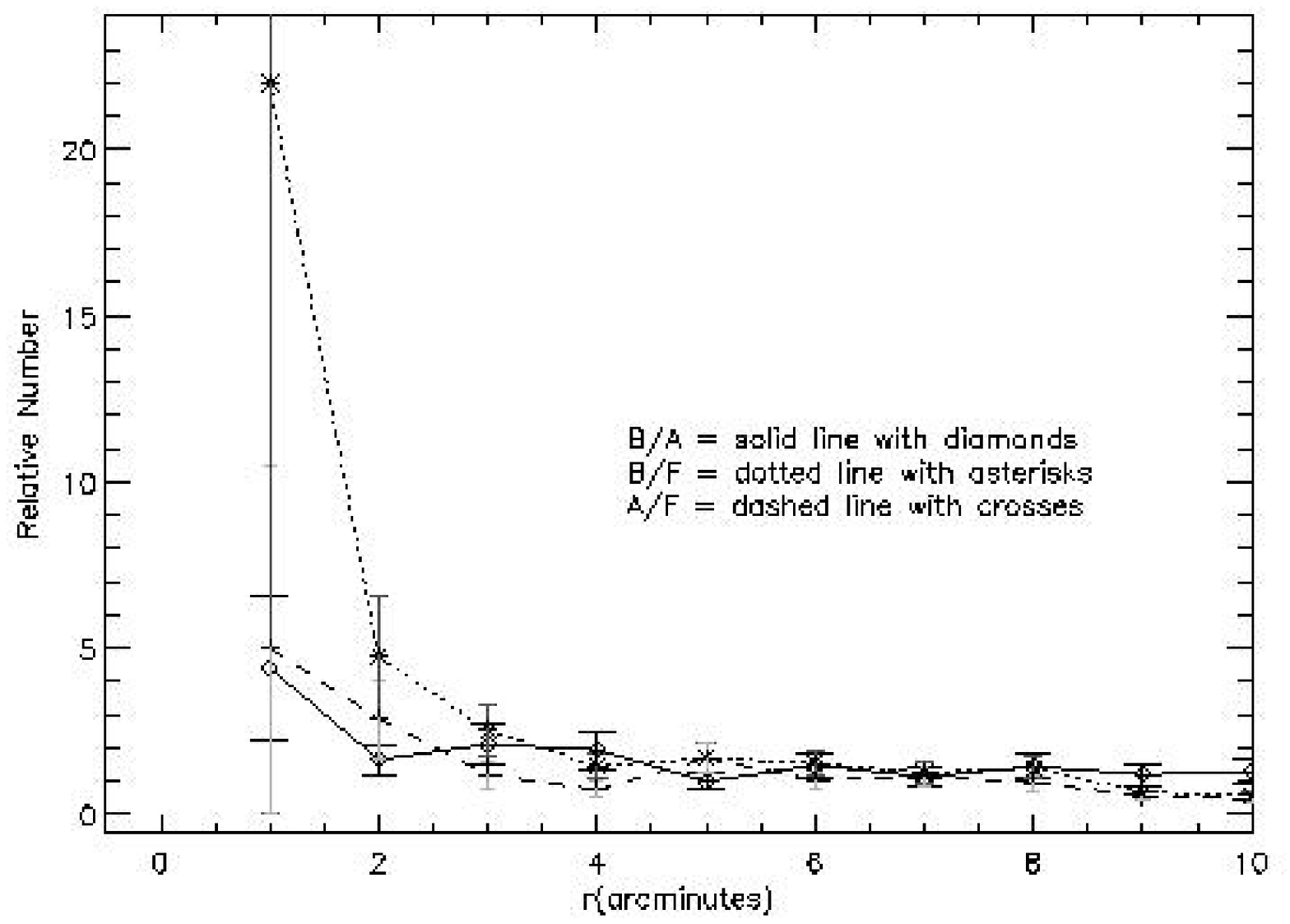}
\plotone{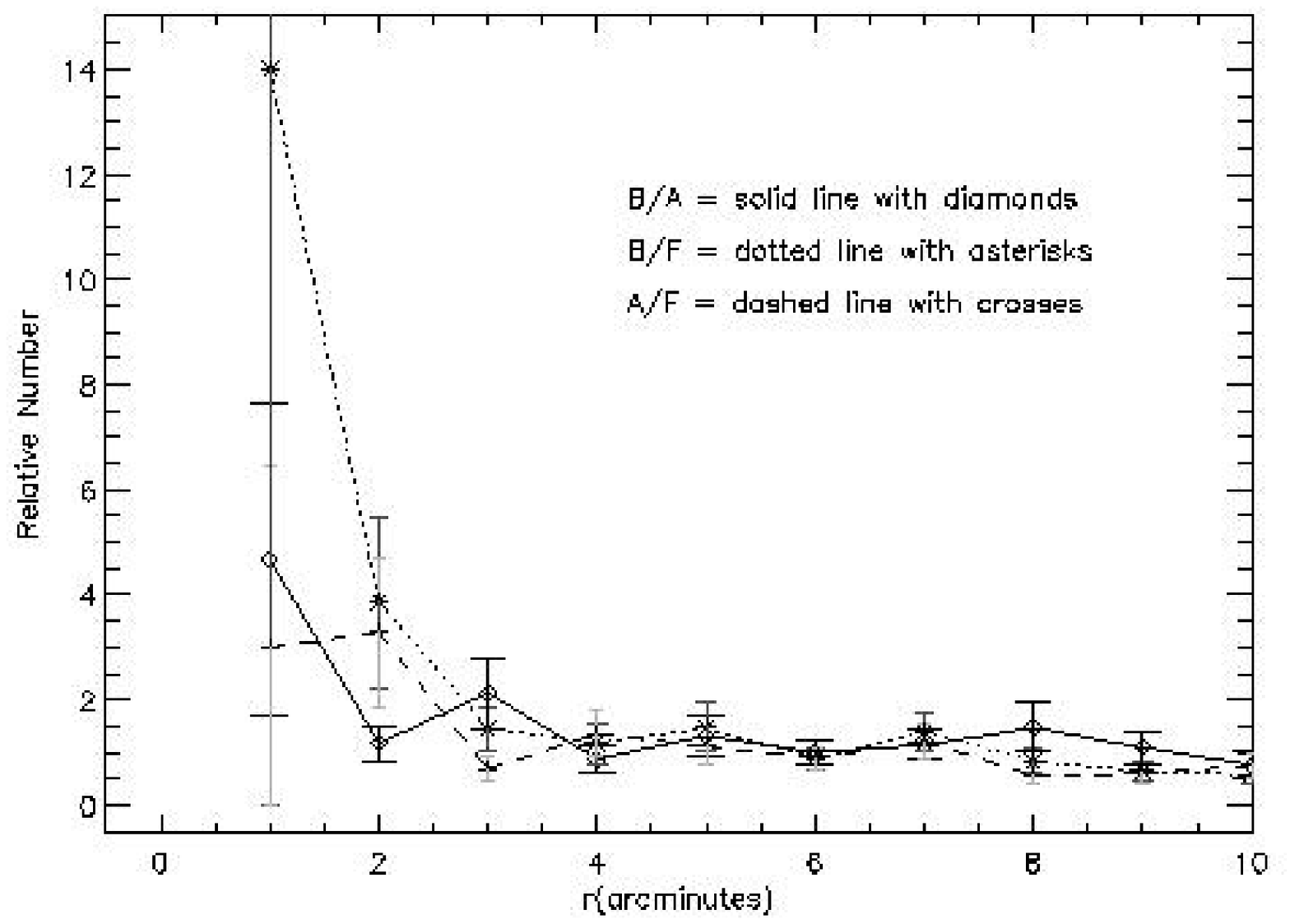}
\caption{The ratio of B to A, B to F, and A to F stars versus distance from the cluster 
center for h Persei (top) and $\chi$ Persei (bottom) illustrating mass segregation in both clusters.  In both plots, the wide, black error bars denote the uncertainty in the ratio of B to A stars, the medium-sized grey error bars identify the B to F star ratio, and the narrow light-grey error bars identify the 
A to F star ratio.}
\label{massseg}
\end{figure}

In addition to advancing our understanding of h and $\chi$ Persei properties,  
this study clearly identifies advances needed to paint a more complete picture 
of the Double Cluster.  We highlight several promising areas of future research below:

\begin{itemize}
\item \textbf{Wide-Field Optical Photometry/Deep Optical Spectroscopy} -- 
Simple wide-field optical photometric surveys can easily expand our 
membership list and amplify the scientific output of our work.  The mismatch 
in survey area between our spectroscopy ( $\sim$ 1 square degree) and 
photometry ($\sim$ 0.6 square degrees) leaves $\sim$ 3500 stars with 
spectroscopy but no photometry.  Our results demonstrate the need 
for high-quality photometry to measure 
reddening and to establish membership.  Because nearly all of these stars are 
bright (V $\lesssim$ 19), they are easily accessible by wide-field cameras 
on 1--2 meter-class telescopes.  

The upper main sequence of halo stars is undersampled largely because of 
our selection criteria for Hectospec targets (J $\ge$ 14).  Added to 
our existing survey, the entire upper main sequence of h and $\chi$ Persei 
can probably be probed after a few additional Hectospec or Hydra fiber settings.  
Because these stars are extremely bright, obtaining high signal-to-noise spectra 
will be trivial.

Furthermore, our optical photometry identifies many probable cluster members
that are beyond the 2MASS detection limit but clearly within range of 
Hectospec for integration times of $\sim$ 1--2 hours (V $\lesssim$ 21--22).
A survey establishing the spectral types for many of these stars would 
provide valuable information on the luminosity and temperatures of 
pre-main sequence stars at $\approx$ 14 Myr.
Other multi-object spectrographs coming online in the near future (e.g. MODS 
on the Large Binocular Telescope, Binospec on the MMT) can probe even further down the cluster 
mass function.

\item \textbf{Robust Membership Determinations from \textit{SIM} and \textit{GAIA}} -- 

Proper motion studies are the most definitive method for determining 
h and $\chi$ Per membership.  Unfortunately, at $\sim$ 2.3 kpc distant, 
h and $\chi$ Per stars exhibit tiny proper motions, the size of which 
render ground-based campaigns to verify our membership list hopeless.
  However, space-based interferometric missions -- specifically \textit{SIM} 
and \textit{GAIA} -- are easily capable of detecting the $\sim$ $\mu$-arc second motions of 
pre-main sequence cluster stars.  \textit{SIM's} and \textit{GAIA's} clear ability to yield 
definitive catalogs of members for h and $\chi$ Persei and other distant, populous clusters 
would have an enormous impact on studies of these clusters and stellar evolution in general.

\item \textbf{Metallicity} -- Clearly, the greatest source of systematic error in 
constraining the clusters' properties is metallicity.  While 
we argue that h and $\chi$ Per likely have a near-solar metallicity, it is possible that 
the sources whose properties support our contention are not indicative of the clusters' 
stars as a whole.  To address metallicity, at least three avenues of research should 
be explored.  First, sophisticated stellar atmosphere modeling of stars in a variety 
of evolutionary states should constrain the photospheric chemical abundances \citep[e.g.][]{Dufton1990, Smartt1997, 
Venn2002} and help to define the typical chemical composition.  Second, high-resolution 
echelle spectra of cluster stars (e.g. with Hectochelle), especially G-type pre-main sequence members, should not 
only constrain the clusters' velocity dispersions (and thus aid membership identification) but will also be capable 
of yielding metallicity estimates.  Third, building upon 
work by \citet{Southworth2004,Southworth2005}, metallicities can be estimated by 
comparing masses and radii of eclipsing binaries to model predictions.
  In this regard, results from the MONITOR program \citep{Ai07} may be crucial for 
ending the debate on metallicity.

\item \textbf{Effective Temperatures} -- As described in Appendix A, the T$_{e}$ scale for 
early B dwarfs and evolved B stars is more poorly constrained than for O stars and A--M stars.  
Non-LTE modeling of the many early B stars in h and $\chi$ Persei may provide better 
constraints on the T$_{e}$ scale.  More specifically, determining T$_{e}$ for individual 
B stars strengthens our ability to derive accurate post-main sequence ages.  As was argued by 
\citet{Sl02} and can be seen in Figures \ref{msturnoffchi}--\ref{msturnoffhalo}, the binning 
of B stars into discrete T$_{e}$ values for their spectral type potentially introduces a spurious
spread in T$_{e}$ at a given V magnitude brightness, which could be misinterpreted as 
a spread in age.  Individual measurements for T$_{e}$ will tighten the locus of cluster stars 
at the main sequence turnoff.  Accurate T$_{e}$ estimates may be be especially important 
for cluster Be stars, which are often difficult to spectral type because of their Balmer 
line emission.  A large campaign to derive atmospheric properties of Be stars in 
h and $\chi$ Persei is underway \citep{Marsh2010}.

\item \textbf{The Halo Population} -- While our results confirm previous claims \citep[e.g.][]{Cu07a}
that h and $\chi$ Persei contains a large halo population, it is possible that 
the halo extends beyond  25--30' from either cluster center, or even 
well beyond the square-degree field covered by our spectroscopic survey.  
A more spatially extended survey of stars would yield a better estimate for the total mass of 
the halo population.  A more complete census of young stars within $\approx$ 5 
degrees of the h and $\chi$ Persei cores better reveals the large star formation history
within which h and $\chi$ Per emerged, including the Double Cluster's relationship to 
other nearby young clusters such as W3, W4, and W5 located several 
degrees away.  Specifically, it may be possible to
address the relationship between h and $\chi$ Persei and the Perseus OB1 association, 
determining if the core and halo populations have properties completely distinct from Per OB1, if
h and $\chi$ Per represents a unique epoch in star formation that propogated through the present day Per OB1 region, or
 if its properties are characteristic of the Per OB1 association as a whole \citep[e.g.][]{Sl02, Lee2008}.

\item \textbf{The Circumstellar Disk Population: Constraints on Planet Formation} -- Finally, 
the large sample of stars now confirmed as h and $\chi$ Persei members can be used 
to investigate planet formation by using Spitzer and (later) the James Webb Space Telescope.
New Spitzer observations reveal an order-of-magnitude increase in the number of stars 
detected at 3.6--8 $\mu m$ \citep{Currie2010}.  Combined with membership information derived here, 
Spitzer data will yield constraints on the disk population from a sample size easily 
dwarfing that of the \textit{FEPS} and \textit{Cores-to-Disks} Legacy Programs combined and thus 
providing far superior statistical reliability.  

This fundamentally different kind of dataset allows robust probes of planet formation that 
are impossible with other samples, providing
estimates of the frequency of long-lived protoplanetary disks; warm, terrestrial planet-forming debris disks; and
 extremely luminous cold debris disks. Combined with data from 5--25 Myr-old 
 clusters -- e.g. NGC 2362, Upper Scorpius, Orion OB1, NGC 1960, and NGC 2232 \citep[][]{CurrieLada2009, Carpenter2006, 
Hernandez2007,Balog2010, Currie2008b} -- it is possible to reconstruct the time history of 
terrestrial, gas giant, and icy planet formation to compare with our solar system's 
chronology and models of planet formation \citep[e.g.][]{KenyonBromley2009, jcast07}.
\end{itemize}

\acknowledgements
We thank the MMT CfA Time Allocation Committee and the MMT director Faith Vilas for their 
enthusiastic, unwavering support for this project 
over the past 3 years.  The MMT staff, telescope operators, and Hectospec instrument team ---
especially Nelson Caldwell and Dan Fabricant -- provided superb technical support.   
We thank Doug Mink for constructing a searchable public archive of our Hectospec data.
Our work is deeply indebted to the sound advice of many experts in the field of stellar atmospheres 
and evolution.  In particular, we thank Phil Massey for patiently and rapidly 
answering many questions posed by the lead author regarding effective temperatures 
of OB main sequence stars, giants, and supergiants.  We also thank Robert Kurucz for many 
detailed, valuable discussions concerning the sensitivities of observed colors and luminosities to
chemical abundances, surface gravities, and evolutionary states.  Finally, we thank Steven Cranmer 
for sharing his nearly encyclopedic knowledge of effective temperature scales from the past 30 years 
of research literature.

\clearpage
\begin{deluxetable}{lllllllllllllllllll}
 \tiny
\setlength{\tabcolsep}{0.02in}
\tabletypesize{\tiny}
\tablecolumns{11}
\tablecaption{Optical Photometry Catalog}
\tiny
\tablehead{{Running Number}&{RA}&{DEC}&{t$_{int., V}$}&{V}&{$\sigma$(V)}&{Photometry Flag (V)}&{t$_{int., I_{c}}$}&
{I$_{c}$}&{$\sigma$(I$_{c}$)}&{Photometry Flag (I$_{c}$)}}
\startdata
 1& 35.6018 &57.1095      &2   &7.8257&  0.0001    &-9   &2   &6.3208  &0.0000 &-9\\
 2& 35.7515 &57.3870      &2   &7.1912&  0.0000    &-9   &2   &6.5893  &0.0000 &-9\\
 3& 34.7691 &57.1355      &2   &7.1731&  0.0000    &-9   &2   &6.6876  &0.0000 &-9\\
 4& 35.4814 &57.2429      &2   &7.2686&  0.0000    &-9   &2   &6.7332  &0.0000 &-9\\
 5& 35.2484 &57.1583      &2   &7.7409&  0.0001    &-9   &2   &6.7185  &0.0000 &-9
\enddata
\tablecomments{The photometry flags have the following meanings: -1 = unsaturated, uncrowded, -2 = crowded, PSF fitting used, -9 = likely saturated.}
\label{opticalcatalog}
\end{deluxetable}

\clearpage
\begin{deluxetable}{lllllllllllllllllll}
 \tiny
\setlength{\tabcolsep}{0.02in}
\tabletypesize{\tiny}
\tablecolumns{12}
\tablecaption{Spectroscopy Catalog}
\tiny
\tablehead{{Running}&{Source}&{RA}&{DEC}&{Numerical}&{Spectral Type}&{J}&{$\sigma$(J)}&{H}&{$\sigma$(H)}&{K$_{s}$} &
{$\sigma$(K$_{s}$)}\\
{Number}&{}&{}&{}&{Spectral Type}&{Uncertainty}}
\startdata
1 & 2 & 35.4850 & 57.1471 & 29.00 & 3.00 & 13.3940 & 0.0450 & 13.2400 & 0.0380 & 13.0114 & 0.1614 \\
2 & 2 & 35.6214 & 57.2132 & 29.00 & 3.00 & 13.8380 & 0.0420 & 13.7570 & 0.0490 & 13.5912 & 0.0256 \\
3 & 2 & 35.0118 & 57.0850 & 38.50 & 2.00 & 11.0890 & 0.0270 & 10.8950 & 0.0210 & 10.7218 & 0.0433 \\
4 & 2 & 34.8023 & 57.1390 & 35.00 & 3.00 & 13.2410 & 0.0420 & 13.1680 & 0.0390 & 12.8050 & 0.0170 \\
5 & 2 & 35.3507 & 57.2118 & 22.00 & 1.90 & 13.3940 & 0.0410 & 13.3100 & 0.0420 & 13.1569 & 0.0821 \\
\enddata
\tablecomments{The numerical spectral type has the following formalism: 10=B0, 11=B1 ... 68=M8.  
The spectral type uncertainty is given in subclasses.  The \textbf{Source} column refers to the source of the spectroscopic
data: Hectospec (1), Hydra (2), FAST (3), or \citet{Sl02} (4).}
\label{speccatalog}
\end{deluxetable}

\clearpage
\begin{deluxetable}{lllllllllllllllllll}
 \tiny
\setlength{\tabcolsep}{0.02in}
\tabletypesize{\tiny}
\tablecolumns{12}
\tablecaption{Combined Catalog for Stars with Optical Photometry and Spectroscopy}
\tiny
\tablehead{{ID}&{RA}&{DEC}&{Numerical}&{Spectral Type}&{E(B-V)}&{Luminosity}&{J}&{$\sigma$(J)}&{H}&{$\sigma$(H)}&{K$_{s}$}&
{$\sigma$(K$_{s}$)}&{V}&{$\sigma$(V)}&{I$_{c}$}&{$\sigma$(I$_{c}$)}\\
{Number}&{}&{}&{Spectral Type}&{Uncertainty}&{}&{Class}}
\startdata
  1 & 35.4684 & 56.9050 & 6.50 & 2.00 & 0.479 & 5 & 8.5170 & 0.0100 & 7.6218 & 0.0001 & 8.1270 & 0.0270 & 8.1320 & 0.0420 & 8.1240 & 0.0260 \\
  2 & 34.7945 & 57.0249 & 10.50 & 1.60 & 0.532 & 1 & 9.8700 & 0.0100 & 8.6622 & 0.0002 & 9.2100 & 0.0270 & 9.1150 & 0.0290 & 9.1350 & 0.0210 \\
  3 & 34.6782 & 57.0726 & 10.50 & 1.50 & 0.610 & 3 & 10.6020 & 0.0100 & 9.6145 & 0.0004 & 9.8270 & 0.0250 & 9.8030 & 0.0290 & 9.7900 & 0.0230 \\
  4 & 34.3963 & 57.0857 & 10.50 & 1.80 & 0.437 & 5 & 9.9250 & 0.0100 & 8.6521 & 0.0002 & 9.5030 & 0.0240 & 9.5510 & 0.0300 & 9.5290 & 0.0220 \\
  5 & 34.7768 & 57.1261 & 11.00 & 1.50 & 0.594 & 5 & 9.8200 & 0.0100 & -99.0000 & 0.0000 & 9.0430 & 0.0320 & 8.9980 & 0.0320 & 8.9740 & 0.0200 \\
\enddata
\tablecomments{The luminosity class has the following meaning: 1= Class I supergiant, 3 = Class III giant, and 5 = Class V dwarf.}
\label{specphotcatalog}
\end{deluxetable}

\clearpage
\begin{deluxetable}{lllllllllll}
 \tiny
\setlength{\tabcolsep}{0.02in}
\tabletypesize{\tiny}
\tablecolumns{8}
\tablecaption{Distance Moduli for h and $\chi$ Persei from this work and from previous work}
\tiny
\tablehead{{Cluster Properties Used}&{h Persei distance}&{$\chi$ Persei distance} & {Halo distance}}
\startdata
\cutinhead{This Work}
V$_{o}$ vs. Spectral Type & 11.80 & 11.85 & 11.85\\
V$_{o}$ vs. log(T$_{e}$) & 11.80 & 11.85 & 11.85\\
V$_{o}$ vs. V$_{o}$-I$_{o}$ & 11.82 & 11.87 & 11.87\\
V$_{o}$ vs. V$_{o}$-J$_{o}$ & 11.80 & 11.85 & 11.85\\
V$_{o}$ vs. V$_{o}$-H$_{o}$ & 11.79 & 11.84 & 11.84\\
V$_{o}$ vs. V$_{o}$-K$_{o}$ & 11.82 & 11.86 & 11.87\\
\\
average value & 11.80 & 11.85 & 11.85\\
\cutinhead{Literature Estimates}
\citet{Mayne2008} & 11.78 & 11.82 & --\\
\citet{Sl02} & 11.85 & 11.85 &11.08--12.42\\
\citet{Uribe2002} & 11.84 & 11.84&--\\
\citet{Capilla2002} & 11.70 & 11.70&--\\
\citet{Ke01} & 11.75 & 11.75& --\\
\citet{Marco2001} & 11.66 & 11.56&--

\enddata
\label{msfits}
\end{deluxetable}

\clearpage
\begin{deluxetable}{lllllllllll}
 \tiny
\setlength{\tabcolsep}{0.02in}
\tabletypesize{\tiny}
\tablecolumns{8}
\tablecaption{Stellar Age Estimates for h and $\chi$ Persei}
\tiny
\tablehead{{Method}&{Isochrones}&{Cluster Properties Used}&{h Persei age (Myr)}&{$\chi$ Persei age (Myr)}& {Halo age (Myr)}}
\startdata
\cutinhead{This Work}
Main Sequence Turnoff (B stars)& Padova post-MS& V$_{o}$ vs. log(T$_{e}$)/Spectral Type & 13.5 $\pm$ 1.5 & 14 $\pm$ 1 & 13 $\pm$ 1\\
M supergiants & Padova post-MS& V$_{o}$ vs. V-J and log(T$_{e}$) & -- & 13.5 $\pm$ 1.5 & 13.5 $\pm$ 1.5\\
Pre-Main Sequence (FGKM stars) & Baraffe pre-MS& V vs. V-I & 14 +2,-1.4  & 14 +2,-1.4 & 14 +2, -1.4\\
\\
median average value& && 13.75 $\pm$ 1 & 14 $\pm$ 1 & 13.5 $\pm$ 1\\
\cutinhead{Literature Estimates}
\citet{Mayne2008}, Main Sequence Turnoff& && 13 & 13\\
\citet{Sl02}, Main Sequence Turnoff& && 12.8 & 12.9 & 10--20\\

\enddata
\tablecomments{For the Baraffe pre-main sequence isochrones we assumed a mixing length parameter of L$_{p}$ = 1.9 H$_{p}$, which is 
required to reproduce the observed luminosity and temperature of the Sun.}
\label{hxPerage}
\end{deluxetable}

\clearpage
\begin{deluxetable}{lllllllllllllllllll}
 \tiny
\setlength{\tabcolsep}{0.02in}
\tabletypesize{\tiny}
\tablecolumns{12}
\tablecaption{Combined List of h and $\chi$ Persei Members}
\tiny
\tablehead{{Member}&{Photometry}&{Spectroscopy}&{Membership Type}&{RA}&{DEC}&{Numerical}&{V}&{$\sigma$(V)}&{I}&{$\sigma$(I$_{c}$)}\\
{ID Number}&{Running Number}&{Running Number}&{}&{}&{}&{Spectral Type}}
\startdata
1 & 3 & 0 & 1 & 34.7691 & 57.1355 & -99.0000 & 7.1731 & 0.0000 & 6.6876 & 0.0000 \\
2 & 12 & 0 & 1 & 34.7946 & 57.1306 & -99.0000 & 7.7561 & 0.0001 & 7.2290 & 0.0001 \\
3 & 16 & 60 & 3 & 34.6173 & 57.2084 & 11.5000 & 7.7866 & 0.0001 & 7.3575 & 0.0001 \\
4 & 17 & 11 & 3 & 34.5962 & 57.0102 & 10.7000 & 7.8288 & 0.0001 & 7.4510 & 0.0001 \\
5 & 21 & 0 & 1 & 34.7917 & 57.1270 & -99.0000 & 8.1324 & 0.0001 & 7.5498 & 0.0001 \\
\enddata
\tablecomments{The membership type has the following meaning: 1 = Spectroscopic Members, 2 = Photometric Members, 3 = Spectroscopic 
and Photometric Members.  A zero for either the photometry or spectroscopy running number means that the source lacks either 
optical photometric or spectroscopic data.}
\label{memberlist}
\end{deluxetable}

\clearpage
\begin{deluxetable}{lllllllllll}
 \tiny
\tabletypesize{\small}
\tablecolumns{8}
\tablecaption{Effective Temperatures and Optical/Infrared Colors for Dwarfs}
\tiny
\tablehead{{Cluster Properties Used}&{h Persei distance}&{$\chi$ Persei distance} & {Halo distance}}
\tablehead{{ST}&{T$_{e}$}&{U-B}&{B-V}&{V-I$_{c}$}&{V-J}&{V-H}&{V-K}}
\startdata
 O5.0 & 41000 & -1.165 & -0.310 & -0.324 & -0.731 & -0.864 & -0.978 \\
  O5.5 & 39500 & -1.160 & -0.306 & -0.321 & -0.727 & -0.859 & -0.972 \\
  O6.0 & 38250 & -1.154 & -0.302 & -0.320 & -0.724 & -0.856 & -0.968 \\
  O6.5 & 37000 & -1.148 & -0.298 & -0.318 & -0.721 & -0.852 & -0.964 \\
  O7.0 & 36000 & -1.143 & -0.295 & -0.317 & -0.719 & -0.850 & -0.960 \\
  O7.5 & 34750 & -1.137 & -0.291 & -0.315 & -0.716 & -0.846 & -0.956 \\
  O8.0 & 33750 & -1.126 & -0.287 & -0.313 & -0.711 & -0.840 & -0.949 \\
  O8.5 & 32750 & -1.114 & -0.282 & -0.312 & -0.706 & -0.834 & -0.942 \\
  O9.0 & 31750 & -1.103 & -0.278 & -0.310 & -0.701 & -0.828 & -0.935 \\
  O9.5 & 30750 & -1.091 & -0.273 & -0.308 & -0.696 & -0.822 & -0.928 \\
B0.0 & 30000 & -1.078 & -0.269 & -0.303 & -0.686 & -0.809 & -0.914 \\
B0.2 & 28100 & -1.044 & -0.257 & -0.291 & -0.658 & -0.776 & -0.876 \\
B0.5 & 25400 & -0.983 & -0.239 & -0.270 & -0.613 & -0.722 & -0.813 \\
B1.0 & 24150 & -0.954 & -0.231 & -0.260 & -0.591 & -0.697 & -0.784 \\
B1.5 & 21800 & -0.882 & -0.214 & -0.238 & -0.543 & -0.637 & -0.715 \\
B2.0 & 20700 & -0.846 & -0.206 & -0.226 & -0.519 & -0.608 & -0.681 \\
B3.0 & 18700 & -0.771 & -0.186 & -0.201 & -0.466 & -0.546 & -0.612 \\
B4.0 & 17200 & -0.707 & -0.168 & -0.181 & -0.422 & -0.495 & -0.554 \\
B5.0 & 15400 & -0.613 & -0.146 & -0.154 & -0.364 & -0.427 & -0.477 \\
B6.0 & 14100 & -0.526 & -0.126 & -0.134 & -0.316 & -0.371 & -0.414 \\
B7.0 & 13000 & -0.434 & -0.107 & -0.115 & -0.268 & -0.316 & -0.353 \\
B8.0 & 11800 & -0.310 & -0.081 & -0.091 & -0.206 & -0.242 & -0.270 \\
B9.0 & 10700 & -0.179 & -0.047 & -0.061 & -0.134 & -0.158 & -0.176 \\
A0.0 & 9886 & -0.078 & -0.005 & -0.027 & -0.064 & -0.076 & -0.082 \\
A1.0 & 9500 & -0.036 & 0.023 & 0.000 & -0.015 & -0.018 & -0.020 \\
A2.0 & 8970 & 0.017 & 0.069 & 0.044 & 0.061 & 0.074 & 0.076 \\
A3.0 & 8720 & 0.029 & 0.098 & 0.076 & 0.110 & 0.133 & 0.138 \\
A4.0 & 8460 & 0.038 & 0.130 & 0.113 & 0.167 & 0.203 & 0.209 \\
A5.0 & 8200 & 0.028 & 0.167 & 0.160 & 0.238 & 0.291 & 0.298 \\
A7.0 & 7850 & 0.009 & 0.219 & 0.228 & 0.341 & 0.422 & 0.431 \\
A8.0 & 7580 & -0.008 & 0.259 & 0.285 & 0.427 & 0.531 & 0.541 \\
A9.0 & 7390 & -0.020 & 0.289 & 0.327 & 0.493 & 0.617 & 0.628 \\
F0.0 & 7200 & -0.031 & 0.318 & 0.370 & 0.558 & 0.702 & 0.715 \\
F1.0 & 7050 & -0.036 & 0.347 & 0.403 & 0.613 & 0.774 & 0.788 \\
F2.0 & 6890 & -0.041 & 0.377 & 0.440 & 0.671 & 0.851 & 0.867 \\
F3.0 & 6740 & -0.039 & 0.409 & 0.474 & 0.727 & 0.927 & 0.944 \\
F4.0 & 6590 & -0.034 & 0.443 & 0.509 & 0.785 & 1.004 & 1.023 \\
F5.0 & 6440 & -0.020 & 0.480 & 0.545 & 0.845 & 1.084 & 1.105 \\
F6.0 & 6360 & -0.011 & 0.501 & 0.564 & 0.877 & 1.128 & 1.150 \\
F7.0 & 6280 & -0.001 & 0.522 & 0.583 & 0.910 & 1.171 & 1.195 \\
F8.0 & 6200 & 0.015 & 0.544 & 0.603 & 0.944 & 1.217 & 1.242 \\
F9.0 & 6115 & 0.039 & 0.569 & 0.625 & 0.981 & 1.267 & 1.294 \\
G0.0 & 6030 & 0.062 & 0.594 & 0.646 & 1.018 & 1.318 & 1.346 \\
G1.0 & 5945 & 0.086 & 0.620 & 0.667 & 1.056 & 1.368 & 1.398 \\
G2.0 & 5860 & 0.113 & 0.645 & 0.689 & 1.094 & 1.420 & 1.452 \\
G3.0 & 5830 & 0.127 & 0.655 & 0.698 & 1.109 & 1.440 & 1.472 \\
G4.0 & 5800 & 0.140 & 0.665 & 0.706 & 1.123 & 1.460 & 1.493 \\
G5.0 & 5770 & 0.153 & 0.675 & 0.714 & 1.138 & 1.480 & 1.513 \\
G6.0 & 5700 & 0.185 & 0.697 & 0.734 & 1.172 & 1.526 & 1.561 \\
G7.0 & 5630 & 0.216 & 0.720 & 0.753 & 1.206 & 1.572 & 1.609 \\
G8.0 & 5520 & 0.276 & 0.757 & 0.787 & 1.266 & 1.653 & 1.693 \\
G9.0 & 5410 & 0.341 & 0.793 & 0.822 & 1.328 & 1.739 & 1.782 \\
K0.0 & 5250 & 0.435 & 0.847 & 0.874 & 1.419 & 1.863 & 1.911 \\
K1.0 & 5080 & 0.548 & 0.903 & 0.944 & 1.537 & 2.024 & 2.079 \\
K2.0 & 4900 & 0.670 & 0.962 & 1.021 & 1.668 & 2.201 & 2.263 \\
K3.0 & 4730 & 0.783 & 1.021 & 1.105 & 1.804 & 2.383 & 2.453 \\
K4.0 & 4590 & 0.873 & 1.072 & 1.186 & 1.932 & 2.551 & 2.629 \\
K5.0 & 4350 & 0.996 & 1.158 & 1.345 & 2.170 & 2.839 & 2.934 \\
K6.0 & 4205 & 1.047 & 1.209 & 1.457 & 2.328 & 3.014 & 3.121 \\
K7.0 & 4060 & 1.066 & 1.247 & 1.588 & 2.505 & 3.180 & 3.302 \\
M0.0 & 3850 & 1.100 & 1.300 & 1.856 & 2.857 & 3.470 & 3.633 \\
M1.0 & 3720 & 1.168 & 1.356 & 2.000 & 3.048 & 3.631 & 3.824 \\
M2.0 & 3580 & 1.295 & 1.445 & 2.137 & 3.238 & 3.812 & 4.038 \\
M3.0 & 3470 & 1.367 & 1.497 & 2.272 & 3.430 & 3.991 & 4.230 \\
M4.0 & 3370 & 1.435 & 1.543 & 2.412 & 3.632 & 4.183 & 4.430 \\
M5.0 & 3240 & 1.471 & 1.623 & 2.655 & 4.014 & 4.572 & 4.837 \\
\enddata
\label{dwarfcolors}
\end{deluxetable}

\clearpage
\begin{deluxetable}{lllllllllll}
 \tiny
\tabletypesize{\small}
\tablecolumns{8}
\tablecaption{Effective Temperatures and Optical/Infrared Colors for Giants (III)}
\tiny
\tablehead{{Cluster Properties Used}&{h Persei distance}&{$\chi$ Persei distance} & {Halo distance}}
\tablehead{{ST}&{T$_{e}$}&{U-B}&{B-V}&{V-I$_{c}$}&{V-J}&{V-H}&{V-K}}
\startdata
  O5.0 & 41000 & -1.162 & -0.304 & -0.320 & -0.724 & -0.855 & -0.968 \\
  O5.5 & 39500 & -1.162 & -0.304 & -0.320 & -0.724 & -0.855 & -0.968 \\
  O6.0 & 38250 & -1.159 & -0.300 & -0.314 & -0.715 & -0.844 & -0.956 \\
  O6.5 & 37000 & -1.155 & -0.296 & -0.310 & -0.709 & -0.837 & -0.948 \\
  O7.0 & 36000 & -1.149 & -0.293 & -0.310 & -0.709 & -0.837 & -0.948 \\
  O7.5 & 34750 & -1.141 & -0.290 & -0.311 & -0.709 & -0.838 & -0.948 \\
  O8.0 & 33750 & -1.134 & -0.287 & -0.311 & -0.709 & -0.838 & -0.948 \\
  O8.5 & 32750 & -1.126 & -0.284 & -0.309 & -0.706 & -0.834 & -0.943 \\
  O9.0 & 31750 & -1.117 & -0.279 & -0.306 & -0.700 & -0.827 & -0.935 \\
  O9.5 & 30750 & -1.114 & -0.272 & -0.299 & -0.686 & -0.811 & -0.918 \\
B0.0 & 30000 & -1.106 & -0.268 & -0.295 & -0.679 & -0.802 & -0.907 \\
B1.0 & 24500 & -1.043 & -0.232 & -0.247 & -0.589 & -0.695 & -0.785 \\
B2.0 & 21050 & -0.969 & -0.203 & -0.208 & -0.515 & -0.605 & -0.684 \\
B3.0 & 16850 & -0.821 & -0.167 & -0.155 & -0.404 & -0.472 & -0.533 \\
B5.0 & 14800 & -0.715 & -0.142 & -0.116 & -0.325 & -0.378 & -0.429 \\
B7.0 & 13700 & -0.672 & -0.117 & -0.085 & -0.271 & -0.313 & -0.358 \\
B8.0 & 13150 & -0.602 & -0.119 & -0.082 & -0.252 & -0.292 & -0.333 \\
B9.0 & 11731 & -0.488 & -0.092 & -0.044 & -0.175 & -0.199 & -0.233 \\
A0.0 & 10000 & -0.273 & -0.054 & 0.005 & -0.054 & -0.057 & -0.077 \\
A2.0 & 9000 & -0.142 & -0.008 & 0.058 & 0.054 & 0.072 & 0.063 \\
A3.0 & 8500 & -0.064 & 0.024 & 0.091 & 0.122 & 0.154 & 0.152 \\
A5.0 & 8000 & 0.030 & 0.061 & 0.128 & 0.202 & 0.251 & 0.257 \\
A7.0 & 7750 & 0.080 & 0.091 & 0.156 & 0.256 & 0.315 & 0.327 \\
A9.0 & 7450 & 0.143 & 0.130 & 0.198 & 0.332 & 0.407 & 0.426 \\
F0.0 & 7350 & 0.165 & 0.142 & 0.215 & 0.363 & 0.445 & 0.466 \\
F1.0 & 7200 & 0.199 & 0.162 & 0.241 & 0.409 & 0.502 & 0.526 \\
F2.0 & 7050 & 0.233 & 0.190 & 0.271 & 0.461 & 0.566 & 0.594 \\
F3.0 & 6850 & 0.281 & 0.263 & 0.331 & 0.554 & 0.682 & 0.714 \\
F5.0 & 6630 & 0.333 & 0.344 & 0.396 & 0.656 & 0.810 & 0.847 \\
F7.0 & 6330 & 0.404 & 0.455 & 0.486 & 0.795 & 0.984 & 1.028 \\
F8.0 & 6220 & 0.430 & 0.495 & 0.519 & 0.846 & 1.048 & 1.094 \\
F9.0 & 6020 & 0.478 & 0.569 & 0.578 & 0.939 & 1.164 & 1.215 \\
G0.0 & 5800 & 0.563 & 0.668 & 0.652 & 1.056 & 1.315 & 1.369 \\
G1.0 & 5700 & 0.624 & 0.724 & 0.691 & 1.119 & 1.398 & 1.454 \\
G2.0 & 5500 & 0.745 & 0.837 & 0.770 & 1.246 & 1.564 & 1.623 \\
G5.0 & 5200 & 0.926 & 1.006 & 0.887 & 1.437 & 1.814 & 1.877 \\
G8.0 & 4950 & 0.843 & 1.019 & 0.996 & 1.623 & 2.107 & 2.171 \\
K0.0 & 4810 & 0.747 & 0.999 & 1.059 & 1.731 & 2.285 & 2.350 \\
K1.0 & 4585 & 0.875 & 1.075 & 1.189 & 1.936 & 2.556 & 2.633 \\
K2.0 & 4390 & 0.982 & 1.145 & 1.315 & 2.126 & 2.790 & 2.881 \\
K3.0 & 4225 & 1.040 & 1.202 & 1.443 & 2.307 & 2.989 & 3.095 \\
K5.0 & 3955 & 1.076 & 1.271 & 1.706 & 2.659 & 3.310 & 3.449 \\
M0.0 & 3845 & 1.102 & 1.302 & 1.863 & 2.866 & 3.477 & 3.642 \\
M1.0 & 3750 & 1.145 & 1.338 & 1.976 & 3.016 & 3.601 & 3.788 \\
M2.0 & 3655 & 1.224 & 1.396 & 2.060 & 3.131 & 3.710 & 3.918 \\
M3.0 & 3560 & 1.314 & 1.458 & 2.159 & 3.267 & 3.839 & 4.070 \\
M4.0 & 3460 & 1.373 & 1.502 & 2.285 & 3.449 & 4.010 & 4.249 \\
M5.0 & 3355 & 1.445 & 1.549 & 2.433 & 3.662 & 4.212 & 4.460 \\
\enddata
\label{giantcolors}
\end{deluxetable}

\clearpage
\begin{deluxetable}{lllllllllll}
 \tiny
\tabletypesize{\small}
\tablecolumns{8}
\tablecaption{Effective Temperatures and Optical/Infrared Colors for Supergiants (I)}
\tiny
\tablehead{{ST}&{T$_{e}$}&{U-B}&{B-V}&{V-I$_{c}$}&{V-J}&{V-H}&{V-K}}
\startdata
B0.0 & 30000 & -1.078 & -0.269 & -0.303 & -0.686 & -0.809 & -0.914 \\
B0.2 & 26300 & -1.014 & -0.247 & -0.275 & -0.628 & -0.741 & -0.835 \\
B0.5 & 23100 & -0.988 & -0.226 & -0.240 & -0.568 & -0.668 & -0.754 \\
B1.0 & 20260 & -0.907 & -0.208 & -0.212 & -0.508 & -0.598 & -0.673 \\
B1.5 & 19400 & -0.881 & -0.200 & -0.201 & -0.485 & -0.570 & -0.643 \\
B2.0 & 18000 & -0.837 & -0.185 & -0.180 & -0.444 & -0.520 & -0.585 \\
B5.0 & 13600 & -0.628 & -0.127 & -0.094 & -0.276 & -0.319 & -0.363 \\
B8.0 & 11000 & -0.423 & -0.073 & -0.020 & -0.125 & -0.141 & -0.168 \\
A0.0 & 9900 & -0.256 & -0.052 & 0.007 & -0.046 & -0.049 & -0.067 \\
A2.0 & 9000 & -0.118 & -0.017 & 0.048 & 0.049 & 0.066 & 0.057 \\
A3.0 & 8400 & -0.005 & 0.013 & 0.085 & 0.134 & 0.168 & 0.168 \\
A5.0 & 8100 & 0.049 & 0.036 & 0.109 & 0.184 & 0.228 & 0.235 \\
A7.0 & 7800 & 0.117 & 0.063 & 0.142 & 0.250 & 0.308 & 0.320 \\
F0.0 & 7200 & 0.223 & 0.160 & 0.249 & 0.430 & 0.528 & 0.552 \\
F1.0 & 7050 & 0.247 & 0.195 & 0.283 & 0.484 & 0.596 & 0.623 \\
F3.0 & 6770 & 0.285 & 0.272 & 0.354 & 0.594 & 0.733 & 0.765 \\
F5.0 & 6570 & 0.302 & 0.349 & 0.417 & 0.689 & 0.855 & 0.890 \\
F7.0 & 6280 & 0.345 & 0.459 & 0.505 & 0.827 & 1.033 & 1.073 \\
F8.0 & 6180 & 0.360 & 0.498 & 0.536 & 0.874 & 1.095 & 1.136 \\
F9.0 & 5980 & 0.422 & 0.577 & 0.596 & 0.967 & 1.212 & 1.259 \\
G0.0 & 5590 & 0.542 & 0.757 & 0.735 & 1.191 & 1.513 & 1.565 \\
G1.0 & 5490 & 0.592 & 0.807 & 0.771 & 1.250 & 1.593 & 1.645 \\
G2.0 & 5250 & 0.734 & 0.932 & 0.858 & 1.398 & 1.791 & 1.849 \\
G5.0 & 5000 & 0.949 & 1.080 & 0.953 & 1.565 & 2.012 & 2.076 \\
G8.0 & 4700 & 1.198 & 1.232 & 1.066 & 1.784 & 2.314 & 2.390 \\
K0.0 & 4500 & 1.426 & 1.336 & 1.158 & 1.958 & 2.549 & 2.640 \\
K1.0 & 4200 & 1.816 & 1.493 & 1.332 & 2.262 & 2.956 & 3.072 \\
K2.0 & 4100 & 1.949 & 1.549 & 1.414 & 2.392 & 3.125 & 3.252 \\
K7.0 & 3840 & 2.180 & 1.686 & 1.695 & 2.829 & 3.664 & 3.821 \\
M0.0 & 3790 & 2.225 & 1.719 & 1.758 & 2.927 & 3.781 & 3.945 \\
M1.0 & 3745 & 2.182 & 1.726 & 1.838 & 3.052 & 3.923 & 4.092 \\
M2.0 & 3660 & 1.858 & 1.686 & 2.082 & 3.436 & 4.341 & 4.513 \\
M3.0 & 3605 & 1.616 & 1.651 & 2.245 & 3.690 & 4.616 & 4.789 \\
M5.0 & 3450 & 1.167 & 1.584 & 2.716 & 4.472 & 5.438 & 5.625 \\
\enddata
\label{sgiantcolors}
\end{deluxetable}

\clearpage
\clearpage
\clearpage
\clearpage

\clearpage
\clearpage

\clearpage
\clearpage

\clearpage


\clearpage


\appendix
\section{Effective Temperature Scale for Dwarfs, Giants, and Supergiants}
Here we describe our adopted effective temperature scales for dwarfs (luminosity class V), 
giants (class III), and supergiants (class I).  As a starting point, we considered the 
effective temperature scales from \citet{Gray2009} because it 
is the most recent and comprehensive; \citet{Bessell1998} and \citet{Kh95}\footnote{Though \citet{Kh95} 
identify \citet{Bessell1988} as a primary source for dwarf stars temperatures and colors, it appears 
that their T$_{e}$ scale is drawn from \citet{Bessell1979} instead, which in turn adopts 
the interferometrically-determined scale from \citet{Code1976}.} because they are
 the most widely cited primary references from the past 15 years; \citet{Humphreys1984} because it is 
also well cited and provides the best sampling with spectral type based on a large 
dataset; and \citet{deJager1987} because it also is a well-cited paper with 
good sampling that treats the spectral type as a continuous, not discrete, variable.

Secondary references provided independent checks on and modifications to these primary 
references.  These references derive temperature scales in one of two 
ways: 1) estimating temperatures using non-LTE atmospheric models \citep[][]{Massey2005, 
Levesque2005} and 2) deriving a purely empirical scale based on 
fitting stellar atmosphere models to synthetic MK spectral standards \citep{Gray1994}.
Where the primary sources disagree, we analyze a range of computed scales from the above 
secondary sources as well as classic references such as \citet{SchmidtKaler1982} and 
\citet{BohmVitense1981} to arrive at a final calibrated scale.

\subsection{Dwarfs}
For O5--B0 dwarfs, sophisticated non-LTE calculations are crucial to accurately determining 
T$_{e}$.  Therefore, we adopt the empirically-derived scale from \citet{Massey2005} for 
these stars.  The \citet{Massey2005} scale departs from the \citet{Gray2009} scale by less
than $\approx$ 500 K for nearly all subtypes.  Because O5--B0 stars have temperatures $\gtrsim$ 
30,000 K, these differences are inconsequential ($\lesssim$ 1.7\%).

For B0.2--A0 stars, there is generally good agreement between \citet{Gray2009}, 
\citet{Bessell1998}, \citet{Humphreys1984} and \citet{deJager1987} for most subclasses.  However, there are 
serious disagreements between these authors for B1--B3 stars; these stars are crucial for 
accurately assessing cluster properties (e.g. the main sequence turnoff) because they 
cover a wide range in V band magnitudes for our sample.  In particular, the \citet{Bessell1998} 
scale is up to 2500 K hotter than the others.  The origin of \citeauthor{Bessell1998}'s scale 
is in a IAU conference proceedings paper, \citet{Crowther1997}.  Figure 1 of \citet{Crowther1997} 
shows that its temperature scale is systematically higher for B0.5--B2 stars than 
other references listed as well as older scales from \citet{SchmidtKaler1982} and \citet{BohmVitense1981}.
  Moreover, the B0.5--B1.5 scale from  
\citet{Humphreys1984}, \citet{deJager1987}, \citet{Gray2009}, and \citet{Gray1994}
 agree against \citet{Crowther1997} by \textit{nearly identical amounts}. 

Given this information, 
we adopt the scale from \citet{Humphreys1984} sampling in spectral type from B0.2 to B1.5.
Similarly for B3 stars, all authors except for \citet{Gray2009} agree.  In this case, 
we adopt the values from \citet{Humphreys1984} and \citet{Kh95}.
For the B2 spectral type, there appears to be complete disagreement with two primary references 
(\citealt{Bessell1998} and \citealt{Kh95}) listing hot temperatures ($\sim$ 21700 K) and 
two listing cooler temperatures ($\sim$ 19500--19700 K; \citealt{Gray2009} and \citealt{Humphreys1984}).  
The average of these values is $\approx$ 20700 K, which is also very close to the value given 
by \citet{deJager1987} so we adopt it.

The temperature scale for B4--A0 stars shows strong agreement between various authors.  
We simply adopt a scale from \citet{Gray2009}, \citet{Bessell1998}, and \citet{Kh95} for 
these spectral types, taking the \citet{Kh95} value by default unless the other two 
disagree against \citet{Kh95} by more than several 100 K.  

For dwarf stars later than A0, only three of our primary references have published values: \citet{Gray2009}, 
\citet{Kh95}, and \citet{deJager1987}.  With the exception of A5--A7 stars, these references show excellent agreement.  
We adopt the \citet{Gray2009} values by default and add the \citet{Kh95} for spectral types 
where \citet{Gray2009} lacks entries.

\subsection{Giants and Supergiants}
As with the dwarf star T$_{e}$ scale, we adopt results from \citet{Massey2005} for all O5--B0 giants 
and supergiants.  For B1--M0 stars, the primary references listing T$_{e}$ vs. spectral type for evolved stars are \citet{deJager1987}, 
\citet{Humphreys1984}, and \citet{Gray2009}.  
  The references show good agreement for stars later than $\sim$ B4; small disagreements 
again show up in the B1--B3 range.  For these stars we adopt the middle value if all three references 
disagree and the more frequent value if two of the three agree against the third.

The T$_{e}$ scale for M supergiants is notoriously hard to calibrate: values from different 
authors often wildly disagree.  To complicate matters, most T$_{e}$ scales yield red supergiant 
that are far redder than post-main sequence isochrones would allow, preventing 
the stars' luminosities and colors from being used to estimate cluster ages.  The most recent calibrations 
from \citet{Levesque2005} substantially revise the temperature scale upward; the post-main sequence 
isochrones easily extend to these new temperatures.  Even though we do not consider \citet{Levesque2005} 
to be a primary reference, we adopt their T$_{e}$ scale for M supergiants almost verbatim for several reasons.  First, 
the data are drawn from high signal-to-noise spectrophotometry of galactic supergiants and compared to 
the NMARCS stellar atmosphere models, which we consider to be robust.  Second, the lead author (R. Humphreys) 
of the primary reference for the T$_{e}$ scale that \citeauthor{Levesque2005}'s replaces refereed the latter 
paper, which increases our confidence that Levesque's scale is an improvement.  Where \citet{Levesque2005}'s sampling 
becomes sparser (e.g. earlier than M0), we adopt the values listed in \citet{Gray2009}, which is are 
drawn from other recent work whose calibration agrees with \citet{Levesque2005}'s where they overlap.

In Tables , we list our entire T$_{e}$ scale along with the Johnson-Cousins-Glass colors drawn from 
the sources discussed in Section 2.  We consider this Table to be an updated and 
more expansive version of similar tables presented by other authors \citep[e.g.][]{Kh95, Bessell1990, 
Bessell1988, Bessell1998, Bessell1979, deJager1987, Humphreys1984, Gray2009}.
\subsection{Further Discussion}
Though our T$_{e}$ scale is not drawn from a uniform sample, we consider it to be robust as it 
is less susceptible to small measurement errors or biases unique to a given paper.  
Moreover, we can identify general trends in different authors' T$_{e}$ calibrations by 
comparing their results to our table.  For example, where we determine \citet{Gray2009} to be 
less accurate, they always predicts cooler temperatures than the ones we adopt; when \citet{Bessell1998} 
is the outlier they almost always predicts hotter temperatures.  Interestingly, \citet{deJager1987} appears 
to be the most accurate source as the majority of other references almost never disagree against it.

Our adopted scales slightly differ from that from \citet{Sl02}.  \citet{Sl02} used \citet{Kilian1991} for 
all stars earlier than B3 and \citet{Humphreys1984} for everything else, a scale that yields 
higher temperatures for early B stars.  If we adopt the calibration used by \citet{Sl02}, our 
derived reddening is slightly higher and MS turnoff ages are younger by $\approx$ 1 Myr.  However, adopting this 
calibration lead to a very slight but perceptible systematic offset in the dereddened V vs. V-J,H, K and 
V vs. spectral type loci of cluster stars.  It also may induce a small but systematic shift in reddening 
for B0--B3 stars compared to later-type stars.  

Recent studies determining effective temperatures for individual B stars from stellar atmosphere modeling 
indicates that our cooler T$_{e}$ scale is more accurate.  In particular, more recent non-LTE stellar atmosphere 
modeling by Crowther, the source of the hotter T$_{e}$ scale for evolved stars used in \citet{Bessell1998}, 
revises the T$_{e}$ for supergiants downwards to almost complete agreement with our adopted scale \citep{Crowther2006}.  The only clear 
systematic difference between our two scales is that \citeauthor{Crowther2006}'s temperatures for B0.5--B1I supergiants are
$\sim$ 2000--2500 K hotter than our entries.   The temperatures are already very hot ($\sim$ 25000 K).  Our 
sample includes only 7 supergiants with spectral types between B0.4 and B1: B1.5--B3 supergiants are far more frequent.  
Therefore, if Crowther's scale is more correct the disagreement should have only a minimal impact on our analysis.   

Other recent determinations for less evolved stars support our adopted scale.
Specifically, \citet{Fitzpatrick2005} determined fundamental 
parameters -- e.g. T$_{e}$, [m/H], R/R$_{\odot}$, and log(\textit{g}) -- for galactic B and early A dwarfs and giants.  
Their derived temperatures for early B stars agree with our adopted scale against the determinations adopted by
\citet{Sl02}.  If anything, our scale is still too hot for B3 V stars by $\approx$ 1000 K.  In summary, we consider our T$_{e}$ scale to be a 
slight improvement over \citet{Sl02}'s, though they used the best available scale at the time, which can 
clearly yield good estimates for h and $\chi$ Per properties.

\section{The metallicity of h and $\chi$ Persei: Previous Estimates, Uncertainties, and Its Effect on Distance Moduli and 
Stellar Ages}

There is significant disagreement over the clusters' metallicities in the literature.  Some authors find
that h and $\chi$ Per stars have a solar metallicity \citep[e.g.][]{Dufton1990, Smartt1997}, while 
 others finding a subsolar metallicity \citep[Z = 0.01][]{Southworth2004,Southworth2005}.  
The \citet{Southworth2004} results present a challenge to our assumption that h and $\chi$ Persei have a near-solar 
metallicity, and we now directly address this issue by explaining why a near-solar metallicity (Z = 0.019) is more likely 
than a subsolar one (Z = 0.01).

\citet{Southworth2004} and \citet{Southworth2005} derive the metallicity for two h Persei eclipsing binary systems, V615 Per and V618 Per, 
and one $\chi$ Persei star (V621 Per) by comparing the derived masses and radii for each binary component with predictions from Granada isochrones 
\citep[][]{Claret1995}.  However, comparisons with the Padova isochrone predictions yield a 
metallicity for 2 of the 4 h Persei stars that is slightly higher than Z=0.01.
 \citet{Venn2002} find that V621 Per has a nearly solar abundance of metals.
In support of their conclusion that h and $\chi$ Persei have a subsolar metallicity, \citeauthor{Southworth2004} 
cites non-LTE modeling of B giants/supergiants by \citet{Vrancken2000}, which yield similar results.  However, 
 \citet{Vrancken2000} note that non-LTE modeling of evolved B stars \textit{systematically} yield low metallicities: it is not clear whether 
whether this difference is physical or whether it reflects uncertainties in the modeling assumptions.

\begin{figure}
\centering
\epsscale{1}
\plottwo{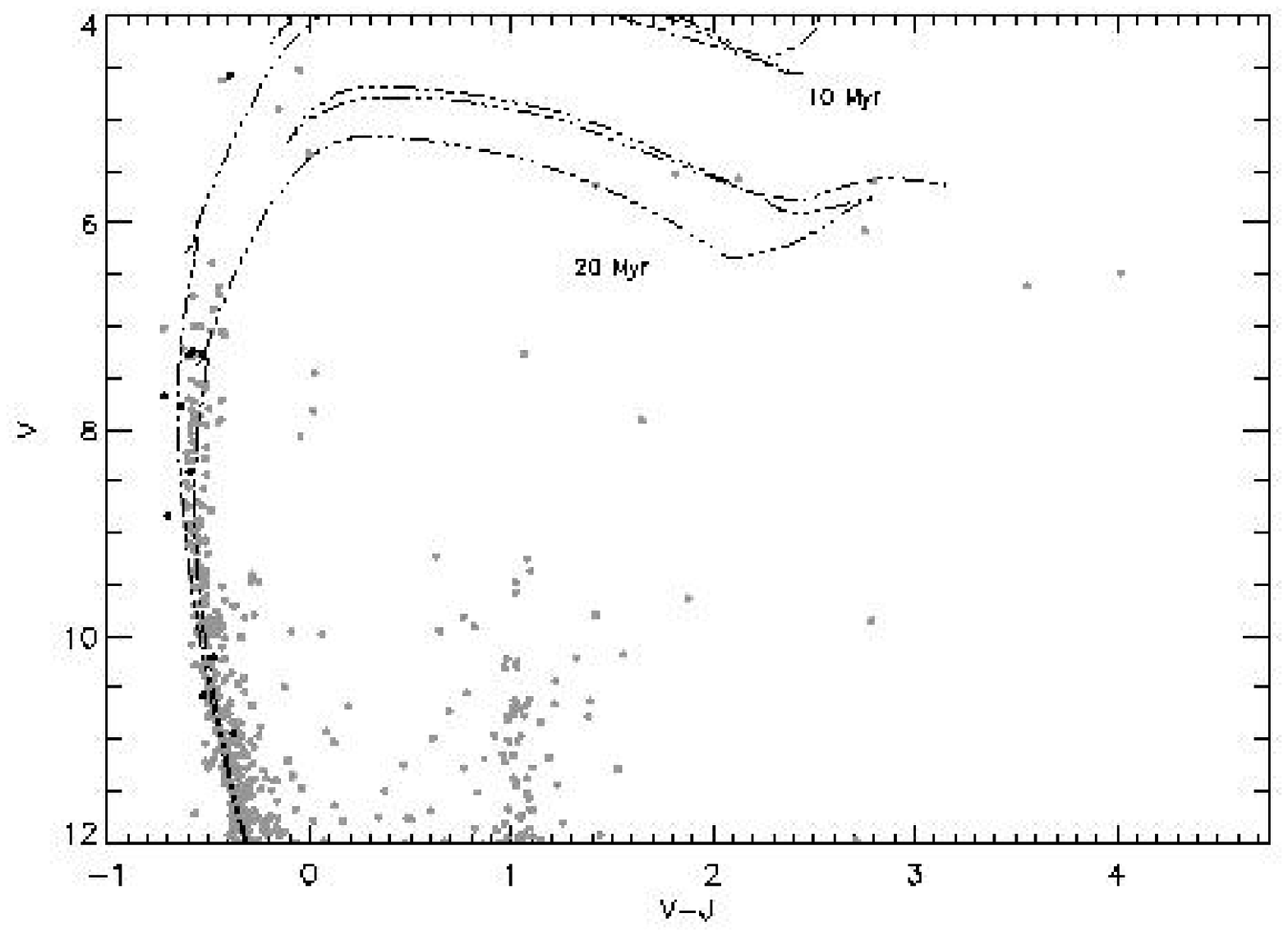}{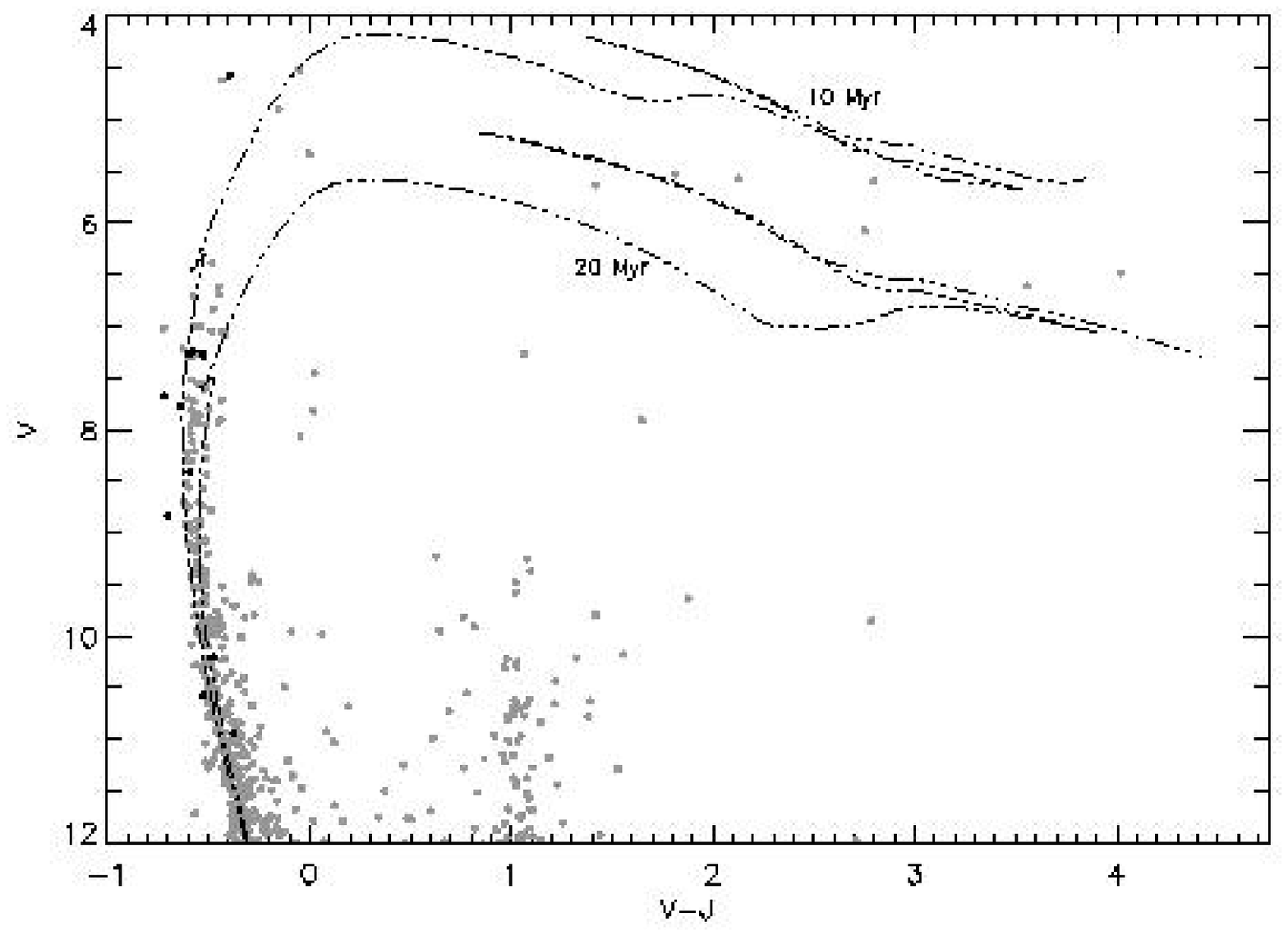}
\plottwo{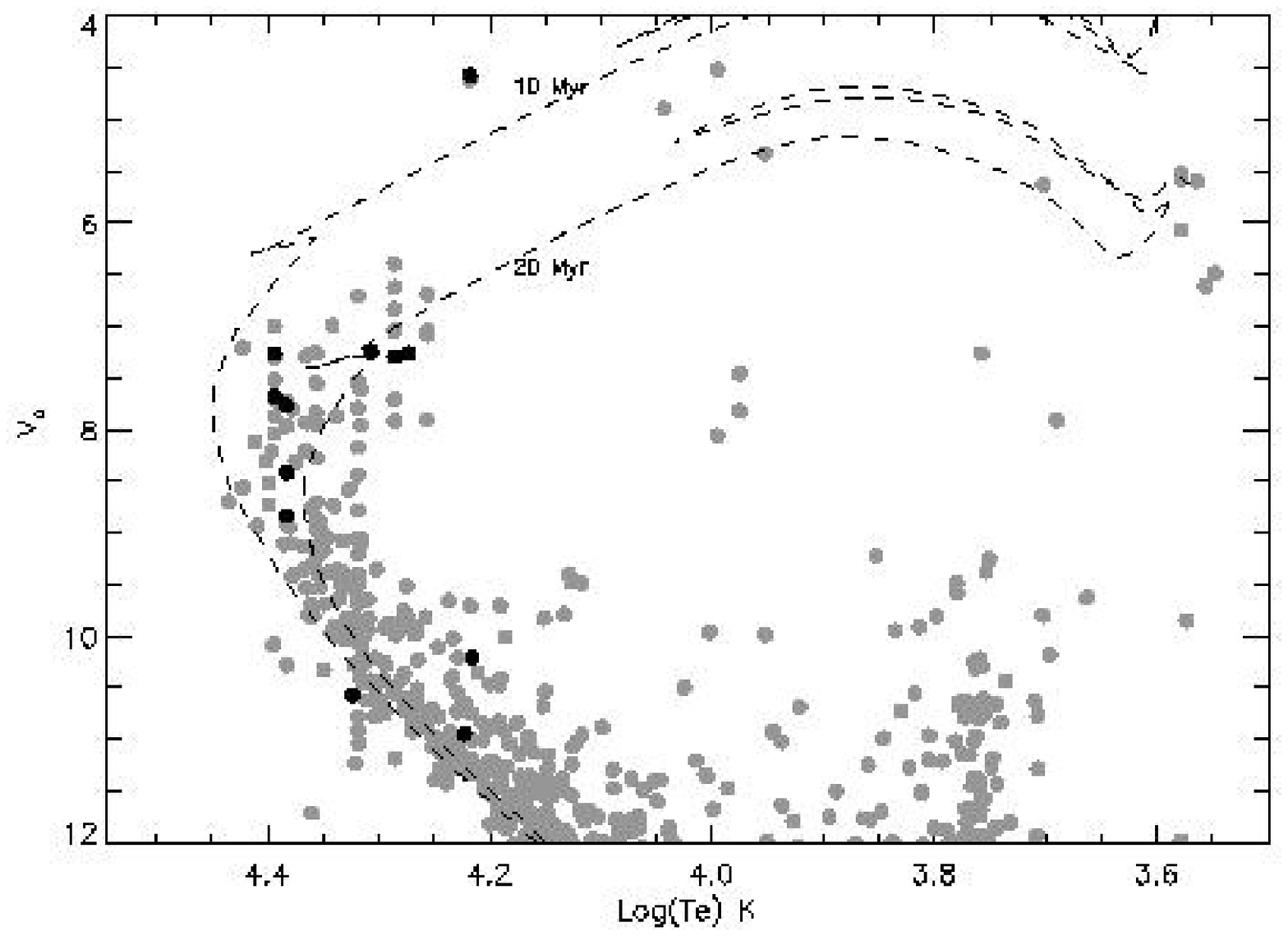}{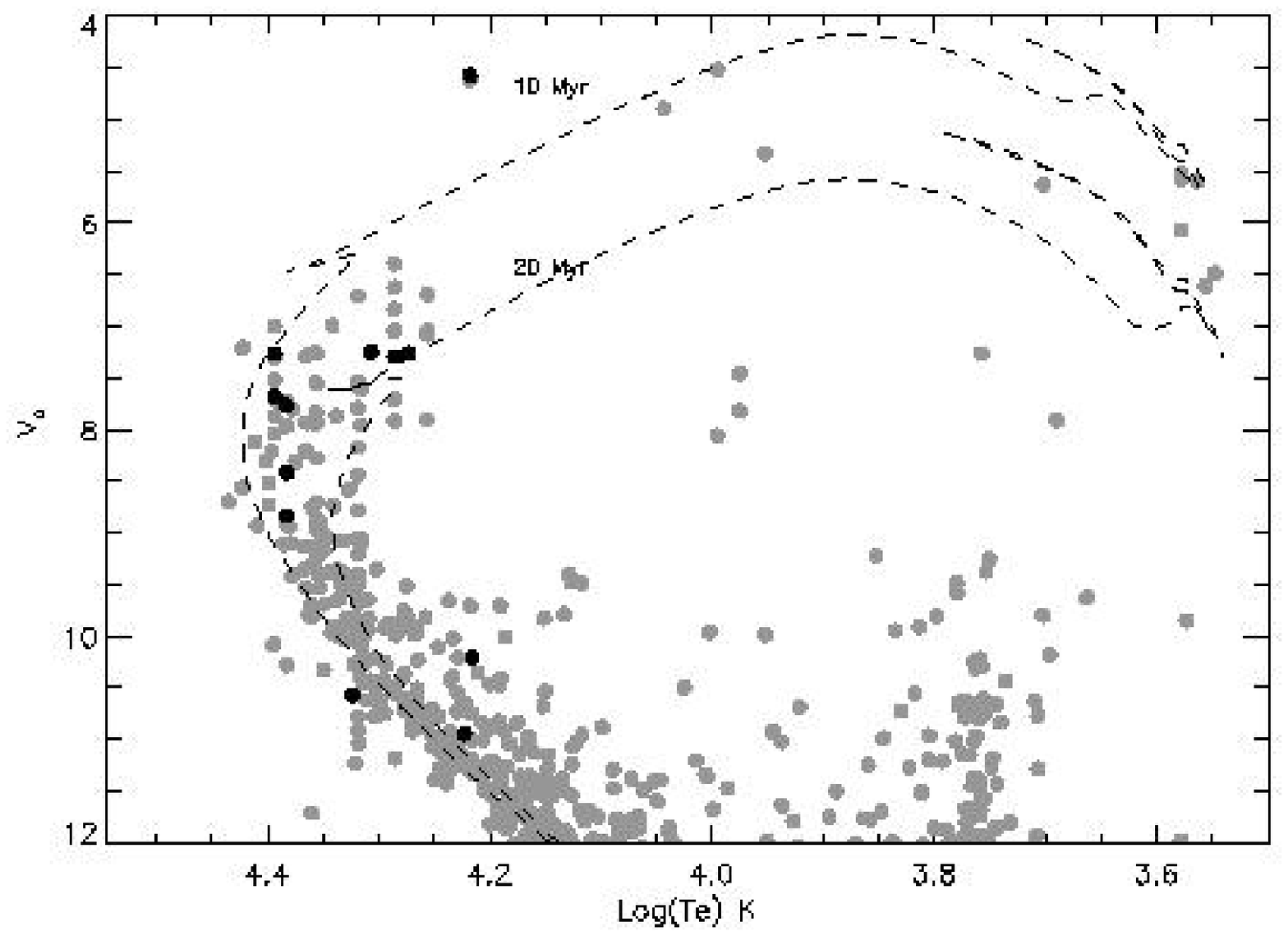}
\caption{V vs. V- J (top panels) and V vs. log(T$_{e}$) (bottom panels) diagrams for all stars with spectra 
and V band photometry, illustrating how subsolar metallicity (left panels) and solar metallicity 
(right panels) isochrones compare to the colors and temperatures of red supergiants.}
\label{metalcompare}
\end{figure}

\begin{figure}
\centering
\plotone{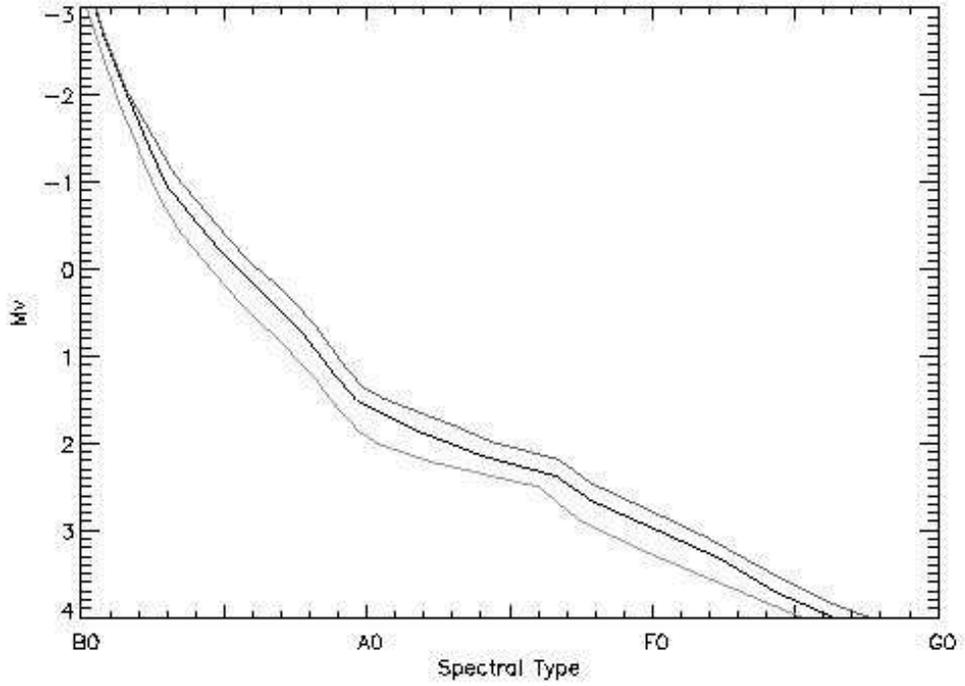}
\caption{Comparison between Padova isochrones with a subsolar, solar, and supersolar metallicity.  In order for 
the main sequence to line up, the distance modulus for the subsolar metallicity isochrone must be decreased by 
0.25 mags.}
\label{metalcompare2}
\end{figure}

The most convincing argument that h and $\chi$ Persei stars typically have a near-solar metallicity comes from analyzing the colors and luminosities 
of their M supergiants.  Figure \ref{metalcompare} shows V vs. log(T$_{e}$) and V vs. V-J diagrams for Z=0.01 isochrones.  The best-fit distance moduli
are systematically smaller by $\approx$ 0.25 mags and range from 11.55 for h Persei to 11.60 for 
$\chi$ Persei and the halo population (Figure \ref{metalcompare2}).  As the Figure \ref{metalcompare} panels clearly show, subsolar metallicity isochrones are 
unable to produce M supergiants with observed red V-J colors and inferred cool effective 
temperatures.  In particular, \textit{none} of the subsolar metallicity isochrones 
are able to reproduce the observed effective temperatures of most M supergiants.  The V-J colors of at least two and perhaps four M supergiants 
are too red for the isochrones by up to 1.5 magnitudes.  In contrast, the solar metallicity isochrones 
are clearly able to produce stars with these properties.  Based on these comparisons, h and $\chi$ Persei is unlikely to have a 
substantially subsolar metallicity.

Even if h and $\chi$ Per did have Z $\sim$ 0.01 as argued by \citet{Southworth2004}, the resulting changes in age estimates are small.  
As can be inferred from Figure \ref{metalcompare}, even the Z = 0.01 isochrones show that h and $\chi$ Persei's main sequence turnoff age is clearly 
older than 10 Myr and younger than 20 Myr.  In fact, the best-fit turnoff age is only $\approx$ 1--2 Myr greater because the overall 
lower luminosity of metal-poor stars is partially offset by their smaller distance moduli.  Thus, the systematic errors in age due to
h and $\chi$ Persei's metallicity are $\lesssim$ 10\% of the clusters' ages.
\clearpage


\end{document}